%% file: main.tex
\documentclass[runningheads]{llncs}
\pdfoutput=1

\input{preamble}

\begin{document}
\title{Writing \acrlong{IOT} applications with \acrlong{TOP}}
\author{Mart Lubbers\orcidID{0000-0002-4015-4878}\and Pieter Koopman\orcidID{0000-0002-3688-0957}\and Rinus Plasmeijer}
\institute{%
	Radboud University, Netherlands\\
	\email{\{mart,pieter,rinus\}@cs.ru.nl}
}

\maketitle

\begin{abstract}
	\input{abstract}
\end{abstract}

\section{Introduction}%
\label{sec:introduction}
\input{1introduction}

\section{\glsentrytext{mTask} system architecture}%
\label{sec:mtask}
\input{2mtask}

\section{\acrshort{IOT} applications with \acrshort{TOP}}%
\label{sec:building}
\input{3building}

\section{Related work}%
\label{sec:related}
\input{4related}

\section{Discussion}%
\label{sec:conclusion}
\input{5conclusion}

\section*{Acknowledgements}
\input{acknowledgements}

\newpage
\appendix%

\section{\Acrlong{EDSL} techniques}%
\label{sec:edsl}
\input{aedsl}

\clearpage
\section{\glsentrytext{iTasks} reference}%
\label{sec:itasks}
\input{aitasks}

\clearpage
\section{How to Install}%
\label{sec:install}
\input{ainstall}

\clearpage
\section{Solutions}%
\label{sec:solutions}
\input{asolutions}

\bibliographystyle{splncs04}
\bibliography{refs}

\end{document}

%% file: preamble.tex
\usepackage[british]{babel}

\usepackage{graphicx}
\graphicspath{{img/}}

\usepackage{textcomp}
\usepackage{gensymb}

\usepackage{booktabs}

\usepackage{colortbl}

\usepackage{threeparttable}

\usepackage{tikz}

\usepackage{stmaryrd}

\usepackage[os=win]{menukeys}

\usepackage{pifont}
\newcommand{\cmark}{\ding{51}}

\usepackage{subcaption}
\usepackage{url}
\newcommand{\fmturl}[1]{\url{#1}}

\usepackage[acronym]{glossaries}
\input{glossaries}

\usepackage{listings}
\lstdefinelanguage{Clean}{%
	alsoletter={ABCDEFGHIJKLMNOPQRSTUVWXYZabcdefghijklmnopqrstuvwxyz_`1234567890},
	alsoletter={~!@\#$\%^\&*-+=?<>:|\\.},
	morekeywords={generic,implementation,definition,dynamic,module,import,from,where,in,of,case,let,infix,infixr,infixl,class,instance,with,if,derive,code,In},
	escapeinside={[+}{+]},
	morecomment=[l]{//},
	morecomment=[n]{/*}{*/},
	morestring=[b]",
	morestring=[s]{['}{']},
	literate=%
		{\\}{{$\lambda\:$}}1
		{->}{{$\shortrightarrow$}}2
		{<-}{{$\shortleftarrow$}}2
		{A.}{{$\forall\;\,$}}1
		{E.}{{$\exists\;\,$}}1
		{'}{{`}}1
		{not}{{$\neg$}}1
}
\newcommand{\Cl}[1]{\lstinline[language=Clean,postbreak=];#1;}

\newcounter{lstex}
\newcounter{lstdef}
\makeatletter
\lstnewenvironment{lstexample}[1][]
	{%
		\lstset{frame=lb, #1}%
		\let\c@lstlisting=\c@lstex%
	}
	{}
\lstnewenvironment{lstdefinition}[1][]
	{%
		\lstset{frame=L, #1}%
		\let\c@lstlisting=\c@lstdef%
	}
	{}
\lstnewenvironment{lstsolution}[1][]{%
		\lstset{frame=single, #1}
	}
	{}
\makeatother

%% file: glossaries.tex
\newacronym{CEFP}{CEFP}{Central European Functional Programming School}
\newacronym{CPM}{\texttt{cpm}}{Clean Project Manager}
\newacronym{I2C}{I\textsuperscript{2}C}{Inter-Intergrated Circuit}
\newacronym{IDE}{IDE}{Integrated Development Environment}
\newacronym{ADT}{ADT}{Algebraic Datatype}
\newacronym{API}{API}{Application Programming Interface}
\newacronym{ARDSL}{ARDSL}{\glsentrytext{Arduino} \acrlong{DSL}}
\newacronym{ARM}{ARM}{Acorn \acrlong{RISC} Machine}
\newacronym{BLE}{BLE}{\acrlong{BT} Low Energy}
\newacronym{BT}{BT}{Bluetooth}
\newacronym{C2}{C2}{Command \& Control}
\newacronym{CPS}{CPS}{Continuation Passing Style}
\newacronym{DHT}{DHT}{Digital Humidity and Temperature sensor}
\newacronym{DSL}{DSL}{Domain Specific Language}
\newacronym{EDSL}{EDSL}{embedded \acrlong{DSL}}
\newacronym{FRP}{FRP}{Functional Reactive Programming}
\newacronym{GADT}{GADT}{Generalized \glsentrytext{ADT}}
\newacronym{GPIO}{GPIO}{General Purpose \acrlong{IO}}
\newacronym{IOT}{IoT}{Internet of Things}
\newacronym{IO}{IO}{Input/Output}
\newacronym{IP}{IP}{Internet Protocol}
\newacronym{LEAN}{LEAN}{Language of East-Anglia and Nijmegen}
\newacronym{LED}{LED}{Lighting Emitting Diode}
\newacronym{MCU}{MCU}{Microcontroller Unit}
\newacronym{OLED}{OLED}{Organic \acrlong{LED}}
\newacronym{OS}{OS}{Operating System}
\newacronym{OTA}{OTA}{Over the Air}
\newacronym{RAM}{RAM}{Random Access Memory}
\newacronym{RFID}{RFID}{Radio-frequency Identification}
\newacronym{RISC}{RISC}{Reduced Instruction Set Computer}
\newacronym{RTOS}{RTOS}{Real-Time \acrlong{OS}}
\newacronym{RTS}{RTS}{Run-time System}
\newacronym{SDS}{SDS}{Shared Data Source}
\newacronym{SN}{SN}{Sensor Network}
\newacronym{TCP}{TCP}{Transmission Control Protocol}
\newacronym{TOP}{TOP}{Task Oriented Programming}
\newacronym{USB}{USB}{Universal Serial Bus}
\newacronym{UDS}{UDS}{Uniform Data Source}
\newacronym{VM}{VM}{Virtual Machine}

\newglossaryentry{AVR}{name=AVR, description={is an architecture for microprocessors used for example in the popular \glsentrytext{Arduino} UNO}}
\newglossaryentry{Arduino}{name=Arduino, description={is an ecosystem and a \glsentrytext{C++} dialect for many different types of \acrlongpl{MCU}}}
\newglossaryentry{C++}{name=C++, description={is an imperative and object oriented low level programming language compatible with \glsentrytext{C} but supporting much more such as classes.}}
\newglossaryentry{Clean}{name=Clean, description={Clean \acrlong{LEAN}, a pure lazy functional programming language based on graph rewriting}}
\newglossaryentry{C}{name=C, description={is an imperative low level system programming language}}
\newglossaryentry{ESP8266}{name=ESP8266, description={is a WiFi chip that can be used as a general purpose microcontroller as well}}
\newglossaryentry{Erlang}{name=Erlang, description={is a concurrent scalable functional programming language designed for soft real-time systems}}
\newglossaryentry{Haskell}{name=Haskell, description={Haskell, a pure lazy functional programming language}}
\newglossaryentry{Java}{name=Java, description={is an imperative and object oriented programming language}}
\newglossaryentry{LOLIN}{name=LOLIN, description={is a prototyping \acrlong{MCU} based on the popular \glsentrytext{ESP8266} chip}}
\newglossaryentry{LUA}{name=LUA, description={is a lightweight high-level scripting language designed for embedded use}}
\newglossaryentry{ML}{name=ML, description={is a functional programming language }}
\newglossaryentry{NodeMCU}{name=NodeMCU, description={is a prototyping \acrlong{MCU} based on the popular \glsentrytext{ESP8266} chip}}
\newglossaryentry{RaspberryPi}{name={Raspberry Pi}, description={is a popular cheap \gls{ARM} single board computer created for education}}
\newglossaryentry{Scheme}{name=Scheme, description={is a functional programming language in the \glsentrytext{ML} family}}
\newglossaryentry{WiFi}{name=WiFi, description={is an open standard for wireless networking}}
\newglossaryentry{ZigBee}{name=ZigBee, description={is an open standard for short-range wireless networking designed to complement \glsentrytext{WiFi} and \acrlong{BT}}}
\newglossaryentry{iTasks}{name=iTask, description={is a \acrlong{TOP} implementation hosted in the purely lazy functional programming language \glsentrytext{Clean}}}
\newglossaryentry{mTask}{name=mTask, description={is an \acrlong{EDSL} for running tasks on \acrlongpl{MCU}}}
\newglossaryentry{xtensa}{name=xtensa, description={is a configurable processor microprocessor core and architecture used for example on the popular \glsentrytext{ESP8266} chip}}

\foreach\x in {API, ARM, I2C, IDE, IP, LED, OS, OLED, TCP, USB, RAM, RISC} {\glsunset{\x}}

%% file: abstract.tex
The \gls{IOT} is growing fast.
In 2018, there was approximately one connected device per person on earth and the number has been growing ever since.
The devices interact with the environment via different modalities at the same time using sensors and actuators making the programs parallel.
Yet, writing this type of programs is difficult because the devices have little computation power and memory, the platforms are heterogeneous and the languages are low level.
\gls{TOP} is a declarative programming language paradigm that is used to express coordination of work, collaboration of users and systems, the distribution of shared data and the human-computer interaction.
The \gls{mTask} language is a specialized, yet full-fledged, multi-backend \gls{TOP} language for \gls{IOT} devices.
With the bytecode interpretation backend and the integration with \gls{iTasks}, tasks can be executed on the device dynamically.
This means that --- according to the current state of affairs --- tasks can be tailor-made at run time, compiled to device-agnostic bytecode and shipped to the device for interpretation.
Tasks sent to the device are fully integrated in \gls{iTasks} to allow every form of interaction with the tasks such as observation of the task value and interaction with \glspl{SDS}.
The entire \gls{IOT} application --- both server and devices --- are programmed in a single language, albeit using two \glspl{EDSL}.

%% file: 1introduction.tex
\subsection{\acrlong{IOT}}
The \gls{IOT} is growing rapidly and it is changing the way people and machines interact with the world.
The term \gls{IOT} was coined around 1999 to describe the communication of \gls{RFID} devices.
\gls{RFID} became more and more popular the years after but the term \gls{IOT} was not associated with it anymore.
Years later, during the rise of novel network technologies, the term \gls{IOT} resurged with a slightly different meaning.
Today, the \gls{IOT} is the term for a system of devices that sense the environment, act upon it and communicate with each other and the world.
At the time of writing, there is about one connected device per person in the world of which many are part of an \gls{IOT} system.
Gartner estimates that of these connected devices, there are about $5.8$ billion \gls{IOT} devices or endpoints connected~\footnote{Gartner (August 2019)}.
They are already in everyone's household in the form of smart electricity meters, smart fridges, smartphones, smart watches, home automation and in the form of much more.
While the number of devices seems to be growing exponentially fast, programming \gls{IOT} applications is difficult.
The devices are a large heterogeneous collection of different platforms, protocols and languages resulting in impedance problems.

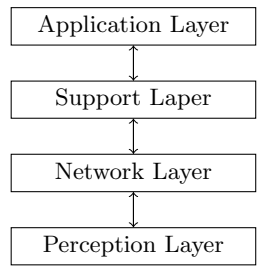
\begin{figure}
	\centering
	\begin{tikzpicture}[node distance=3em]
		\node (1) [rectangle,draw,minimum width=10em] {Application Layer};
		\node (2) [rectangle,draw,minimum width=10em,below of=1] {Support Laper};
		\node (3) [rectangle,draw,minimum width=10em,below of=2] {Network Layer};
		\node (4) [rectangle,draw,minimum width=10em,below of=3] {Perception Layer};

		\draw [->] (1) -- (2);
		\draw [->] (2) -- (1);
		\draw [->] (2) -- (3);
		\draw [->] (3) -- (2);
		\draw [->] (3) -- (4);
		\draw [->] (4) -- (3);
	\end{tikzpicture}
	\caption{The four layered \gls{IOT} architecture as described by the ITU-T.}%
	\label{fig:layers}
\end{figure}

The devices in \gls{IOT} systems are equipped with various sensors and actuators.
These range from external ones such as positioning, temperature and humidity to more internal ones like heartbeat and respiration~\cite{da_xu_internet_2014}.
When describing \gls{IOT} systems, a layered architecture is often used to compartmentalize the technology.
For the intents and purposes of this paper the four layer architecture defined by ITU-T (International Telecommunications Union {-} Telecommunication Standardization Sector) will be used as visualized in Figure~\ref{fig:layers}.

The first layer is called the perception layer and contains the actual endpoints with their peripherals.
For example in home automation, the sensors reading the room and the actuators opening the curtains are in the perception layer.
As a special type of device, it may also contain a \gls{SN}.
A \gls{SN} is a collection of sensors connected by a mesh network or central hub.
The network layer is the second layer and it consists of the hardware and software to connect the perception layer to the world.
In home automation, this layer may consist of a specialized \gls{IOT} technology such as \gls{BLE} or \gls{ZigBee} but it may also use existing technologies such as WiFi or wired connections.
The third layer is named support layer and is responsible for the servicing and business rules surrounding the application.
One of its goals is to provide the \gls{API}, interfaces and data storage.
In home automation this provides the server storing the data.
The fourth and final layer in this architecture is the application layer.
The application layer provides the interaction between the user and the \gls{IOT} system.
In home automation, this layer contains the apps for to read the measurements and control the devices.

The perception layer often is a heterogeneous collections of microcontrollers, each having their own peculiarities, language of choice and hardware interfaces.
The hardware needs to be cheap, small-scale and energy efficient.
As a result, the \glspl{MCU} used to power these devices do not have a lot of computational power, a soup\c{c}on of memory, and little communication bandwidth.
Typically the devices do not run a full fledged \gls{OS} but a compiled firmware.
This firmware is often written in an imperative language that needs to be flashed to the program memory.
It is possible to dynamically send the program to the program memory using \gls{OTA} programming~\cite{baccelli_scripting_2018,baccelli_reprogramming_2018}.
Program memory typically is flash based and only lasts a couple of thousand writes before it wears out\textsuperscript{\ref{ftn:flash_memory}}.
While devices are getting a bit faster, smaller, and cheaper, they keep these properties to an extent.
The properties of the device greatly reduce the flexibility for dynamic systems where tasks are created on the fly, executed on demand and require parallel execution.
These problems can be mitigated by dynamically sending code to be interpreted to the \gls{MCU}.
With interpretation, a specialized interpreter is flashed in the program memory once that receives the program code to execute at runtime.

\subsection{\acrlong{TOP}}
\Gls{TOP} is a declarative programming paradigm designed to model interactive systems~\cite{plasmeijer_task-oriented_2012}.
A task is an abstract representation of a piece of work that needs to be done.
It provides an intuitive abstraction over work in the real world.
Just as with real-life tasks and workflow, tasks can be combined in various ways such as in parallel or in sequence.
Furthermore, tasks are observable which means it is possible to observe a --- partial --- result during execution and act upon it by for example starting new tasks.
Examples of tasks are filling in a form, sending an email, reading a sensor or even doing a physical task.
The task itself abstracts away from implementation details such as the interface, the communication and the sharing of data.

\begin{figure}
	\centering
	\begin{tikzpicture}[node distance=7em]
		\node (1) {$No Value$};
		\node (2) [right of=1] {$Unstable$};
		\node (3) [right of=2] {$Stable$};

		\draw [->] (1) -- (2);
		\draw [->] (2) -- (1);
		\draw [->] (2) -- (3);
		\draw [->] (1) to [out=12,in=165] (3);
	\end{tikzpicture}
	\caption{State diagram for the legal transitions of task values}\label{fig:taskvalue}
\end{figure}
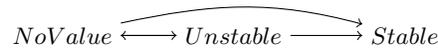

In many implementations the value observable in a task is a three state value that adheres to the transition diagram seen in Figure~\ref{fig:taskvalue}.
If a task emits no value, it means that the task has not made sufficient progress to produce a complete value.
It might be the case that some work has been done but just not quite enough (e.g.\ an open serial port with a partial message).
An unstable value means that a complete value is present but it may change in the future (i.e.\ a side effect).
A web editor for filling in a form is an example of a task that always emits an unstable value since the contents may change over time.
Stable values never change.
When the continue button has been pressed, the contents of the web editor is relayed, the values can never change, hence it is stable.

Tasks can communicate using task values but this imposes a problem in many collaboration patterns where tasks that are not necessarily related need to share data.
Tasks can also share data using \glspl{SDS}.
\Glspl{SDS} are an abstraction over any data.
An \gls{SDS} can represent typed data stored in a file, a chunk of memory, a database etc.
\Glspl{SDS} can also represent external impure data such as the time, random numbers or sensory data.
Similar to tasks, transformation and combination of \glspl{SDS} is possible.
In this architecture, tasks function as lightweight communicating threads

\subsection{\glsentrytext{iTasks}}\label{ssec:itasks}
The \gls{iTasks} system originated as a system for developing distributed collaborative interactive web applications and the \gls{TOP} paradigm grew from it~\cite{plasmeijer_itasks:_2007}.
It is suitable to model collaboration in almost any domain (see Subsection~\ref{ssec:related_top}).

The \gls{iTasks} system is implemented as an \gls{EDSL} hosted in \gls{Clean}~\cite{brus_clean_1987}.
Compiling the embedded \gls{TOP} specification results in a multi-user distributed webserver offering an interface to users for actually doing the work.
By default, implementation details such as the graphical user interface, serialization and communication are automatically generated.
Section~\ref{sec:itasks} gives a non-comprehensive overview that is sufficient for the exercises and examples in this paper.

In \gls{iTasks} a task is implemented as an event-driven stateful rewrite function.
This means that, when there is an event, the function is executed with the current state of the system and the event as arguments.
As a result, it produces a new state and either a value or an exception.
If a value is produced, it consists of a task value, an update to the user interface and a rewritten function.
The current state of a task can be represented by the structure of the tasks and their combinators and is dubbed the task tree~\cite{lijnse_itasks_2009}.

\Glspl{SDS} in \gls{iTasks} are based on \glspl{UDS}.
\Glspl{UDS} are a type safe, uniform and composable abstraction over arbitrary data through a read/write interface~\cite{michels_uniform_2012}.
This interface is extended with parametric lenses to also allow fine-grained control over accessing subsets of the data and filtering notifications~\cite{domoszlai_parametric_2014}.
Any type in the host language \gls{Clean} is an \gls{SDS} when it implements the \Cl{RWShared} class collection that contains the read, write and notification functions.
The \gls{iTasks} library contains \glspl{SDS} for storing data in files, databases, memory but also to provide access to system information such as date, time and random streams.
Furthermore it contains combinators to apply all types of transformations to \glspl{SDS}.
Multiple \glspl{SDS} can be combined to form new \gls{SDS}, \glspl{SDS} modelling collections can be filtered, information of an \gls{SDS} can determine the lens on another one and the data modelled by an \gls{SDS} can be transformed.

\subsubsection{Examples}
\begin{figure}
	\includegraphics[width=.75\linewidth]{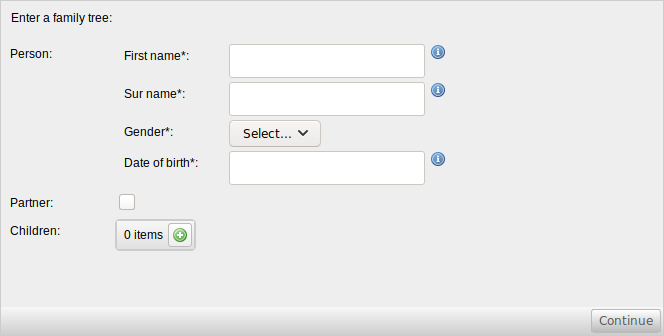}\\
	\includegraphics[width=.75\linewidth]{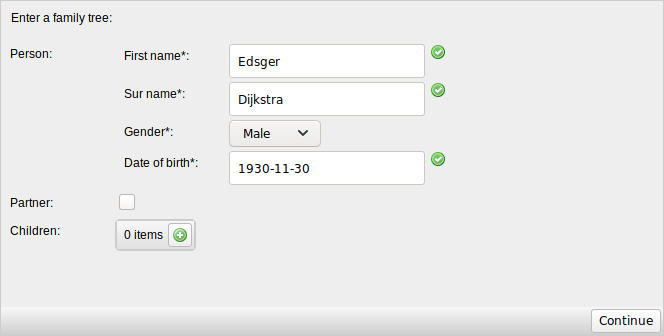}\\
	\caption{The initial user interface and the enabling of the continue button for the example application.}%
	\label{fig:itasksexample1}
\end{figure}

\begin{figure}
	\includegraphics[width=.75\linewidth]{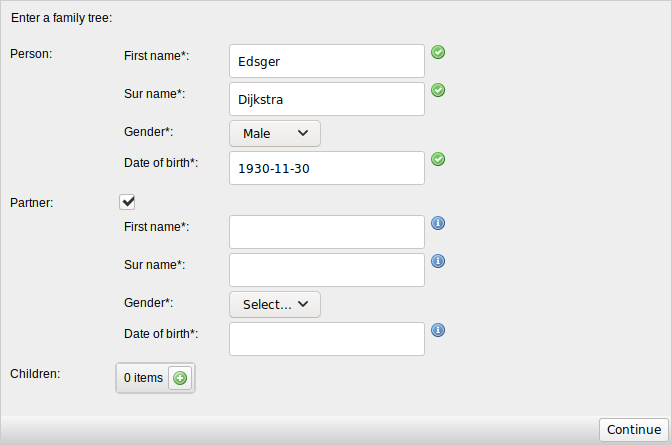}\\
	\caption{The user interface after the user ticks the Partner box.}%
	\label{fig:itasksexample2}
\end{figure}

Example~\ref{lst:itasksexample} shows a simple example of an \gls{iTasks} application, more examples are available in Section~\ref{sec:itasks}.
In the application, the user can enter a family tree and when they are finished, view the result.
The screenshots in Figure~\ref{fig:itasksexample1}~and~\ref{fig:itasksexample2} show this workflow.
Lines~\ref{lst:itasksex:data:fro}~to~\ref{lst:itasksex:data:to} define the data types, \Cl{Family} and \Cl{Person} are record types with named fields and \Cl{Gender} is an algebraic data type.
For any first order type, the necessary machinery housed in the \gls{iTasks} generic function collection can be derived automatically~\cite{alimarine_generic_2005}.
The collection contains functions for deserialization, serialization, editors, pretty printing and equality.
Line~\ref{lst:itasksex:derive} shows the derivation of the generic functions for the types in this example.
The actual task is of type \Cl{Task Family} and shown at Line~\ref{lst:itasksex:task}.
The workflow consists of two tasks, the first task is for entering (Line~\ref{lst:itasksex:enter}) and the second one for viewings (Line~\ref{lst:itasksex:view}).
They are combined using a sequential task combinator (\Cl{>>=}) that results in a continue button being shown to the user.
At the start of the workflow, the form is empty, and thus the continue button is disabled.
When the user enters some information, the continue button enables when there is a complete value.
However, the value may still change, as can be seen in the third figure when the partner tickbox is ticked and a recursive editor appears.

\begin{lstexample}[language=Clean,caption={Source code for some example \gls{iTasks} tasks.},label={lst:itasksexample},numbers=left]
:: Family = { person    :: Person,  partner     :: Maybe Person [+\label{lst:itasksex:data:fro}+]
            , children  :: [Family]
            }
:: Person = { firstName :: String,  surName     :: String
            , gender    :: Gender,  dateOfBirth :: Date
            }
:: Gender = Male | Female | Other String [+\label{lst:itasksex:data:to}+]

derive class iTask Family, Person, Gender [+\label{lst:itasksex:derive}+]

enterFamily :: Task Family [+\label{lst:itasksex:task}+]
enterFamily
	=         Hint "Enter a family tree:" @>> enterInformation []     [+\label{lst:itasksex:enter}+]
	>>= \res->Hint "You Entered:"         @>> viewInformation  [] res [+\label{lst:itasksex:view}+]
\end{lstexample}

\subsection{\Acrshort{TOP} for the \Acrshort{IOT}}
\gls{IOT} devices are often doing loosely related things in parallel.
For example, they are reading sensors, doing some processing on the data, operating actuators and communicating with the world.
The \gls{TOP} paradigm is an intuitive description language for theses tasks.
Furthermore, due to the execution semantics of tasks, seemingly parallel operation due to interleaving comes for free.
Unfortunately, running \gls{iTasks} tasks on the device is not an option due to the high memory requirements of the software.
Therefore, \gls{mTask} has been created, a \gls{TOP} language for small memory environments like \gls{IOT} devices that also contains constructions to interact with the peripherals as well.
It compiles the tasks to bytecode and sends them to the \gls{IOT} device at run time.
This allows the creation of dynamic applications, i.e.\ applications where tasks for the \gls{IOT} devices are tailor-made at runtime and scheduled when needed.

\subsection{Structure of the paper}
This section contains the introduction to \gls{IOT}, \gls{TOP} and \gls{iTasks}.
The \gls{mTask} ecosystem is explained in Section~\ref{sec:mtask} followed by a language overview in Section~\ref{sec:language}.
Section~\ref{sec:building} contains gradually introduces more \gls{mTask} concepts and provides a step by step tutorial for creating more interesting \gls{IOT} applications.
Section~\ref{sec:related} contains the related work and Section~\ref{sec:conclusion} concludes with discussions.
Background material on \gls{EDSL} techniques is available in Section~\ref{sec:edsl}.
An \gls{iTasks} reference manual containing all the tasks and functions required for the exercises can be found in Section~\ref{sec:itasks} and Section~\ref{sec:install} contains detailed instructions on setting up an \gls{mTask} development distribution.

Inline code snippets are typeset using a \Cl{teletype} font.

\setcounter{lstlisting}{-1}
\begin{lstdefinition}[caption={This is an example definition.}]
Program definitons are typeset in listings with a double left vertical border
\end{lstdefinition}

\setcounter{lstlisting}{-1}
\begin{lstexample}[caption={This is an example example.}]
Program examples are typeset in listings with a single left and bottom border
\end{lstexample}
\setcounter{lstlisting}{1}

\setcounter{exercise}{-1}
\begin{exercise}[The title of the example exercise]
	Exercises are numbered and typeset like this.
	The filename of the skeleton --- located in the distribution, see Section~\ref{sec:install} --- is typeset in teletype and placed between brackets (\texttt{fileName}).
\end{exercise}

%% file: 2mtask.tex
\subsection{Blink}\label{ssec:blink}
Traditionally, the first program that one writes when trying a new language is the so called \emph{Hello World!} program.
This program has the single task of printing the text \emph{Hello World!} to the screen and exiting again.
On microcontrollers, there often is no screen for displaying text.
Nevertheless, almost always there is a rudimentary single pixel screen, namely an --- often builtin --- \gls{LED}.
The \emph{Hello World} equivalent on microcontrollers blinks this \gls{LED}.

Example~\ref{lst:arduinoBlink} shows how the logic of a blink program might look when using the \gls{Arduino} \gls{C++} dialect.
The main event loop of the \gls{Arduino} language continuously calls the user defined \Cl{loop} function.
Blink's \Cl{loop} function alternates the state of the pin representing the \gls{LED} between \Cl{HIGH} and \Cl{LOW}, turning the \gls{LED} off and on respectively.
In between it waits for 500 milliseconds so that the blinking is actually visible for the human eye.
Compiling this results in a binary firmware that needs to be flashed onto the program memory.

Translating the traditional blink program to \gls{mTask} can almost be done by simply substituting some syntax as seen in Example~\ref{lst:blinkImp}.
E.g.\ \Cl{digitalWrite} becomes \Cl{writeD}, literals are prefixed with \Cl{lit} and the pin to blink is changed to represent the actual pin for the builtin \gls{LED} of the device used in the exercises.
In contrast to the imperative \gls{Arduino} \gls{C++} dialect, \gls{mTask} is a \gls{TOP} language and therefore there is no such thing as a loop, only task combinators to combine tasks.
To simulate this, the \Cl{rpeat} task can be used, this task executes the argument task and, when stable, reinstates it.
The body of the \Cl{rpeat} contains similarly named tasks to write to the pins and to wait in between.
The tasks are connected using the sequential \Cl{>>|.} combinator that for all intents and purposes executes the tasks after each other.

\begin{exercise}[Blink the builtin \glsentrytext{LED}]\label{ass:blink}
	Compile and run the blink program to test your \gls{mTask} setup (\texttt{blinkImp}).
	Instructions on how to install \gls{mTask} and how to find the example code can be found in Section~\ref{sec:install}.
\end{exercise}

\begin{figure}
	\begin{subfigure}[b]{.45\linewidth}
		\begin{lstexample}[language=C++,caption={Blink in \gls{Arduino}.},label={lst:arduinoBlink}]
void loop() {
	digitalWrite(BUILTIN_LED, HIGH);
	delay(500);
	digitalWrite(BUILTIN_LED, LOW);
	delay(500);
}\end{lstexample}
	\end{subfigure}%
	\hfill
	\begin{subfigure}[b]{.45\linewidth}
		\begin{lstexample}[language=Clean,caption={Blink in \gls{mTask}},label={lst:blinkImp}]
blink :: Main (MTask v ()) | mtask v
blink = {main = rpeat (
		     writeD d2 (lit True)
		>>|. delay (lit 500)
		>>|. writeD d2 (lit False)
		>>|. delay (lit 500)
	)}\end{lstexample}
	\end{subfigure}
\end{figure}

\subsection{Language}
The \gls{mTask} language is a \gls{TOP} \gls{EDSL} hosted in the pure lazy functional programming language \gls{Clean}~\cite{brus_clean_1987}.
An \gls{EDSL} is a language embedded in a host language created for a specific domain~\cite{hickey_building_2014}.
The two main techniques for embedding are deep embedding --- representing the language as data --- and shallow embedding --- representing the languages as function.
Depending on the embedding technique, \glspl{EDSL} support one or multiple backends or views.
Commonly used views are pretty printing, compiling, simulating, verifying and proving properties of the program.
Deep and shallow embedding have their own advantages and disadvantages in terms of extendability, type safety and view support that are described in more detail in Section~\ref{sec:edsl}.

\subsection{Class based shallow embedding}
There are also some hybrid approaches that try to mitigate the downsides of the standard embedding techniques.
The \gls{mTask} language is using class-based --- or tagless --- shallow embedding that has both the advantages of shallow and deep embedding while keeping the disadvantages to a minimum~\cite{carette_finally_2009}.
This embedding technique is chosen because it allows adding backends and functionality orthogonally, i.e.\ without touching old code.
E.g.\ adding functionality orthogonally is useful to add constructions for interact with new peripherals without requiring other backends to implement them.
At the time of writing there is bytecode generation, symbolic simulation and pretty printing available as a backend.

Definition~\ref{lst:exclassshallow} shows an illustrative example of this embedding technique using a multi backend expression language.
In class-based shallow embedding the language constructs are defined as type classes (\Cl{intArith} and \Cl{boolArith}).
In contrast to regular shallow embedding, functions in class based shallow embedding are overloaded in the backend and in the types.
Furthermore, the functions can be overloaded and contain class constraints, i.e.\ type safety is inherited from the host language.
Lastly, extensions can be added easily, just as in shallow embedding.
When an extension is made in an existing class, all views must be updated accordingly to prevent possible runtime errors.
But when an extension is added in a new class, this problem does not arise and views can choose to implement only parts of the collection of classes.

\begin{lstdefinition}[language=Clean,label={lst:exclassshallow},caption={A minimal class based shallow \gls{EDSL}.}]
class intArith v where
	lit :: t -> v t             | toString t
	add :: (v t) (v t) -> (v t) | + t
	sub :: (v t) (v t) -> (v t) | - t

class boolArith v where
	and :: (v Bool) (v Bool) -> (v Bool)
	eq  :: (v t)    (v t)    -> (v Bool) | == t
\end{lstdefinition}

A backend in a class based shallowly \gls{EDSL} is just a type implementing some of the classes which makes adding backends relatively easy.
It is even possible to create partial backends that do not support all classes from the language.
The type of the backend are often --- e.g.\ in the \Cl{PrettyPrinter} type --- phantom types, only there to the resulting expression type safe.
Example~\ref{lst:exclassshallowbackend} shows an example of two backends implementing the expression \gls{DSL}.

\begin{lstexample}[language=Clean,label={lst:exclassshallowbackend},caption={A minimal class based shallow \gls{EDSL}.}]
:: PrettyPrinter a = PP String
runPrinter :: (PrettyPrinter t) -> String
runPrinter (PrettyPrinter s) = s

instance intArith PrettyPrinter where
	lit x = PP (toString x)
	add (PP x) (PP y) = PP (x +++ "+" +++ y)
	...
instance boolArith PrettyPrinter where ...

:: Evaluator a = Eval a
runEval :: (Evaluator a) -> a
runEval (Eval a) = a

instance intArith Evaluator where ...
instance boolArith Evaluator where ...
\end{lstexample}

A downside of using classes instead of functions is that the more flexible implementation technique makes the type errors more complicated.
Also, as a consequence of using classes instead of data, a program wanting to use the same expression twice has to play some tricks (see Example~\ref{lst:exclassshallow}).
If the language supports rank-2 polymorphism, it can use the same expression for multiple backends.
Another solution is to create a combinator backend that combines the two argument backends in a single structure.

\begin{lstexample}[language=Clean,label={lst:exclassshallowmulti},caption={Using multiple backends simultaneously in a shallow \gls{EDSL}.}]
printAndEval :: ( A.v: v t | intArith, boolArith v) -> (String, t)
printAndEval c = (runPrinter c, runEval c)

:: Two l r a = Two (l a) (r a)
printAndEval` :: (Two PrettyPrinter Evaluator t) -> (String, t)
printAndEval` (Two (PP t) (Eval a)) = (t, a)

instance intArith (Two l r) | intArith l & intArith r where
	lit x = Two (lit x) (lit x)
	add (Two lx rx) (Two ly ry) = Two (add lx ly) (add rx ry)
instance boolArith (Two l r) | boolArith l & boolArith r where
	eq (Two lx rx) (Two ly ry) = Two (eq lx ly) (eq rx ry)
\end{lstexample}

\subsection{\glsentrytext{DSL} design}
To leverage the type checker of the host language, types in the \gls{mTask} language are expressed as types in the host language, to make the language type safe.
However, not all types in the host language are suitable for microcontrollers that may only have 2KiB of \gls{RAM} so class constraints are therefore added to the \gls{EDSL} functions (see Definition~\ref{lst:constraints}).
The most used class constraint is the \Cl{type} class collection containing functions for serialization, printing, \gls{iTasks} constraints etc.
Many of these functions can be derived using generic programming.
An even stronger restriction on types is defined for types that have a stack representation.
This \Cl{basicType} class has instances for many \gls{Clean} basic types such as \Cl{Int}, \Cl{Real} and \Cl{Bool} but also for tuples.
The class constraints for values in \gls{mTask} are omnipresent in all functions and therefore often omitted throughout this paper for brevity and clarity.

Furthermore, expressions overloaded in backend add all \gls{mTask} classes as constraints.
To shorten this, a class collection is defined that contains all standard \gls{mTask} classes to relieve this strain.
However, classes for peripherals --- or other non standard classes that not all backends have --- need to be added still.

\begin{lstdefinition}[language=Clean,caption={Classes and class collections for the \gls{mTask} \gls{EDSL}.},label={lst:constraints}]
class type t | iTask, ... ,fromByteCode, toByteCode t
class basicType t | type t where ...

class mtask v | arith, ..., cond v

someExpr :: v Int | mtask v
readTempClass :: v Bool | mtask, dht v
\end{lstdefinition}

The \gls{mTask} language is a \gls{TOP} language and therefore supports tasks.
For seamless integration, the \Cl{TaskValue} type from \gls{iTasks} is used for task values in \gls{mTask} as well (see Definition~\ref{lst:task_types}).
The leafs are basic tasks (i.e.\ editors) and the forks are task combinators.
Every evaluation step, the task tree is traversed from the root up and nodes are rewritten while at the mean time keeping track of the task value of the tree as a whole.
This means that there is a difference in execution between expressions and tasks.
Expressions are always evaluated completely and therefore block the execution.
Tasks on the other hand have small evaluation steps to allow seemingly parallel execution when interleaved.

\begin{lstdefinition}[language=Clean,caption={The \gls{mTask} task types.},label={lst:task_types}]
:: TaskValue t = NoValue | Value a Bool //from iTasks
:: MTask v t :== v (TaskValue t)
\end{lstdefinition}

\subsection{Backends}
The classes are just a description of the language.
It is the backend that actually gives meaning to the language.
There are many backends possible for a \gls{TOP} programming language for tiny computers.
At the time of writing, there is a pretty printing, symbolic simulation and bytecode generation backend.
These lecture notes only regard the bytecode generation backend but the other backends will be briefly discussed for completeness sake.

\subsubsection{Pretty printer}
The pretty printing backend produces a pretty printer for the given program.
The only function exposed is the \Cl{showMain} (Definition~\ref{lst:showmain}) function which runs the pretty printer and returns a list of strings containing the pretty printed result as shown in Example~\ref{lst:show}.
The pretty printing function does the best it can but obviously cannot reproduce the layout, curried functions and variable names.

\begin{lstdefinition}[language=Clean,caption={The entrypoint for the pretty printing backend.},label={lst:showmain}]
:: Show a // from the mTask Show library
showMain :: (Main (Show a)) -> [String] | type a
\end{lstdefinition}

\begin{lstexample}[language=Clean,caption={Pretty printing backend example.},label={lst:show}]
blink :: Main (MTask v Bool) | mtask v
blink =
	fun \blink = (\state->
		     writeD d13 state
		>>|. delay (lit 500)
		>>=. blink o Not)
	In {main = blink true}

Start :: [String]
Start = showMain blink

// output:
// let f0 a1 = writeD(D13, a1) >>= \a2.(delay 1000) >>| (f0 (Not a1)) in (f0 True)
\end{lstexample}

\subsubsection{Simulator}
\begin{figure}
	\centering
	\includegraphics[width=.85\linewidth]{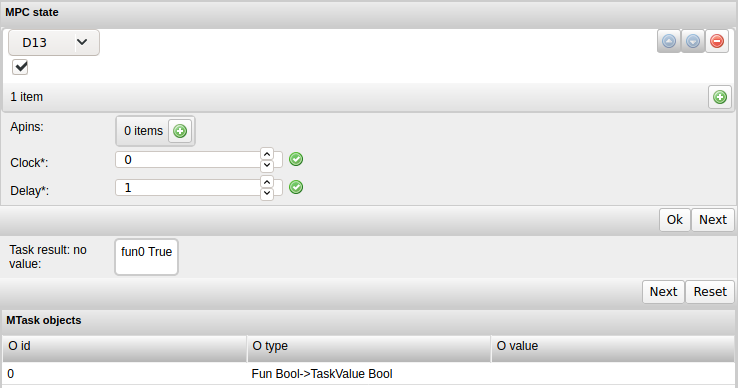}
	\caption{Simulator interface for the blink program.}\label{fig:sim}
\end{figure}

The simulation backend produces a symbolic simulator embedded in \gls{iTasks} for the given program.
When task resulting from the \Cl{simulate} function presents the user with an interactive simulation environment (see Definition~\ref{lst:simulatemain}, Example~\ref{lst:simulate} and Figure~\ref{fig:sim}).
From within the environment, tasks can be rewritten, peripheral states changed and \glspl{SDS} interacted with.

\begin{lstdefinition}[language=Clean,caption={The entrypoint for the simulation backend.},label={lst:simulatemain}]
:: TraceTask a // from the mTask Show library
simulate :: (Main (TraceTask a)) -> [String] | type a
\end{lstdefinition}

\begin{lstexample}[language=Clean,caption={Simulation backend example.},label={lst:simulate}]
Start :: *World -> *World
Start w = doTasks (simulate blink) w
\end{lstexample}

\subsection{Bytecode}\label{ssec:bytecode}
Programs written in \gls{mTask} are compiled to bytecode to be integrated in \gls{iTasks}.
The microcontroller stores the tasks in their \gls{RAM}, leaving the program memory untouched.
In \gls{TOP}, it is not uncommon to create tasks every minute.
Writing the program memory of an \gls{MCU} every minute would wear a typical \gls{MCU} out within a week\footnote{%
	Atmel, the producer of AVR microprocessors, specifies the flash memory of the \gls{MCU} in the \gls{Arduino} {UNO} to about 10,000 cycles.
	This specification is a minimal specification and most likely the memory will be able to sustain many more writes.
	However, even if the memory can sustain ten times the amount, it is still a short time.%
	\label{ftn:flash_memory}
}.

\begin{figure}
	\centering
	\resizebox{.6\linewidth}{!}{\input{arch}}
	\normalsize
	\caption{The world of \gls{TOP} applications that supports devices.}%
	\label{fig:architecture}
\end{figure}
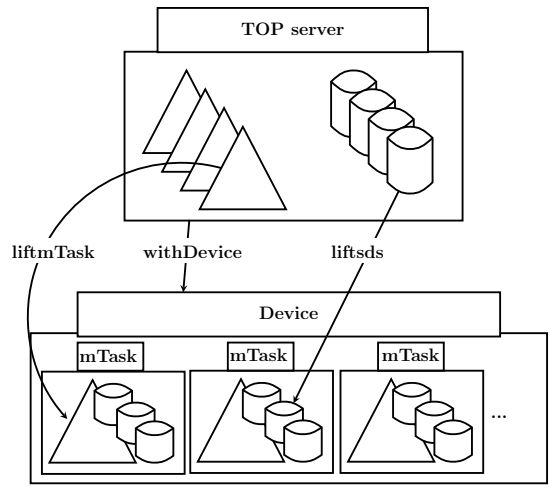

A complete specification of an \gls{mTask} program --- including the \glspl{SDS} and peripherals --- of type \Cl{t} has the following type in the host language: \Cl{:: Main (MTask BCInterpret t)}.
Under the hood, \Cl{BCInterpret} is a monad stack that generates the bytecode when executed.
Interplay between \gls{mTask} and \gls{iTasks} happens via three different constructions that are visualized in Figure~\ref{fig:architecture}.

\subsubsection{Connecting devices}
For a device to be suitable for \gls{mTask}, it needs to be able to run the \gls{RTS}.
At the time of writing, the \gls{RTS} is ported to \gls{Arduino} compatible \gls{xtensa} boards such as the \gls{LOLIN} D1 Mini and \gls{NodeMCU}, \gls{Arduino} compatible \gls{AVR} boards such as the \gls{Arduino} {UNO}, and for platforms running \glspl{OS} such as Linux, Windows or MacOS regardless of the architecture.

The \Cl{withDevice} function offers access to a specific device given the communication specification (see Definition~\ref{lst:withdevice}).
The first argument of the function contains the information about the connection that is used to communicate with the device.
Any reliable sequential character based connection is suitable as a means of communication between the device and the server.
In the \gls{mTask} system at the time of writing, \Cl{channelSync} instances are available for \gls{TCP} connections and serial port connections.

The second argument is a function that --- given a device handle --- produces a task that can do something with the device.
The task resulting from the \Cl{withDevice} function will first setup a connection to the device and exchange specifications.
After the initialization, the task retrieved from the function in the second argument is executed.
When this task is finished, the connection with the device is closed down again.

\begin{lstdefinition}[language=Clean,caption={Connecting \gls{mTask} devices to an \gls{iTasks} server},label={lst:withdevice}]
:: MTDevice  // Abstract device representation
:: Channels  // Communication channels

class channelSync a :: a (Shared Channels) -> Task ()
withDevice :: a (MTDevice -> Task b)
	-> Task b | iTask b & channelSync, iTask a

instance channelSync TCPSettings, TTYSettings
\end{lstdefinition}

\subsubsection{Lifting tasks}
Sending a task to a device always occurs from within \gls{iTasks} and is called \emph{lifting} a task from \gls{mTask} to \gls{iTasks}.
The function for this is called \Cl{liftmTask} (see Definition~\ref{lst:liftmtask}).
The first argument is the \gls{mTask} program and the second argument is the device handle.
The resulting task is an \gls{iTasks} proxy task representing the exact state of the \gls{mTask} task on the device.

Under the hood it first generates the bytecode of the \gls{mTask} task by evaluating the monad.
This bytecode is bundled with metadata of the (lifted) \glspl{SDS} and peripherals and sent to the device.
The device executes the task and notifies the server on any changes in task value or when it writes a lifted \gls{SDS}.
These changes are immediately reflected in the server resulting either in a changed observable task value or a server side write to the \gls{SDS} to which the lifted \gls{SDS} was connected.
On the server side, the \Cl{liftmTask} task also subscribes to all lifted \gls{SDS} so that when the \gls{SDS} on the server changes, the device can be notified as well.
The result is that this lifted task reflects the exact state of the \gls{mTask} task.

\begin{lstdefinition}[language=Clean,caption={Lifting an \gls{mTask} to an \gls{iTasks} task.},label={lst:liftmtask}]
liftmTask :: (Main (MTask BCInterpret u)) MTDevice -> Task u | iTask, type u
\end{lstdefinition}

\subsection{Skeleton}
Subsection~\ref{ssec:blink} showed an example \gls{mTask} task that blinks the builtin \gls{LED}.
This is not yet a complete \gls{Clean}/\gls{iTasks} program that can be executed.
A skeleton follows that can be used as a basis for the exercises that is explained line by line.
Future snippets will again only give the \gls{mTask} code for brevity.

\begin{lstexample}[language=Clean,caption={An \gls{mTask} skeleton program.},label={lst:blink},numbers=left]
module blink [+\label{blink:modname}+]

import StdEnv, iTasks                  //iTasks imports [+\label{blink:importsfro}+]
import Interpret, Interpret.Device.TCP //mTask imports  [+\label{blink:importsto}+]

Start :: *World -> *World [+\label{blink:startfro}+]
Start w = doTasks main w  [+\label{blink:startto}+]

main :: Task Bool
main = enterDevice >>= \spec->withDevice spec    [+\label{blink:editor}+]
	\dev->liftmTask blink dev -|| viewDevice dev [+\label{blink:fun}+]
where
	blink :: Main (MTask v Bool) | mtask v       [+\label{blink:mtaskfro}+]
	blink = ... //e.g. blink from [+Listing~\ref{lst:blinkImp}+] [+\label{blink:mtaskto}+]
\end{lstexample}

Line~\ref{blink:modname} declares the name of the module, this has to match the name of the filename.
Line~\ref{blink:importsfro} import \Cl{StdEnv} and \Cl{iTasks} libraries, these imports are required when using \Cl{iTasks}.
Line~\ref{blink:importsto} imports the \Cl{Interpret} --- the \gls{mTask} bytecode backend --- and \Cl{Interpret.Device.TCP} --- the \gls{TCP} device connectivity modules.
Both imports are always required for these exercises.
Line~\ref{blink:startfro}~and~\ref{blink:startto} gives the \Cl{Start} function, the entry point for a \gls{Clean} program.
This start function always calls the \gls{iTasks} specific entry point called \Cl{doTasks} that starts up the \gls{iTasks} machinery and launches the task \Cl{main}.

\begin{figure}
	\centering
	\includegraphics[width=.85\linewidth]{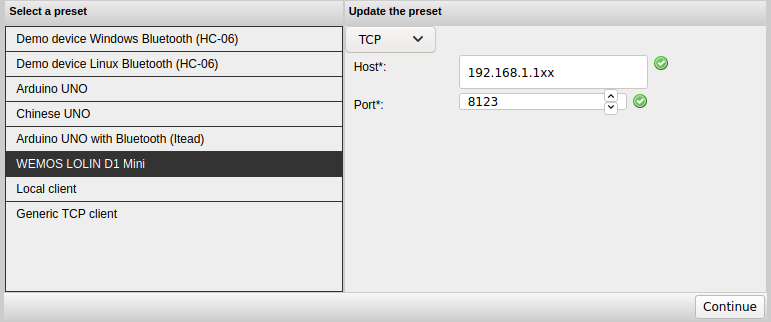}
	\caption{The interface for the \Cl{enterDevice} task.}\label{fig:device_selection}
\end{figure}

The \Cl{main} task first starts with an editor on Line~\ref{blink:editor}.
This editor presents an interface to the user connecting to the server for it to select a device as seen in Figure~\ref{fig:device_selection}.
The \Cl{enterDevice} task allows selecting devices from presets and allows changing the parameters to select a custom device.
After entering the \gls{IP} address the device shows, the task continues with connecting the device \Cl{withDevice} that takes a function requiring a device and resulting in a task.
This function (Line~\ref{blink:fun}) executes the \Cl{blink} task and shows some information about the device at the same time.
Line~\ref{blink:mtaskfro}~and~\ref{blink:mtaskto} contain the actual task, for example the task shown in Example~\ref{lst:blink}.

\section{\gls{mTask} language}\label{sec:language}
\subsection{Expressions}
The classes for expressions --- i.e.\ arithmetic functions, conditional expressions and tuples --- are listed in Definition~\ref{lst:arith_cond}.
Some of the class members are oddly named (e.g.\ \Cl{+.}) to make sure there is no name conflict with \gls{Clean}'s builtin overloaded functions that are of a different kind (\Cl{*} instead of \Cl{*->*}).
There is no need for loop control due to support for tail call optimized recursive functions and tasks.
The types speak for themselves but there are a few functions to explain.
The \Cl{lit} function lifts a value from the host language to the \gls{mTask} domain.
For tuples there is a useful macro (\Cl{topen}) to convert a function with an \gls{mTask} tuple as an argument to a function with a tuple of \gls{mTask} values as an argument.

\begin{lstdefinition}[language=Clean,caption={The \gls{mTask} classes for arithmetic, conditional and tuple expressions.},label={lst:arith_cond}]
class arith v where
	lit :: t -> v t | type t
	(+.) infixl 6 :: (v t) (v t) -> v t    | basicType, +, zero, t
	...
	(==.) infix 4 :: (v a) (v a) -> v Bool | basicType, == a
	...
class cond v where
	If :: (v Bool) (v t) (v t) -> v t | type t
class tupl v where
	first  :: (v (a, b))  -> v a      | type a & type b
	second :: (v (a, b))  -> v b      | type a & type b
	tupl   :: (v a) (v b) -> v (a, b) | type a & type b
\end{lstdefinition}

\subsection{Functions}
Functions are supported in the \gls{EDSL}, albeit with some limitations.
All user defined \gls{mTask} functions are typed by \gls{Clean} functions so that they are type-safe and are first class citizens in the \gls{DSL}.
They are defined using the multi-parameter typeclass \Cl{fun}.
The first parameter (\Cl{a}) of the typeclass is the shape of the argument and the second parameter (\Cl{v}) is the backend (see Definition~\ref{lst:dec_fun}).
Functions may only be defined at the top level and to constrain this, the \Cl{main} type is introduced to box a program.

\begin{lstdefinition}[language=Clean,caption={The \gls{mTask} classes for functions definitions.},label={lst:dec_fun}]
:: Main a = {main :: a}
:: In a b = In infix 0 a b
class fun a v where
	fun :: ((a -> v s) -> In (a -> v s) (Main (v u))) -> Main (v u) | ...
\end{lstdefinition}

For every possible arity of the function, a separate implementation for the \Cl{fun} class has to be defined (see Example~\ref{lst:dec_fun_arity})
The listing gives example instances for arities zero to two for backend \Cl{T}.
Defining the different arities as tuples of arguments instead of a more general definition forbids the use of curried functions.
All functions are therefore known at compile time and when a function is called, all arguments are always known which is beneficial for keeping the memory requirements low.

\begin{lstexample}[language=Clean,caption={Different class instances for different arities in \gls{mTask} functions.},label={lst:dec_fun_arity}]
:: T a // a backend
instance fun () T
instance fun (T a) T | type a
instance fun (T a, T b) T | type a & type b
\end{lstexample}

To demonstrate the use, Example~\ref{lst:function_ex} shows examples for two functions.
The \Cl{type} constraint on the function arguments forbid the use of higher order functions because functions do not have instances for all classes of the collection.
The functions (\Cl{sum}, \Cl{factorial}) constructs the program that calculates the result of the arguments.
In the bytecode backend, there is full tailcall optimization and therefore, writing \Cl{factorial} as \Cl{factorial`} pays off in memory usage.

\begin{lstexample}[language=Clean,caption={Example \glsentrytext{mTask} functions.},label={lst:function_ex}]
sum :: Int Int -> Main (v Int)
sum x y =
	fun \sum = (\(l, r)->l +. r) In
	{main = sum (lit x, lit y)}

factorial :: Int -> Main (v Int) | mtask v
factorial x =
	fun \fac = (\i->
		If (i ==. lit 0) (lit 1) (i *. fac (i -. lit 1)))
	In {main = fac (lit i)}

factorial` :: Int -> Main (v Int) | mtask v
factorial` x =
	fun \facacc = (\(n,a)->
		If (n ==. lit 0) a (facacc (n -. lit 0, n *. a)))
	In fun \fac = (\i->
		facacc (i, lit 1))
	In {main = fac (lit i)}
\end{lstexample}

\subsubsection{Functional blinking}\label{ssec:fblink}
The \gls{mTask} blink implementation does not show the advantage of function or \gls{TOP}.
With functions, the blink behaviour can be lifted to a function to make the program more functional and composable (see Example~\ref{lst:fblink}).
The function takes a single argument, the state and recursively calls itself.
It creates an infinite task that first waits 500 milliseconds.
Then it will write the current state to the pin followed by a recursive call to with the inverse of the state.

\begin{lstexample}[language=Clean,caption={A functional \gls{mTask} translatation of Hello World! (\texttt{blink})},label={lst:fblink}]
blinkTask :: Main (MTask v Bool) | mtask v
blinkTask
	= fun \blink = (\x->
		     delay (lit 500)
		>>|. writeD d2 x
		>>=. blink o Not)
	In {main = blink (lit True)}
\end{lstexample}

\begin{exercise}[Blink the builtin \glsentrytext{LED} with a different interval]\label{ass:blinkint}
	Change the blinking interval of the functional blink program (\texttt{blink}).
\end{exercise}

\subsection{Basic tasks}
Definition~\ref{lst:def_basic} shows the classes for the basic tasks in \gls{mTask}.
Interaction with peripherals also occurs through basic tasks and they are shown later.
To lift a value in the expression domain to the task domain, the basic task \Cl{rtrn} is used.
The resulting task will forever yield the given value as a stable task value.
The \Cl{rpeat} task continuously executes the argument task, restarting it when it yields a stable value.
The resulting compound task itself never yields a value.
The \Cl{delay} task emits no value while waiting for the elapsed number of milliseconds.
When enough time elapsed, it returns the number of milliseconds that it overshot the target time as a stable value.

\begin{lstdefinition}[language=Clean,caption={The \gls{mTask} classes for basic tasks.},label={lst:def_basic}]
class rtrn v where
	rtrn :: (v t) -> MTask v t | type t
class rpeat v where
	rpeat :: (MTask v a) -> MTask v () | type a
class delay v
	delay :: (v Int) -> MTask v Int | type n
\end{lstdefinition}

\subsection{Parallel task combinators}
Task combinators can be divided into two categories, namely parallel and sequential combinators.
In parallel combination, the evaluation of the two tasks are interleaved, resulting in seemingly parallel execution.
In contrast to \gls{iTasks}, there are only two parallel combinators available in \gls{mTask}.
Definition~\ref{lst:def_parallel} shows the class definitions.
Both combinators execute the two argument tasks in an interleaved fashion resulting in parallel execution.

\begin{lstdefinition}[language=Clean,caption={The \gls{mTask} classes for parallel task combinators and the rules for combining the value.},label={lst:def_parallel}]
class .&&. v where
	(.&&.) infixr 4 v :: (MTask v a) (MTask v b) -> MTask v (a, b) | ...
class .||. v where
	(.||.) infixr 3 v :: (MTask v a) (MTask v a) -> MTask v a | ...
\end{lstdefinition}

The resulting task value for the conjunction combinator \Cl{.&&.} is a pair of the task values of the children.
The resulting task value for the disjunction combinator \Cl{.||.} is a single task value, giving preference to the most stable one.
The exact task value production is explained as a \gls{Clean} function in the listing below.

\begin{lstdefinition}[language=Clean,caption={The rules for the task value of the parallel combinators.},label={lst:sem_parallel}]
(.&&.) :: (TaskValue a) (TaskValue b) -> TaskValue (a, b)
(.&&.) (Value l s1) (Value r s2) = Value (l, r) (s1 && s2)
(.&&.) _              _          = NoValue

(.||.) :: (TaskValue a) (TaskValue a) -> TaskValue a
(.||.) (Value _ True) _              = Value l True
(.||.) (Value _ _)    (Value r True) = Value r True
(.||.) NoValue        r              = r
(.||.) l              _              = l
\end{lstdefinition}

When using the parallel combinator \Cl{.&&.} the result is something of type \Cl{v (a, b)}.
This means that it is a tuple in the \gls{mTask} language and not in the host language and therefore pattern matching the tuple directly is not possible.
For that, the \Cl{topen} macro is defined as can be seen in the listing together with an example of the usage.

\begin{lstexample}[language=Clean,caption={An example of the usage of the parallel combinators.},label={lst:ex_parallel}]
topen :: (v (a, b) -> c) (v a, v b) -> c | tupl v
topen f x :== f (first x, second x)

firstPinToYield :: MTask v Int
firstPinToYield = readA A0 .||. readA A1 >>~. rtrn

sumpins :: MTask v Int
sumpins = readA A0 .&&. readA A1 >>~. topen \(x, y)->rtrn (x +. y)
\end{lstexample}

\subsection{Threaded blinking}
Now say that we want to blink multiple blinking patterns on different \glspl{LED} concurrently.
Intuitively we want to lift the blinking behaviour to a function and call this function three times with different parameters as done in Example~\ref{lst:arduinoBlinkMulti}.

\begin{lstexample}[language=C++,caption={Naive approach to multiple blinking patterns in \gls{Arduino} \gls{C++}.},label={lst:arduinoBlinkMulti}]
void blink (int pin, int wait) {
	digitalWrite(pin, HIGH);
	delay(wait);
	digitalWrite(pin, LOW);
	delay(wait);
}

void loop() {
	blink (1, 500);
	blink (2, 300);
	blink (3, 800);
}
\end{lstexample}

Unfortunately, this does not work because the \Cl{delay} function blocks all further execution.
The resulting program will blink the \glspl{LED} after each other instead of at the same time.
To overcome this, it is necessary to slice up the blinking behaviour in very small fragments so it can be manually interleaved~\cite{feijs_multi-tasking_2013}.
Example~\ref{lst:blinkthread} shows how to implement three different blinking patterns in \gls{Arduino} using the slicing method.
If we want the blink function to be a separate parametrizable function we need to explicitly provide all references to the required state.
Furthermore, the \Cl{delay} function can not be used and polling \Cl{millis} is required.
The \Cl{millis} function returns the number of milliseconds that have passed since the boot of the \gls{MCU}.
Some devices use very little energy when in \Cl{delay} or sleep state.
Resulting in \Cl{millis} potentially affects power consumption since the processor is basically busy looping all the time.
In the simple case of blinking three \glspl{LED} on fixed intervals, it might be possible to calculate the delays in advance using static analysis and generate the appropriate \Cl{delay} code.
Unfortunately, this is very difficult in general when the thread timings are determined at run time.
Manual interleaving is very error prone, requires a lot of pointer juggling and generally results in spaghetti code.
Furthermore, it is very difficult to represent dependencies between threads, often state machines have to be explicitly programmed by hand to achieve this.

\begin{lstexample}[language=C++,label={lst:blinkthread},caption={Threading three blinking patterns in \gls{Arduino}.}]
long led1 = 0, led2 = 0, led3 = 0;
bool st1 = false, st2 = false, st3 = false;

void blink(int pin, int delay, long *lastrun, bool *st) {
	if (millis() - *lastrun > delay) {
		digitalWrite(pin, *st = !*st);
		*lastrun += delay;
	}
}

void loop() {
	blink(1, 500, &led1, &st1);
	blink(2, 300, &led2, &st1);
	blink(3, 800, &led3, &st1);
}
\end{lstexample}

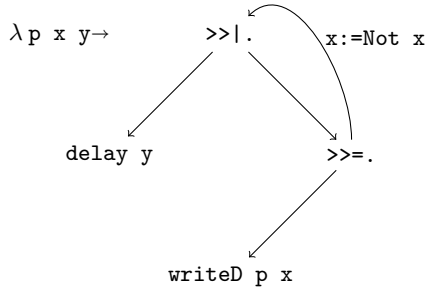
\begin{figure}
	\begin{tikzpicture}[node distance=7em]
		\node (0) [left of=1] {\Cl{\\p x y->}};
		\node (1) {\Cl{>>|.}};
		\node (2) [below left of=1] {\Cl{delay y}};
		\node (3) [below right of=1] {\Cl{>>=.}};
		\node (4) [below left of=3] {\Cl{writeD p x}};

		\draw [->] (1) to (2);
		\draw [->] (1) to (3);
		\draw [->] (3) to (4);
		\draw [->] (3) to [out=90,in=45] node [midway,right] {\Cl{x:=Not x}} (1);
	\end{tikzpicture}
	\caption{The task tree for the blink task.}\label{fig:blink_tree}
\end{figure}

Blinking multiple patterns in \gls{mTask} is as simple as combining several calls to an adapted version of the \Cl{blink} function from Example~\ref{lst:blink} with a parallel combinator as shown in Example~\ref{lst:blinkThread}.
The resulting task tree of a single blink function call can then be visualized as in Figure~\ref{fig:blink_tree}.

\begin{lstexample}[language=Clean,caption={An \gls{mTask} program for blinking multple patterns. (\texttt{blinkThread})},label={lst:blinkThread},numbers=left]
blink :: Main (MTask v Bool) | mtask v
blink
	= fun \blink = (\(p, x, y)->
		     delay y
		>>|. writeD p x
		>>=. \x->blink (p, Not x, y))
	In {main = blink (d1, true, lit 500)
		  .||. blink (d2, true, lit 300)
		  .||. blink (d3, true, lit 800)}
 \end{lstexample}

\begin{exercise}[Blink the builtin \acrshort{LED} with two patterns]\label{ass:blinkThread}
	Adapt the program in Example~\ref{lst:blinkThread} so that it blinks the builtin \gls{LED} with two different patterns concurrently.
	The times for the patterns are queried from the user.

	The function signature for \Cl{blink} becomes (\texttt{blinkThread})\\

	\begin{lstexample}[language=Clean,postbreak=]
blink :: Int Int ->  Main (MTask v Bool) | mtask v
	\end{lstexample}

	You should \Cl{enterInformation} to get the information from the user (see Section~\ref{sec:editor}).
\end{exercise}

\subsection{Sequential task combinators}
The second way of combining tasks is sequential combination in which tasks are executed after each other.
Similar to \gls{iTasks}, there is one Swiss army knife sequential combinator (\Cl{>>*.}) which is listed in Definition~\ref{lst:seq_comb}.
The task value yielded by the left-hand side is matched against all task continuations (\Cl{Step v t u}) on the right-hand side, i.e.\ the right-hand side tasks observes the task value.
When one of the continuations yields a new task, the combined task continues with it, pruning the left-hand side.
All other sequential combinators are derived from the step combinator as default class member instances.
Their implementation can therefore be overridden to provide a more efficient implementation.
For example, the \Cl{>>=.} combinator is very similar to the monadic bind, it continues if and only if a stable value is yielded with the task resulting from the function.
The \Cl{>>~.} combinator continues when any value, stable or unstable, is yielded.
The \Cl{>>|.} and \Cl{>>..} combinators are variants that do not take the value into account of the aforementioned combinators.

\begin{lstdefinition}[language=Clean,caption={The \gls{mTask} classes for sequential task combinators.},label={lst:seq_comb}]
class step v where
	(>>*.) infixl 1 :: (MTask v t) [Step v t u] -> MTask v u | ...

	(>>=.) infixl 0 :: (MTask v t) ((v t) -> MTask v u) -> MTask v u | ...
	(>>=.) m f = m >>*. [IfStable (\_->lit True) f]
	(>>~.) infixl 0 :: (MTask v t) ((v t) -> MTask v u) -> MTask v u | ...
	(>>~.) m f = m >>*. [IfValue (\_->lit True) f]
	(>>|.) infixl 0 :: (MTask v t) (MTask v u) -> MTask v u | ...
	(>>|.) m f = m >>=. \_->f
	(>>..) infixl 0 :: (MTask v t) (MTask v u) -> MTask v u | ...
	(>>..) m f = m >>~. \_->f

:: Step v t u
  = IfValue    ((v t) -> v Bool) ((v t) -> MTask v u)
  | IfStable   ((v t) -> v Bool) ((v t) -> MTask v u)
  | IfUnstable ((v t) -> v Bool) ((v t) -> MTask v u)
  | IfNoValue                             (MTask v u)
  | Always                                (MTask v u)
\end{lstdefinition}

The following listing shows an example of a step in action.
The \Cl{readPinBin} function will produce an \gls{mTask} task that will classify the value of an analog pin into four bins.
It also shows how the nature of embedding allows the host language to be used as a macro language.

\begin{lstexample}[language=Clean,caption={An example task using sequential combinators.},label={lst:readpinbin}]
readPinBin :: Main (MTask v Int) | mtask v
readPinBin = {main = readA A2 >>*.
	[  IfValue (\x->x <. lim) \_->rtrn (lit bin)
	\\ lim<-[64,128,192,256]
	&  bin<-[0..]]}\end{lstexample}

\subsection{\acrlong{SDS}}
In \gls{mTask} it is also possible to share data between tasks type safely using \glspl{SDS}.
Similar to functions, \glspl{SDS} can only be defined at the top level.

The \Cl{sds} class contains the function for defining and accessing \glspl{SDS}.
With the \Cl{sds} construction function, local \glspl{SDS} can be defined that are typed by functions in the host language to assure type safety.
The other functions in the class are for creating \Cl{get} and \Cl{set} tasks.
The \Cl{getSds} returns a task that constantly emits the value of the \gls{SDS} as an unstable task value.
\Cl{setSds} writes the given value to the task and re-emits it as a stable task value when it is done.

Definition~\ref{lst:def_sds_ex} and Example~\ref{lst:def_sds} present the definitions and an example.
The artificial example shows a task that mirrors a pin value to another pin using an \gls{SDS}.

\begin{lstdefinition}[language=Clean,caption={The \gls{mTask} class for \gls{SDS} tasks.},label={lst:def_sds}]
:: Sds a
class sds v where
	sds    :: ((v (Sds t)) -> In t (Main (MTask v u)))
		-> Main (MTask v u) | type t & type u
	getSds :: (v (Sds t))       -> MTask v t | type t
	setSds :: (v (Sds t)) (v t) -> MTask v t | type t
\end{lstdefinition}

\begin{lstexample}[language=Clean,caption={An example \gls{mTask} task using \glspl{SDS}.},label={lst:def_sds_ex}]
localvar :: Main (MTask v ()) | mtask v
localvar = sds \x=42 In {main =  rpeat (readA D13 >>~. setSds x)
                            .||. rpeat (getSds x  >>~. writeD D1)}
\end{lstexample}

\subsection{Lifted \acrlongpl{SDS}}\label{ssec:liftsds}
The \Cl{liftsds} class is defined to allow \gls{iTasks} \glspl{SDS} to be accessed from within \gls{mTask} tasks.
The function has a similar type as \Cl{sds} and creates an \gls{mTask} \gls{SDS} from an \gls{iTasks} \gls{SDS} so that it can be accessed using the class functions from the \Cl{sds} class.
Definition~\ref{lst:def_liftsds} and Example~\ref{lst:def_liftsds_ex} show an example of this where an \gls{iTasks} \gls{SDS} is used to control an \gls{LED} on a device.
When used, the server automatically notifies the device if the \gls{SDS} is written to and vice versa.
The \Cl{liftsds} class only makes sense in the context of actually executing backends.
Therefore this class is excluded from the \Cl{mtask} class collection.

\begin{lstdefinition}[language=Clean,caption={The \gls{mTask} class for \gls{iTasks} \glspl{SDS}.},label={lst:def_liftsds}]
:: Shared a // an iTasks SDS
class liftsds v | sds v where
	liftsds :: ((v (Sds t)) -> In (Shared t) (Main (MTask v u)))
		-> Main (MTask v u) | type t & type u
\end{lstdefinition}

\begin{lstexample}[language=Clean,caption={An example \gls{mTask} task using \gls{iTasks} \glspl{SDS}.},label={lst:def_liftsds_ex}]
lightSwitch :: (Shared Bool) -> Main (MTask v ()) | mtask v & liftsds v
lightSwitch s = liftsds \x=s In {main = rpeat (getSds x >>~. writeD D13)}
\end{lstexample}

\subsection{Interactive blinking}
Example~\ref{lst:readpinbin} showed that \gls{Clean} can be used as a macro language for \gls{mTask}, customizing the tasks using runtime values when needed.
\Glspl{SDS} can also be used to interact with the \gls{mTask} tasks during execution.
This can for example be used to let the user control the blinking frequency.
Example~\ref{lst:blinkInteractive} shows how the blinking frequency can be controlled by the user using \glspl{SDS}.

\begin{lstexample}[language=Clean,caption={An \gls{mTask} program for interactively changing the blinking frequency. (\texttt{blinkInteractive})},label={lst:blinkInteractive},numbers=left]
main :: Task Bool
main = enterDevice >>= \spec->withDevice spec
	\dev->withShared 500 \delayShare->                                 [+\label{blinkInteractive:share}+]
			liftmTask (blink delayShare) dev                           [+\label{blinkInteractive:devfunfro}+]
		-|| updateSharedInformation [] delayShare <<@ Title "Interval" [+\label{blinkInteractive:devfunto}+]
where
	blink :: (Shared s Int) -> Main (MTask v Bool) | mtask, liftsds v & RWShared s
	blink delayShare =
		liftsds \delaysh=delayShare [+\label{blinkInteractive:liftsds}+]
		In fun \blink = (\x->
			     writeD d2 x
			>>|. getSds delaysh     [+\label{blinkInteractive:getsds}+]
			>>~. delay
			>>|. blink (Not x))
		In {main = blink (lit True)}
\end{lstexample}

Line~\ref{blinkInteractive:share} shows the creation of the controlling \gls{iTasks} \gls{SDS} using \Cl{withShared} (see Section~\ref{sec:sds}).

Line~\ref{blinkInteractive:devfunfro}~and~\ref{blinkInteractive:devfunto} compromise the device function for \Cl{withDevice}.
It lifts the \Cl{blink} task to \gls{iTasks} and provides the user with an \Cl{updateSharedInformation} for the delay \gls{SDS}.
The \Cl{blink} task itself is hardly modified.
Line~\ref{blinkInteractive:liftsds} lifts the \gls{SDS} to an \gls{mTask} \gls{SDS} using \Cl{liftsds} (see Subsection~\ref{ssec:liftsds}).
Note that the \Cl{>>~.} combinator is used since the \Cl{getSds} task always yields an unstable value.
The lifted \gls{SDS} can be accessed as usual using the \Cl{getSds} task (Line~\ref{blinkInteractive:getsds}).
The value this yields is immediately fed to \Cl{delay}.
The \gls{mTask} machinery takes care of synchronising the \glspl{SDS}, when the user changes the delay, it is automatically reported to the device as well.

\begin{exercise}[Blink the builtin \acrshort{LED} on demand]\label{ass:blinkInteractive}
	Adapt the program in Example~\ref{lst:blinkInteractive} so that the user can control whether the \gls{LED} blinks or not.

	The \Cl{blink} function will then have the following type signature\\
	(\texttt{blinkInteractive}):

	\begin{lstexample}[language=Clean,postbreak=]
blink :: (Shared s Bool) -> Main (MTask v Bool) | mtask, liftsds v & RWShared s
	\end{lstexample}
\end{exercise}

\subsection{Peripherals}\label{ssec:peripherals}
Interaction with the \gls{GPIO} pins, and other peripherals for that matter, is also captured in basic tasks.
Some peripherals need initialization parameters and they are defined on the top level using host language functions similar to \glspl{SDS} and functions.
Typically from tasks reading peripherals such as sensors an unstable value can be observed.

\subsubsection{\acrlong{GPIO}}\label{ssec:gpio}
For each type of pin, there is a function that creates a task that --- given the pin --- either reads or writes the pin.
The class for \gls{GPIO} pin access is shown in Definition~\ref{lst:def_peripherals}.
The \Cl{readA}/\Cl{readD} task constantly yields the value of the analog pin as an unstable task value.
The \Cl{writeA}/\Cl{writeD} writes the given value to the given pin once and returns the written value as a stable task value.
Note that the digital \gls{GPIO} class is overloaded in the type of pin because analog pins can be used as digital ones as well.

\begin{lstdefinition}[language=Clean,caption={The \gls{mTask} classes for \gls{GPIO} tasks.},label={lst:def_peripherals}]
class aio v where
	readA  :: (v APin) -> MTask v Int
	writeA :: (v APin) (v Int) -> MTask v Int

class dio p v | pin p where
	readD  :: (v p) -> MTask v Bool
	writeD :: (v p) (v Bool) -> MTask v Bool

:: Pin = AnalogPin APin | DigitalPin DPin
class pin p :: p -> Pin | type p
instance pin APin, DPin
\end{lstdefinition}

\subsubsection{Peripherals}\label{ssec:dht}
All sensors have the same general structure in their classes and to illustrate this, the \gls{DHT} and \gls{LED} matrix are shown.
Using the \Cl{DHT} function, the device can be initialized with the correct parameters and used safely within the task.
The \Cl{temperature} and \Cl{humidity} task respectively query the temperature and the relative humidity from the sensor and yield it as an unstable task value.
This interface matches the \gls{C++} interface very closely but the semantics have been transformed to be suitable as a task.
Note that this class is not part of the \Cl{mtask} class collection and needs to be added as a separate constraint.
At the time of writing, \gls{mTask} supports in a similar fashion \glspl{DHT}, \gls{LED} matrices, ambient light sensors, passive infrared sensors, sound level sensors and air quality sensors.

\begin{lstdefinition}[language=Clean,caption={The \gls{mTask} classes for the \gls{DHT}.}]
:: DHT
:: DHTtype = DHT11 | DHT21 | DHT22

class dht v where
	DHT         :: p DHTtype ((v DHT) -> Main (v b)) -> Main (v b) | pin p & ...
	temperature :: (v DHT) -> MTask v Real
	humidity    :: (v DHT) -> MTask v Real
\end{lstdefinition}

\begin{lstdefinition}[language=Clean,caption={The \gls{mTask} classes for the \gls{LED} matrix.},label={lst:ledmatrix}]
:: LEDMatrix

class LEDMatrix v where
	ledmatrix   :: DPin DPin ((v LEDMatrix) -> Main (v b)) -> Main (v b) | type b
	LMDot       :: (v LEDMatrix) (v Int) (v Int) (v Bool) -> MTask v ()
	LMIntensity :: (v LEDMatrix) (v Int) -> MTask v ()
	LMClear     :: (v LEDMatrix) -> MTask v ()
	LMDisplay   :: (v LEDMatrix) -> MTask v ()
\end{lstdefinition}

%% file: arch.tex
\ifx\du\undefined%
  \newlength{\du}
\fi
\setlength{\du}{15\unitlength}
\Large\bf%
\begin{tikzpicture}[even odd rule]
\pgftransformxscale{1.000000}
\pgftransformyscale{-1.000000}
\definecolor{dialinecolor}{rgb}{0.000000, 0.000000, 0.000000}
\pgfsetstrokecolor{dialinecolor}
\pgfsetstrokeopacity{1.000000}
\definecolor{diafillcolor}{rgb}{1.000000, 1.000000, 1.000000}
\pgfsetfillcolor{diafillcolor}
\pgfsetfillopacity{1.000000}
\pgfsetlinewidth{0.100000\du}
\pgfsetdash{}{0pt}
\pgfsetmiterjoin
\pgfsetbuttcap
{\pgfsetcornersarced{\pgfpoint{0.000000\du}{0.000000\du}}\definecolor{dialinecolor}{rgb}{0.000000, 0.000000, 0.000000}
\pgfsetstrokecolor{dialinecolor}
\pgfsetstrokeopacity{1.000000}
\draw (15.000000\du,-9.000000\du)--(15.000000\du,0.000000\du)--(33.000000\du,0.000000\du)--(33.000000\du,-9.000000\du)--cycle;
}\pgfsetlinewidth{0.100000\du}
\pgfsetdash{}{0pt}
\pgfsetbuttcap
\pgfsetmiterjoin
\pgfsetlinewidth{0.100000\du}
\pgfsetbuttcap
\pgfsetmiterjoin
\pgfsetdash{}{0pt}
\definecolor{diafillcolor}{rgb}{1.000000, 1.000000, 1.000000}
\pgfsetfillcolor{diafillcolor}
\pgfsetfillopacity{1.000000}
\definecolor{dialinecolor}{rgb}{0.000000, 0.000000, 0.000000}
\pgfsetstrokecolor{dialinecolor}
\pgfsetstrokeopacity{1.000000}
\pgfpathmoveto{\pgfpoint{26.000000\du}{-7.406819\du}}
\pgfpathcurveto{\pgfpoint{26.480381\du}{-7.851705\du}}{\pgfpoint{26.720572\du}{-8.000000\du}}{\pgfpoint{27.200954\du}{-8.000000\du}}
\pgfpathcurveto{\pgfpoint{27.681335\du}{-8.000000\du}}{\pgfpoint{27.921526\du}{-7.851705\du}}{\pgfpoint{28.401907\du}{-7.406819\du}}
\pgfpathlineto{\pgfpoint{28.401907\du}{-5.034093\du}}
\pgfpathcurveto{\pgfpoint{27.921526\du}{-4.589207\du}}{\pgfpoint{27.681335\du}{-4.440911\du}}{\pgfpoint{27.200954\du}{-4.440911\du}}
\pgfpathcurveto{\pgfpoint{26.720572\du}{-4.440911\du}}{\pgfpoint{26.480381\du}{-4.589207\du}}{\pgfpoint{26.000000\du}{-5.034093\du}}
\pgfpathlineto{\pgfpoint{26.000000\du}{-7.406819\du}}
\pgfpathclose
\pgfusepath{fill,stroke}
\pgfsetbuttcap
\pgfsetmiterjoin
\pgfsetdash{}{0pt}
\definecolor{dialinecolor}{rgb}{0.000000, 0.000000, 0.000000}
\pgfsetstrokecolor{dialinecolor}
\pgfsetstrokeopacity{1.000000}
\pgfpathmoveto{\pgfpoint{26.000000\du}{-7.406819\du}}
\pgfpathcurveto{\pgfpoint{26.480381\du}{-6.961933\du}}{\pgfpoint{26.720572\du}{-6.813637\du}}{\pgfpoint{27.200954\du}{-6.813637\du}}
\pgfpathcurveto{\pgfpoint{27.681335\du}{-6.813637\du}}{\pgfpoint{27.921526\du}{-6.961933\du}}{\pgfpoint{28.401907\du}{-7.406819\du}}
\pgfusepath{stroke}
\definecolor{dialinecolor}{rgb}{0.000000, 0.000000, 0.000000}
\pgfsetstrokecolor{dialinecolor}
\pgfsetstrokeopacity{1.000000}
\definecolor{diafillcolor}{rgb}{0.000000, 0.000000, 0.000000}
\pgfsetfillcolor{diafillcolor}
\pgfsetfillopacity{1.000000}
\node[anchor=base,inner sep=0pt, outer sep=0pt,color=dialinecolor] at (27.200954\du,-5.723865\du){};
\pgfsetlinewidth{0.100000\du}
\pgfsetdash{}{0pt}
\pgfsetbuttcap
\pgfsetmiterjoin
\pgfsetlinewidth{0.100000\du}
\pgfsetbuttcap
\pgfsetmiterjoin
\pgfsetdash{}{0pt}
\definecolor{dialinecolor}{rgb}{0.000000, 0.000000, 0.000000}
\pgfsetstrokecolor{dialinecolor}
\pgfsetstrokeopacity{1.000000}
\draw (16.000000\du,-3.600000\du)--(20.600000\du,-3.600000\du)--(18.300000\du,-8.000000\du)--cycle;
\definecolor{dialinecolor}{rgb}{0.000000, 0.000000, 0.000000}
\pgfsetstrokecolor{dialinecolor}
\pgfsetstrokeopacity{1.000000}
\definecolor{diafillcolor}{rgb}{0.000000, 0.000000, 0.000000}
\pgfsetfillcolor{diafillcolor}
\pgfsetfillopacity{1.000000}
\node[anchor=base,inner sep=0pt, outer sep=0pt,color=dialinecolor] at (18.300000\du,-4.500000\du){};
\pgfsetlinewidth{0.100000\du}
\pgfsetdash{}{0pt}
\pgfsetmiterjoin
\pgfsetbuttcap
{\pgfsetcornersarced{\pgfpoint{0.000000\du}{0.000000\du}}\definecolor{dialinecolor}{rgb}{0.000000, 0.000000, 0.000000}
\pgfsetstrokecolor{dialinecolor}
\pgfsetstrokeopacity{1.000000}
\draw (10.000000\du,6.000000\du)--(10.000000\du,14.000000\du)--(37.500000\du,14.000000\du)--(37.500000\du,6.000000\du)--cycle;
}\pgfsetlinewidth{0.100000\du}
\pgfsetdash{}{0pt}
\pgfsetmiterjoin
{\pgfsetcornersarced{\pgfpoint{0.000000\du}{0.000000\du}}\definecolor{diafillcolor}{rgb}{1.000000, 1.000000, 1.000000}
\pgfsetfillcolor{diafillcolor}
\pgfsetfillopacity{1.000000}
\fill (10.600000\du,8.050000\du)--(10.600000\du,13.500000\du)--(18.200000\du,13.500000\du)--(18.200000\du,8.050000\du)--cycle;
}{\pgfsetcornersarced{\pgfpoint{0.000000\du}{0.000000\du}}\definecolor{dialinecolor}{rgb}{0.000000, 0.000000, 0.000000}
\pgfsetstrokecolor{dialinecolor}
\pgfsetstrokeopacity{1.000000}
\draw (10.600000\du,8.050000\du)--(10.600000\du,13.500000\du)--(18.200000\du,13.500000\du)--(18.200000\du,8.050000\du)--cycle;
}
\definecolor{dialinecolor}{rgb}{0.000000, 0.000000, 0.000000}
\pgfsetstrokecolor{dialinecolor}
\pgfsetstrokeopacity{1.000000}
\definecolor{diafillcolor}{rgb}{0.000000, 0.000000, 0.000000}
\pgfsetfillcolor{diafillcolor}
\pgfsetfillopacity{1.000000}
\node[anchor=base,inner sep=0pt, outer sep=0pt,color=dialinecolor] at (14.400000\du,10.970000\du){};
\pgfsetlinewidth{0.100000\du}
\pgfsetdash{}{0pt}
\pgfsetbuttcap
\pgfsetmiterjoin
\pgfsetlinewidth{0.100000\du}
\pgfsetbuttcap
\pgfsetmiterjoin
\pgfsetdash{}{0pt}
\definecolor{dialinecolor}{rgb}{0.000000, 0.000000, 0.000000}
\pgfsetstrokecolor{dialinecolor}
\pgfsetstrokeopacity{1.000000}
\draw (11.000000\du,12.900000\du)--(15.600000\du,12.900000\du)--(13.300000\du,8.500000\du)--cycle;
\definecolor{dialinecolor}{rgb}{0.000000, 0.000000, 0.000000}
\pgfsetstrokecolor{dialinecolor}
\pgfsetstrokeopacity{1.000000}
\definecolor{diafillcolor}{rgb}{0.000000, 0.000000, 0.000000}
\pgfsetfillcolor{diafillcolor}
\pgfsetfillopacity{1.000000}
\node[anchor=base,inner sep=0pt, outer sep=0pt,color=dialinecolor] at (13.300000\du,12.000000\du){};
\pgfsetlinewidth{0.100000\du}
\pgfsetdash{}{0pt}
\pgfsetbuttcap
\pgfsetmiterjoin
\pgfsetlinewidth{0.100000\du}
\pgfsetbuttcap
\pgfsetmiterjoin
\pgfsetdash{}{0pt}
\definecolor{diafillcolor}{rgb}{1.000000, 1.000000, 1.000000}
\pgfsetfillcolor{diafillcolor}
\pgfsetfillopacity{1.000000}
\definecolor{dialinecolor}{rgb}{0.000000, 0.000000, 0.000000}
\pgfsetstrokecolor{dialinecolor}
\pgfsetstrokeopacity{1.000000}
\pgfpathmoveto{\pgfpoint{13.400000\du}{9.066667\du}}
\pgfpathcurveto{\pgfpoint{13.800000\du}{8.791667\du}}{\pgfpoint{14.000000\du}{8.700000\du}}{\pgfpoint{14.400000\du}{8.700000\du}}
\pgfpathcurveto{\pgfpoint{14.800000\du}{8.700000\du}}{\pgfpoint{15.000000\du}{8.791667\du}}{\pgfpoint{15.400000\du}{9.066667\du}}
\pgfpathlineto{\pgfpoint{15.400000\du}{10.533333\du}}
\pgfpathcurveto{\pgfpoint{15.000000\du}{10.808333\du}}{\pgfpoint{14.800000\du}{10.900000\du}}{\pgfpoint{14.400000\du}{10.900000\du}}
\pgfpathcurveto{\pgfpoint{14.000000\du}{10.900000\du}}{\pgfpoint{13.800000\du}{10.808333\du}}{\pgfpoint{13.400000\du}{10.533333\du}}
\pgfpathlineto{\pgfpoint{13.400000\du}{9.066667\du}}
\pgfpathclose
\pgfusepath{fill,stroke}
\pgfsetbuttcap
\pgfsetmiterjoin
\pgfsetdash{}{0pt}
\definecolor{dialinecolor}{rgb}{0.000000, 0.000000, 0.000000}
\pgfsetstrokecolor{dialinecolor}
\pgfsetstrokeopacity{1.000000}
\pgfpathmoveto{\pgfpoint{13.400000\du}{9.066667\du}}
\pgfpathcurveto{\pgfpoint{13.800000\du}{9.341667\du}}{\pgfpoint{14.000000\du}{9.433333\du}}{\pgfpoint{14.400000\du}{9.433333\du}}
\pgfpathcurveto{\pgfpoint{14.800000\du}{9.433333\du}}{\pgfpoint{15.000000\du}{9.341667\du}}{\pgfpoint{15.400000\du}{9.066667\du}}
\pgfusepath{stroke}
\definecolor{dialinecolor}{rgb}{0.000000, 0.000000, 0.000000}
\pgfsetstrokecolor{dialinecolor}
\pgfsetstrokeopacity{1.000000}
\definecolor{diafillcolor}{rgb}{0.000000, 0.000000, 0.000000}
\pgfsetfillcolor{diafillcolor}
\pgfsetfillopacity{1.000000}
\node[anchor=base,inner sep=0pt, outer sep=0pt,color=dialinecolor] at (14.400000\du,10.183333\du){};
\pgfsetlinewidth{0.100000\du}
\pgfsetdash{}{0pt}
\pgfsetbuttcap
\pgfsetmiterjoin
\pgfsetlinewidth{0.100000\du}
\pgfsetbuttcap
\pgfsetmiterjoin
\pgfsetdash{}{0pt}
\definecolor{diafillcolor}{rgb}{1.000000, 1.000000, 1.000000}
\pgfsetfillcolor{diafillcolor}
\pgfsetfillopacity{1.000000}
\definecolor{dialinecolor}{rgb}{0.000000, 0.000000, 0.000000}
\pgfsetstrokecolor{dialinecolor}
\pgfsetstrokeopacity{1.000000}
\pgfpathmoveto{\pgfpoint{14.600000\du}{10.066667\du}}
\pgfpathcurveto{\pgfpoint{15.000000\du}{9.791667\du}}{\pgfpoint{15.200000\du}{9.700000\du}}{\pgfpoint{15.600000\du}{9.700000\du}}
\pgfpathcurveto{\pgfpoint{16.000000\du}{9.700000\du}}{\pgfpoint{16.200000\du}{9.791667\du}}{\pgfpoint{16.600000\du}{10.066667\du}}
\pgfpathlineto{\pgfpoint{16.600000\du}{11.533333\du}}
\pgfpathcurveto{\pgfpoint{16.200000\du}{11.808333\du}}{\pgfpoint{16.000000\du}{11.900000\du}}{\pgfpoint{15.600000\du}{11.900000\du}}
\pgfpathcurveto{\pgfpoint{15.200000\du}{11.900000\du}}{\pgfpoint{15.000000\du}{11.808333\du}}{\pgfpoint{14.600000\du}{11.533333\du}}
\pgfpathlineto{\pgfpoint{14.600000\du}{10.066667\du}}
\pgfpathclose
\pgfusepath{fill,stroke}
\pgfsetbuttcap
\pgfsetmiterjoin
\pgfsetdash{}{0pt}
\definecolor{dialinecolor}{rgb}{0.000000, 0.000000, 0.000000}
\pgfsetstrokecolor{dialinecolor}
\pgfsetstrokeopacity{1.000000}
\pgfpathmoveto{\pgfpoint{14.600000\du}{10.066667\du}}
\pgfpathcurveto{\pgfpoint{15.000000\du}{10.341667\du}}{\pgfpoint{15.200000\du}{10.433333\du}}{\pgfpoint{15.600000\du}{10.433333\du}}
\pgfpathcurveto{\pgfpoint{16.000000\du}{10.433333\du}}{\pgfpoint{16.200000\du}{10.341667\du}}{\pgfpoint{16.600000\du}{10.066667\du}}
\pgfusepath{stroke}
\definecolor{dialinecolor}{rgb}{0.000000, 0.000000, 0.000000}
\pgfsetstrokecolor{dialinecolor}
\pgfsetstrokeopacity{1.000000}
\definecolor{diafillcolor}{rgb}{0.000000, 0.000000, 0.000000}
\pgfsetfillcolor{diafillcolor}
\pgfsetfillopacity{1.000000}
\node[anchor=base,inner sep=0pt, outer sep=0pt,color=dialinecolor] at (15.600000\du,11.183333\du){};
\pgfsetlinewidth{0.100000\du}
\pgfsetdash{}{0pt}
\pgfsetbuttcap
\pgfsetmiterjoin
\pgfsetlinewidth{0.100000\du}
\pgfsetbuttcap
\pgfsetmiterjoin
\pgfsetdash{}{0pt}
\definecolor{diafillcolor}{rgb}{1.000000, 1.000000, 1.000000}
\pgfsetfillcolor{diafillcolor}
\pgfsetfillopacity{1.000000}
\definecolor{dialinecolor}{rgb}{0.000000, 0.000000, 0.000000}
\pgfsetstrokecolor{dialinecolor}
\pgfsetstrokeopacity{1.000000}
\pgfpathmoveto{\pgfpoint{15.600000\du}{11.066667\du}}
\pgfpathcurveto{\pgfpoint{16.000000\du}{10.791667\du}}{\pgfpoint{16.200000\du}{10.700000\du}}{\pgfpoint{16.600000\du}{10.700000\du}}
\pgfpathcurveto{\pgfpoint{17.000000\du}{10.700000\du}}{\pgfpoint{17.200000\du}{10.791667\du}}{\pgfpoint{17.600000\du}{11.066667\du}}
\pgfpathlineto{\pgfpoint{17.600000\du}{12.533333\du}}
\pgfpathcurveto{\pgfpoint{17.200000\du}{12.808333\du}}{\pgfpoint{17.000000\du}{12.900000\du}}{\pgfpoint{16.600000\du}{12.900000\du}}
\pgfpathcurveto{\pgfpoint{16.200000\du}{12.900000\du}}{\pgfpoint{16.000000\du}{12.808333\du}}{\pgfpoint{15.600000\du}{12.533333\du}}
\pgfpathlineto{\pgfpoint{15.600000\du}{11.066667\du}}
\pgfpathclose
\pgfusepath{fill,stroke}
\pgfsetbuttcap
\pgfsetmiterjoin
\pgfsetdash{}{0pt}
\definecolor{dialinecolor}{rgb}{0.000000, 0.000000, 0.000000}
\pgfsetstrokecolor{dialinecolor}
\pgfsetstrokeopacity{1.000000}
\pgfpathmoveto{\pgfpoint{15.600000\du}{11.066667\du}}
\pgfpathcurveto{\pgfpoint{16.000000\du}{11.341667\du}}{\pgfpoint{16.200000\du}{11.433333\du}}{\pgfpoint{16.600000\du}{11.433333\du}}
\pgfpathcurveto{\pgfpoint{17.000000\du}{11.433333\du}}{\pgfpoint{17.200000\du}{11.341667\du}}{\pgfpoint{17.600000\du}{11.066667\du}}
\pgfusepath{stroke}
\definecolor{dialinecolor}{rgb}{0.000000, 0.000000, 0.000000}
\pgfsetstrokecolor{dialinecolor}
\pgfsetstrokeopacity{1.000000}
\definecolor{diafillcolor}{rgb}{0.000000, 0.000000, 0.000000}
\pgfsetfillcolor{diafillcolor}
\pgfsetfillopacity{1.000000}
\node[anchor=base,inner sep=0pt, outer sep=0pt,color=dialinecolor] at (16.600000\du,12.183333\du){};
\definecolor{dialinecolor}{rgb}{0.000000, 0.000000, 0.000000}
\pgfsetstrokecolor{dialinecolor}
\pgfsetstrokeopacity{1.000000}
\definecolor{diafillcolor}{rgb}{0.000000, 0.000000, 0.000000}
\pgfsetfillcolor{diafillcolor}
\pgfsetfillopacity{1.000000}
\node[anchor=base west,inner sep=0pt,outer sep=0pt,color=dialinecolor] at (12.600000\du,7.550000\du){mTask};
\definecolor{dialinecolor}{rgb}{0.000000, 0.000000, 0.000000}
\pgfsetstrokecolor{dialinecolor}
\pgfsetstrokeopacity{1.000000}
\definecolor{diafillcolor}{rgb}{0.000000, 0.000000, 0.000000}
\pgfsetfillcolor{diafillcolor}
\pgfsetfillopacity{1.000000}
\node[anchor=base west,inner sep=0pt,outer sep=0pt,color=dialinecolor] at (34.500000\du,10.500000\du){...};
\pgfsetlinewidth{0.100000\du}
\pgfsetdash{}{0pt}
\pgfsetmiterjoin
{\pgfsetcornersarced{\pgfpoint{0.000000\du}{0.000000\du}}\definecolor{diafillcolor}{rgb}{1.000000, 1.000000, 1.000000}
\pgfsetfillcolor{diafillcolor}
\pgfsetfillopacity{1.000000}
\fill (26.500000\du,8.000000\du)--(26.500000\du,13.450000\du)--(34.100000\du,13.450000\du)--(34.100000\du,8.000000\du)--cycle;
}{\pgfsetcornersarced{\pgfpoint{0.000000\du}{0.000000\du}}\definecolor{dialinecolor}{rgb}{0.000000, 0.000000, 0.000000}
\pgfsetstrokecolor{dialinecolor}
\pgfsetstrokeopacity{1.000000}
\draw (26.500000\du,8.000000\du)--(26.500000\du,13.450000\du)--(34.100000\du,13.450000\du)--(34.100000\du,8.000000\du)--cycle;
}
\definecolor{dialinecolor}{rgb}{0.000000, 0.000000, 0.000000}
\pgfsetstrokecolor{dialinecolor}
\pgfsetstrokeopacity{1.000000}
\definecolor{diafillcolor}{rgb}{0.000000, 0.000000, 0.000000}
\pgfsetfillcolor{diafillcolor}
\pgfsetfillopacity{1.000000}
\node[anchor=base,inner sep=0pt, outer sep=0pt,color=dialinecolor] at (30.300000\du,10.920000\du){};
\pgfsetlinewidth{0.100000\du}
\pgfsetdash{}{0pt}
\pgfsetbuttcap
\pgfsetmiterjoin
\pgfsetlinewidth{0.100000\du}
\pgfsetbuttcap
\pgfsetmiterjoin
\pgfsetdash{}{0pt}
\definecolor{dialinecolor}{rgb}{0.000000, 0.000000, 0.000000}
\pgfsetstrokecolor{dialinecolor}
\pgfsetstrokeopacity{1.000000}
\draw (26.900000\du,12.850000\du)--(31.500000\du,12.850000\du)--(29.200000\du,8.450000\du)--cycle;
\definecolor{dialinecolor}{rgb}{0.000000, 0.000000, 0.000000}
\pgfsetstrokecolor{dialinecolor}
\pgfsetstrokeopacity{1.000000}
\definecolor{diafillcolor}{rgb}{0.000000, 0.000000, 0.000000}
\pgfsetfillcolor{diafillcolor}
\pgfsetfillopacity{1.000000}
\node[anchor=base,inner sep=0pt, outer sep=0pt,color=dialinecolor] at (29.200000\du,11.950000\du){};
\pgfsetlinewidth{0.100000\du}
\pgfsetdash{}{0pt}
\pgfsetbuttcap
\pgfsetmiterjoin
\pgfsetlinewidth{0.100000\du}
\pgfsetbuttcap
\pgfsetmiterjoin
\pgfsetdash{}{0pt}
\definecolor{diafillcolor}{rgb}{1.000000, 1.000000, 1.000000}
\pgfsetfillcolor{diafillcolor}
\pgfsetfillopacity{1.000000}
\definecolor{dialinecolor}{rgb}{0.000000, 0.000000, 0.000000}
\pgfsetstrokecolor{dialinecolor}
\pgfsetstrokeopacity{1.000000}
\pgfpathmoveto{\pgfpoint{29.300000\du}{9.016667\du}}
\pgfpathcurveto{\pgfpoint{29.700000\du}{8.741667\du}}{\pgfpoint{29.900000\du}{8.650000\du}}{\pgfpoint{30.300000\du}{8.650000\du}}
\pgfpathcurveto{\pgfpoint{30.700000\du}{8.650000\du}}{\pgfpoint{30.900000\du}{8.741667\du}}{\pgfpoint{31.300000\du}{9.016667\du}}
\pgfpathlineto{\pgfpoint{31.300000\du}{10.483333\du}}
\pgfpathcurveto{\pgfpoint{30.900000\du}{10.758333\du}}{\pgfpoint{30.700000\du}{10.850000\du}}{\pgfpoint{30.300000\du}{10.850000\du}}
\pgfpathcurveto{\pgfpoint{29.900000\du}{10.850000\du}}{\pgfpoint{29.700000\du}{10.758333\du}}{\pgfpoint{29.300000\du}{10.483333\du}}
\pgfpathlineto{\pgfpoint{29.300000\du}{9.016667\du}}
\pgfpathclose
\pgfusepath{fill,stroke}
\pgfsetbuttcap
\pgfsetmiterjoin
\pgfsetdash{}{0pt}
\definecolor{dialinecolor}{rgb}{0.000000, 0.000000, 0.000000}
\pgfsetstrokecolor{dialinecolor}
\pgfsetstrokeopacity{1.000000}
\pgfpathmoveto{\pgfpoint{29.300000\du}{9.016667\du}}
\pgfpathcurveto{\pgfpoint{29.700000\du}{9.291667\du}}{\pgfpoint{29.900000\du}{9.383333\du}}{\pgfpoint{30.300000\du}{9.383333\du}}
\pgfpathcurveto{\pgfpoint{30.700000\du}{9.383333\du}}{\pgfpoint{30.900000\du}{9.291667\du}}{\pgfpoint{31.300000\du}{9.016667\du}}
\pgfusepath{stroke}
\definecolor{dialinecolor}{rgb}{0.000000, 0.000000, 0.000000}
\pgfsetstrokecolor{dialinecolor}
\pgfsetstrokeopacity{1.000000}
\definecolor{diafillcolor}{rgb}{0.000000, 0.000000, 0.000000}
\pgfsetfillcolor{diafillcolor}
\pgfsetfillopacity{1.000000}
\node[anchor=base,inner sep=0pt, outer sep=0pt,color=dialinecolor] at (30.300000\du,10.133333\du){};
\pgfsetlinewidth{0.100000\du}
\pgfsetdash{}{0pt}
\pgfsetbuttcap
\pgfsetmiterjoin
\pgfsetlinewidth{0.100000\du}
\pgfsetbuttcap
\pgfsetmiterjoin
\pgfsetdash{}{0pt}
\definecolor{diafillcolor}{rgb}{1.000000, 1.000000, 1.000000}
\pgfsetfillcolor{diafillcolor}
\pgfsetfillopacity{1.000000}
\definecolor{dialinecolor}{rgb}{0.000000, 0.000000, 0.000000}
\pgfsetstrokecolor{dialinecolor}
\pgfsetstrokeopacity{1.000000}
\pgfpathmoveto{\pgfpoint{30.500000\du}{10.016667\du}}
\pgfpathcurveto{\pgfpoint{30.900000\du}{9.741667\du}}{\pgfpoint{31.100000\du}{9.650000\du}}{\pgfpoint{31.500000\du}{9.650000\du}}
\pgfpathcurveto{\pgfpoint{31.900000\du}{9.650000\du}}{\pgfpoint{32.100000\du}{9.741667\du}}{\pgfpoint{32.500000\du}{10.016667\du}}
\pgfpathlineto{\pgfpoint{32.500000\du}{11.483333\du}}
\pgfpathcurveto{\pgfpoint{32.100000\du}{11.758333\du}}{\pgfpoint{31.900000\du}{11.850000\du}}{\pgfpoint{31.500000\du}{11.850000\du}}
\pgfpathcurveto{\pgfpoint{31.100000\du}{11.850000\du}}{\pgfpoint{30.900000\du}{11.758333\du}}{\pgfpoint{30.500000\du}{11.483333\du}}
\pgfpathlineto{\pgfpoint{30.500000\du}{10.016667\du}}
\pgfpathclose
\pgfusepath{fill,stroke}
\pgfsetbuttcap
\pgfsetmiterjoin
\pgfsetdash{}{0pt}
\definecolor{dialinecolor}{rgb}{0.000000, 0.000000, 0.000000}
\pgfsetstrokecolor{dialinecolor}
\pgfsetstrokeopacity{1.000000}
\pgfpathmoveto{\pgfpoint{30.500000\du}{10.016667\du}}
\pgfpathcurveto{\pgfpoint{30.900000\du}{10.291667\du}}{\pgfpoint{31.100000\du}{10.383333\du}}{\pgfpoint{31.500000\du}{10.383333\du}}
\pgfpathcurveto{\pgfpoint{31.900000\du}{10.383333\du}}{\pgfpoint{32.100000\du}{10.291667\du}}{\pgfpoint{32.500000\du}{10.016667\du}}
\pgfusepath{stroke}
\definecolor{dialinecolor}{rgb}{0.000000, 0.000000, 0.000000}
\pgfsetstrokecolor{dialinecolor}
\pgfsetstrokeopacity{1.000000}
\definecolor{diafillcolor}{rgb}{0.000000, 0.000000, 0.000000}
\pgfsetfillcolor{diafillcolor}
\pgfsetfillopacity{1.000000}
\node[anchor=base,inner sep=0pt, outer sep=0pt,color=dialinecolor] at (31.500000\du,11.133333\du){};
\pgfsetlinewidth{0.100000\du}
\pgfsetdash{}{0pt}
\pgfsetbuttcap
\pgfsetmiterjoin
\pgfsetlinewidth{0.100000\du}
\pgfsetbuttcap
\pgfsetmiterjoin
\pgfsetdash{}{0pt}
\definecolor{diafillcolor}{rgb}{1.000000, 1.000000, 1.000000}
\pgfsetfillcolor{diafillcolor}
\pgfsetfillopacity{1.000000}
\definecolor{dialinecolor}{rgb}{0.000000, 0.000000, 0.000000}
\pgfsetstrokecolor{dialinecolor}
\pgfsetstrokeopacity{1.000000}
\pgfpathmoveto{\pgfpoint{31.500000\du}{11.016667\du}}
\pgfpathcurveto{\pgfpoint{31.900000\du}{10.741667\du}}{\pgfpoint{32.100000\du}{10.650000\du}}{\pgfpoint{32.500000\du}{10.650000\du}}
\pgfpathcurveto{\pgfpoint{32.900000\du}{10.650000\du}}{\pgfpoint{33.100000\du}{10.741667\du}}{\pgfpoint{33.500000\du}{11.016667\du}}
\pgfpathlineto{\pgfpoint{33.500000\du}{12.483333\du}}
\pgfpathcurveto{\pgfpoint{33.100000\du}{12.758333\du}}{\pgfpoint{32.900000\du}{12.850000\du}}{\pgfpoint{32.500000\du}{12.850000\du}}
\pgfpathcurveto{\pgfpoint{32.100000\du}{12.850000\du}}{\pgfpoint{31.900000\du}{12.758333\du}}{\pgfpoint{31.500000\du}{12.483333\du}}
\pgfpathlineto{\pgfpoint{31.500000\du}{11.016667\du}}
\pgfpathclose
\pgfusepath{fill,stroke}
\pgfsetbuttcap
\pgfsetmiterjoin
\pgfsetdash{}{0pt}
\definecolor{dialinecolor}{rgb}{0.000000, 0.000000, 0.000000}
\pgfsetstrokecolor{dialinecolor}
\pgfsetstrokeopacity{1.000000}
\pgfpathmoveto{\pgfpoint{31.500000\du}{11.016667\du}}
\pgfpathcurveto{\pgfpoint{31.900000\du}{11.291667\du}}{\pgfpoint{32.100000\du}{11.383333\du}}{\pgfpoint{32.500000\du}{11.383333\du}}
\pgfpathcurveto{\pgfpoint{32.900000\du}{11.383333\du}}{\pgfpoint{33.100000\du}{11.291667\du}}{\pgfpoint{33.500000\du}{11.016667\du}}
\pgfusepath{stroke}
\definecolor{dialinecolor}{rgb}{0.000000, 0.000000, 0.000000}
\pgfsetstrokecolor{dialinecolor}
\pgfsetstrokeopacity{1.000000}
\definecolor{diafillcolor}{rgb}{0.000000, 0.000000, 0.000000}
\pgfsetfillcolor{diafillcolor}
\pgfsetfillopacity{1.000000}
\node[anchor=base,inner sep=0pt, outer sep=0pt,color=dialinecolor] at (32.500000\du,12.133333\du){};
\definecolor{dialinecolor}{rgb}{0.000000, 0.000000, 0.000000}
\pgfsetstrokecolor{dialinecolor}
\pgfsetstrokeopacity{1.000000}
\definecolor{diafillcolor}{rgb}{0.000000, 0.000000, 0.000000}
\pgfsetfillcolor{diafillcolor}
\pgfsetfillopacity{1.000000}
\node[anchor=base west,inner sep=0pt,outer sep=0pt,color=dialinecolor] at (28.700000\du,7.500000\du){mTask};
\pgfsetlinewidth{0.100000\du}
\pgfsetdash{}{0pt}
\pgfsetmiterjoin
{\pgfsetcornersarced{\pgfpoint{0.000000\du}{0.000000\du}}\definecolor{diafillcolor}{rgb}{1.000000, 1.000000, 1.000000}
\pgfsetfillcolor{diafillcolor}
\pgfsetfillopacity{1.000000}
\fill (18.500000\du,8.000000\du)--(18.500000\du,13.450000\du)--(26.100000\du,13.450000\du)--(26.100000\du,8.000000\du)--cycle;
}{\pgfsetcornersarced{\pgfpoint{0.000000\du}{0.000000\du}}\definecolor{dialinecolor}{rgb}{0.000000, 0.000000, 0.000000}
\pgfsetstrokecolor{dialinecolor}
\pgfsetstrokeopacity{1.000000}
\draw (18.500000\du,8.000000\du)--(18.500000\du,13.450000\du)--(26.100000\du,13.450000\du)--(26.100000\du,8.000000\du)--cycle;
}
\definecolor{dialinecolor}{rgb}{0.000000, 0.000000, 0.000000}
\pgfsetstrokecolor{dialinecolor}
\pgfsetstrokeopacity{1.000000}
\definecolor{diafillcolor}{rgb}{0.000000, 0.000000, 0.000000}
\pgfsetfillcolor{diafillcolor}
\pgfsetfillopacity{1.000000}
\node[anchor=base,inner sep=0pt, outer sep=0pt,color=dialinecolor] at (22.300000\du,10.920000\du){};
\pgfsetlinewidth{0.100000\du}
\pgfsetdash{}{0pt}
\pgfsetbuttcap
\pgfsetmiterjoin
\pgfsetlinewidth{0.100000\du}
\pgfsetbuttcap
\pgfsetmiterjoin
\pgfsetdash{}{0pt}
\definecolor{dialinecolor}{rgb}{0.000000, 0.000000, 0.000000}
\pgfsetstrokecolor{dialinecolor}
\pgfsetstrokeopacity{1.000000}
\draw (18.900000\du,12.850000\du)--(23.500000\du,12.850000\du)--(21.200000\du,8.450000\du)--cycle;
\definecolor{dialinecolor}{rgb}{0.000000, 0.000000, 0.000000}
\pgfsetstrokecolor{dialinecolor}
\pgfsetstrokeopacity{1.000000}
\definecolor{diafillcolor}{rgb}{0.000000, 0.000000, 0.000000}
\pgfsetfillcolor{diafillcolor}
\pgfsetfillopacity{1.000000}
\node[anchor=base,inner sep=0pt, outer sep=0pt,color=dialinecolor] at (21.200000\du,11.950000\du){};
\pgfsetlinewidth{0.100000\du}
\pgfsetdash{}{0pt}
\pgfsetbuttcap
\pgfsetmiterjoin
\pgfsetlinewidth{0.100000\du}
\pgfsetbuttcap
\pgfsetmiterjoin
\pgfsetdash{}{0pt}
\definecolor{diafillcolor}{rgb}{1.000000, 1.000000, 1.000000}
\pgfsetfillcolor{diafillcolor}
\pgfsetfillopacity{1.000000}
\definecolor{dialinecolor}{rgb}{0.000000, 0.000000, 0.000000}
\pgfsetstrokecolor{dialinecolor}
\pgfsetstrokeopacity{1.000000}
\pgfpathmoveto{\pgfpoint{21.300000\du}{9.016667\du}}
\pgfpathcurveto{\pgfpoint{21.700000\du}{8.741667\du}}{\pgfpoint{21.900000\du}{8.650000\du}}{\pgfpoint{22.300000\du}{8.650000\du}}
\pgfpathcurveto{\pgfpoint{22.700000\du}{8.650000\du}}{\pgfpoint{22.900000\du}{8.741667\du}}{\pgfpoint{23.300000\du}{9.016667\du}}
\pgfpathlineto{\pgfpoint{23.300000\du}{10.483333\du}}
\pgfpathcurveto{\pgfpoint{22.900000\du}{10.758333\du}}{\pgfpoint{22.700000\du}{10.850000\du}}{\pgfpoint{22.300000\du}{10.850000\du}}
\pgfpathcurveto{\pgfpoint{21.900000\du}{10.850000\du}}{\pgfpoint{21.700000\du}{10.758333\du}}{\pgfpoint{21.300000\du}{10.483333\du}}
\pgfpathlineto{\pgfpoint{21.300000\du}{9.016667\du}}
\pgfpathclose
\pgfusepath{fill,stroke}
\pgfsetbuttcap
\pgfsetmiterjoin
\pgfsetdash{}{0pt}
\definecolor{dialinecolor}{rgb}{0.000000, 0.000000, 0.000000}
\pgfsetstrokecolor{dialinecolor}
\pgfsetstrokeopacity{1.000000}
\pgfpathmoveto{\pgfpoint{21.300000\du}{9.016667\du}}
\pgfpathcurveto{\pgfpoint{21.700000\du}{9.291667\du}}{\pgfpoint{21.900000\du}{9.383333\du}}{\pgfpoint{22.300000\du}{9.383333\du}}
\pgfpathcurveto{\pgfpoint{22.700000\du}{9.383333\du}}{\pgfpoint{22.900000\du}{9.291667\du}}{\pgfpoint{23.300000\du}{9.016667\du}}
\pgfusepath{stroke}
\definecolor{dialinecolor}{rgb}{0.000000, 0.000000, 0.000000}
\pgfsetstrokecolor{dialinecolor}
\pgfsetstrokeopacity{1.000000}
\definecolor{diafillcolor}{rgb}{0.000000, 0.000000, 0.000000}
\pgfsetfillcolor{diafillcolor}
\pgfsetfillopacity{1.000000}
\node[anchor=base,inner sep=0pt, outer sep=0pt,color=dialinecolor] at (22.300000\du,10.133333\du){};
\pgfsetlinewidth{0.100000\du}
\pgfsetdash{}{0pt}
\pgfsetbuttcap
\pgfsetmiterjoin
\pgfsetlinewidth{0.100000\du}
\pgfsetbuttcap
\pgfsetmiterjoin
\pgfsetdash{}{0pt}
\definecolor{diafillcolor}{rgb}{1.000000, 1.000000, 1.000000}
\pgfsetfillcolor{diafillcolor}
\pgfsetfillopacity{1.000000}
\definecolor{dialinecolor}{rgb}{0.000000, 0.000000, 0.000000}
\pgfsetstrokecolor{dialinecolor}
\pgfsetstrokeopacity{1.000000}
\pgfpathmoveto{\pgfpoint{22.500000\du}{10.016667\du}}
\pgfpathcurveto{\pgfpoint{22.900000\du}{9.741667\du}}{\pgfpoint{23.100000\du}{9.650000\du}}{\pgfpoint{23.500000\du}{9.650000\du}}
\pgfpathcurveto{\pgfpoint{23.900000\du}{9.650000\du}}{\pgfpoint{24.100000\du}{9.741667\du}}{\pgfpoint{24.500000\du}{10.016667\du}}
\pgfpathlineto{\pgfpoint{24.500000\du}{11.483333\du}}
\pgfpathcurveto{\pgfpoint{24.100000\du}{11.758333\du}}{\pgfpoint{23.900000\du}{11.850000\du}}{\pgfpoint{23.500000\du}{11.850000\du}}
\pgfpathcurveto{\pgfpoint{23.100000\du}{11.850000\du}}{\pgfpoint{22.900000\du}{11.758333\du}}{\pgfpoint{22.500000\du}{11.483333\du}}
\pgfpathlineto{\pgfpoint{22.500000\du}{10.016667\du}}
\pgfpathclose
\pgfusepath{fill,stroke}
\pgfsetbuttcap
\pgfsetmiterjoin
\pgfsetdash{}{0pt}
\definecolor{dialinecolor}{rgb}{0.000000, 0.000000, 0.000000}
\pgfsetstrokecolor{dialinecolor}
\pgfsetstrokeopacity{1.000000}
\pgfpathmoveto{\pgfpoint{22.500000\du}{10.016667\du}}
\pgfpathcurveto{\pgfpoint{22.900000\du}{10.291667\du}}{\pgfpoint{23.100000\du}{10.383333\du}}{\pgfpoint{23.500000\du}{10.383333\du}}
\pgfpathcurveto{\pgfpoint{23.900000\du}{10.383333\du}}{\pgfpoint{24.100000\du}{10.291667\du}}{\pgfpoint{24.500000\du}{10.016667\du}}
\pgfusepath{stroke}
\definecolor{dialinecolor}{rgb}{0.000000, 0.000000, 0.000000}
\pgfsetstrokecolor{dialinecolor}
\pgfsetstrokeopacity{1.000000}
\definecolor{diafillcolor}{rgb}{0.000000, 0.000000, 0.000000}
\pgfsetfillcolor{diafillcolor}
\pgfsetfillopacity{1.000000}
\node[anchor=base,inner sep=0pt, outer sep=0pt,color=dialinecolor] at (23.500000\du,11.133333\du){};
\pgfsetlinewidth{0.100000\du}
\pgfsetdash{}{0pt}
\pgfsetbuttcap
\pgfsetmiterjoin
\pgfsetlinewidth{0.100000\du}
\pgfsetbuttcap
\pgfsetmiterjoin
\pgfsetdash{}{0pt}
\definecolor{diafillcolor}{rgb}{1.000000, 1.000000, 1.000000}
\pgfsetfillcolor{diafillcolor}
\pgfsetfillopacity{1.000000}
\definecolor{dialinecolor}{rgb}{0.000000, 0.000000, 0.000000}
\pgfsetstrokecolor{dialinecolor}
\pgfsetstrokeopacity{1.000000}
\pgfpathmoveto{\pgfpoint{23.500000\du}{11.016667\du}}
\pgfpathcurveto{\pgfpoint{23.900000\du}{10.741667\du}}{\pgfpoint{24.100000\du}{10.650000\du}}{\pgfpoint{24.500000\du}{10.650000\du}}
\pgfpathcurveto{\pgfpoint{24.900000\du}{10.650000\du}}{\pgfpoint{25.100000\du}{10.741667\du}}{\pgfpoint{25.500000\du}{11.016667\du}}
\pgfpathlineto{\pgfpoint{25.500000\du}{12.483333\du}}
\pgfpathcurveto{\pgfpoint{25.100000\du}{12.758333\du}}{\pgfpoint{24.900000\du}{12.850000\du}}{\pgfpoint{24.500000\du}{12.850000\du}}
\pgfpathcurveto{\pgfpoint{24.100000\du}{12.850000\du}}{\pgfpoint{23.900000\du}{12.758333\du}}{\pgfpoint{23.500000\du}{12.483333\du}}
\pgfpathlineto{\pgfpoint{23.500000\du}{11.016667\du}}
\pgfpathclose
\pgfusepath{fill,stroke}
\pgfsetbuttcap
\pgfsetmiterjoin
\pgfsetdash{}{0pt}
\definecolor{dialinecolor}{rgb}{0.000000, 0.000000, 0.000000}
\pgfsetstrokecolor{dialinecolor}
\pgfsetstrokeopacity{1.000000}
\pgfpathmoveto{\pgfpoint{23.500000\du}{11.016667\du}}
\pgfpathcurveto{\pgfpoint{23.900000\du}{11.291667\du}}{\pgfpoint{24.100000\du}{11.383333\du}}{\pgfpoint{24.500000\du}{11.383333\du}}
\pgfpathcurveto{\pgfpoint{24.900000\du}{11.383333\du}}{\pgfpoint{25.100000\du}{11.291667\du}}{\pgfpoint{25.500000\du}{11.016667\du}}
\pgfusepath{stroke}
\definecolor{dialinecolor}{rgb}{0.000000, 0.000000, 0.000000}
\pgfsetstrokecolor{dialinecolor}
\pgfsetstrokeopacity{1.000000}
\definecolor{diafillcolor}{rgb}{0.000000, 0.000000, 0.000000}
\pgfsetfillcolor{diafillcolor}
\pgfsetfillopacity{1.000000}
\node[anchor=base,inner sep=0pt, outer sep=0pt,color=dialinecolor] at (24.500000\du,12.133333\du){};
\definecolor{dialinecolor}{rgb}{0.000000, 0.000000, 0.000000}
\pgfsetstrokecolor{dialinecolor}
\pgfsetstrokeopacity{1.000000}
\definecolor{diafillcolor}{rgb}{0.000000, 0.000000, 0.000000}
\pgfsetfillcolor{diafillcolor}
\pgfsetfillopacity{1.000000}
\node[anchor=base west,inner sep=0pt,outer sep=0pt,color=dialinecolor] at (20.600000\du,7.500000\du){mTask};
\pgfsetlinewidth{0.100000\du}
\pgfsetdash{}{0pt}
\pgfsetmiterjoin
\pgfsetbuttcap
{\pgfsetcornersarced{\pgfpoint{0.000000\du}{0.000000\du}}\definecolor{dialinecolor}{rgb}{0.000000, 0.000000, 0.000000}
\pgfsetstrokecolor{dialinecolor}
\pgfsetstrokeopacity{1.000000}
\draw (12.500000\du,6.500000\du)--(12.500000\du,8.000000\du)--(16.000000\du,8.000000\du)--(16.000000\du,6.500000\du)--cycle;
}\pgfsetlinewidth{0.100000\du}
\pgfsetdash{}{0pt}
\pgfsetmiterjoin
\pgfsetbuttcap
{\pgfsetcornersarced{\pgfpoint{0.000000\du}{0.000000\du}}\definecolor{dialinecolor}{rgb}{0.000000, 0.000000, 0.000000}
\pgfsetstrokecolor{dialinecolor}
\pgfsetstrokeopacity{1.000000}
\draw (20.500000\du,6.500000\du)--(20.500000\du,8.000000\du)--(24.000000\du,8.000000\du)--(24.000000\du,6.500000\du)--cycle;
}\pgfsetlinewidth{0.100000\du}
\pgfsetdash{}{0pt}
\pgfsetmiterjoin
\pgfsetbuttcap
{\pgfsetcornersarced{\pgfpoint{0.000000\du}{0.000000\du}}\definecolor{dialinecolor}{rgb}{0.000000, 0.000000, 0.000000}
\pgfsetstrokecolor{dialinecolor}
\pgfsetstrokeopacity{1.000000}
\draw (28.500000\du,6.500000\du)--(28.500000\du,8.000000\du)--(32.000000\du,8.000000\du)--(32.000000\du,6.500000\du)--cycle;
}\pgfsetlinewidth{0.100000\du}
\pgfsetdash{}{0pt}
\pgfsetmiterjoin
{\pgfsetcornersarced{\pgfpoint{0.000000\du}{0.000000\du}}\definecolor{diafillcolor}{rgb}{1.000000, 1.000000, 1.000000}
\pgfsetfillcolor{diafillcolor}
\pgfsetfillopacity{1.000000}
\fill (12.500000\du,3.820644\du)--(12.500000\du,6.179356\du)--(35.000000\du,6.179356\du)--(35.000000\du,3.820644\du)--cycle;
}{\pgfsetcornersarced{\pgfpoint{0.000000\du}{0.000000\du}}\definecolor{dialinecolor}{rgb}{0.000000, 0.000000, 0.000000}
\pgfsetstrokecolor{dialinecolor}
\pgfsetstrokeopacity{1.000000}
\draw (12.500000\du,3.820644\du)--(12.500000\du,6.179356\du)--(35.000000\du,6.179356\du)--(35.000000\du,3.820644\du)--cycle;
}
\definecolor{dialinecolor}{rgb}{0.000000, 0.000000, 0.000000}
\pgfsetstrokecolor{dialinecolor}
\pgfsetstrokeopacity{1.000000}
\definecolor{diafillcolor}{rgb}{0.000000, 0.000000, 0.000000}
\pgfsetfillcolor{diafillcolor}
\pgfsetfillopacity{1.000000}
\node[anchor=base,inner sep=0pt, outer sep=0pt,color=dialinecolor] at (23.750000\du,5.258333\du){Device};
\pgfsetlinewidth{0.100000\du}
\pgfsetdash{}{0pt}
\pgfsetbuttcap
\pgfsetmiterjoin
\pgfsetlinewidth{0.100000\du}
\pgfsetbuttcap
\pgfsetmiterjoin
\pgfsetdash{}{0pt}
\definecolor{diafillcolor}{rgb}{1.000000, 1.000000, 1.000000}
\pgfsetfillcolor{diafillcolor}
\pgfsetfillopacity{1.000000}
\fill (17.000000\du,-2.600000\du)--(21.600000\du,-2.600000\du)--(19.300000\du,-7.000000\du)--cycle;
\definecolor{dialinecolor}{rgb}{0.000000, 0.000000, 0.000000}
\pgfsetstrokecolor{dialinecolor}
\pgfsetstrokeopacity{1.000000}
\draw (17.000000\du,-2.600000\du)--(21.600000\du,-2.600000\du)--(19.300000\du,-7.000000\du)--cycle;
\definecolor{dialinecolor}{rgb}{0.000000, 0.000000, 0.000000}
\pgfsetstrokecolor{dialinecolor}
\pgfsetstrokeopacity{1.000000}
\definecolor{diafillcolor}{rgb}{0.000000, 0.000000, 0.000000}
\pgfsetfillcolor{diafillcolor}
\pgfsetfillopacity{1.000000}
\node[anchor=base,inner sep=0pt, outer sep=0pt,color=dialinecolor] at (19.300000\du,-3.500000\du){};
\pgfsetlinewidth{0.100000\du}
\pgfsetdash{}{0pt}
\pgfsetbuttcap
\pgfsetmiterjoin
\pgfsetlinewidth{0.100000\du}
\pgfsetbuttcap
\pgfsetmiterjoin
\pgfsetdash{}{0pt}
\definecolor{diafillcolor}{rgb}{1.000000, 1.000000, 1.000000}
\pgfsetfillcolor{diafillcolor}
\pgfsetfillopacity{1.000000}
\fill (18.000000\du,-1.600000\du)--(22.600000\du,-1.600000\du)--(20.300000\du,-6.000000\du)--cycle;
\definecolor{dialinecolor}{rgb}{0.000000, 0.000000, 0.000000}
\pgfsetstrokecolor{dialinecolor}
\pgfsetstrokeopacity{1.000000}
\draw (18.000000\du,-1.600000\du)--(22.600000\du,-1.600000\du)--(20.300000\du,-6.000000\du)--cycle;
\definecolor{dialinecolor}{rgb}{0.000000, 0.000000, 0.000000}
\pgfsetstrokecolor{dialinecolor}
\pgfsetstrokeopacity{1.000000}
\definecolor{diafillcolor}{rgb}{0.000000, 0.000000, 0.000000}
\pgfsetfillcolor{diafillcolor}
\pgfsetfillopacity{1.000000}
\node[anchor=base,inner sep=0pt, outer sep=0pt,color=dialinecolor] at (20.300000\du,-2.500000\du){};
\pgfsetlinewidth{0.100000\du}
\pgfsetdash{}{0pt}
\pgfsetbuttcap
\pgfsetmiterjoin
\pgfsetlinewidth{0.100000\du}
\pgfsetbuttcap
\pgfsetmiterjoin
\pgfsetdash{}{0pt}
\definecolor{diafillcolor}{rgb}{1.000000, 1.000000, 1.000000}
\pgfsetfillcolor{diafillcolor}
\pgfsetfillopacity{1.000000}
\fill (19.000000\du,-0.600000\du)--(23.600000\du,-0.600000\du)--(21.300000\du,-5.000000\du)--cycle;
\definecolor{dialinecolor}{rgb}{0.000000, 0.000000, 0.000000}
\pgfsetstrokecolor{dialinecolor}
\pgfsetstrokeopacity{1.000000}
\draw (19.000000\du,-0.600000\du)--(23.600000\du,-0.600000\du)--(21.300000\du,-5.000000\du)--cycle;
\definecolor{dialinecolor}{rgb}{0.000000, 0.000000, 0.000000}
\pgfsetstrokecolor{dialinecolor}
\pgfsetstrokeopacity{1.000000}
\definecolor{diafillcolor}{rgb}{0.000000, 0.000000, 0.000000}
\pgfsetfillcolor{diafillcolor}
\pgfsetfillopacity{1.000000}
\node[anchor=base,inner sep=0pt, outer sep=0pt,color=dialinecolor] at (21.300000\du,-1.500000\du){};
\pgfsetlinewidth{0.100000\du}
\pgfsetdash{}{0pt}
\pgfsetbuttcap
\pgfsetmiterjoin
\pgfsetlinewidth{0.100000\du}
\pgfsetbuttcap
\pgfsetmiterjoin
\pgfsetdash{}{0pt}
\definecolor{diafillcolor}{rgb}{1.000000, 1.000000, 1.000000}
\pgfsetfillcolor{diafillcolor}
\pgfsetfillopacity{1.000000}
\definecolor{dialinecolor}{rgb}{0.000000, 0.000000, 0.000000}
\pgfsetstrokecolor{dialinecolor}
\pgfsetstrokeopacity{1.000000}
\pgfpathmoveto{\pgfpoint{27.000000\du}{-6.406819\du}}
\pgfpathcurveto{\pgfpoint{27.480381\du}{-6.851705\du}}{\pgfpoint{27.720572\du}{-7.000000\du}}{\pgfpoint{28.200954\du}{-7.000000\du}}
\pgfpathcurveto{\pgfpoint{28.681335\du}{-7.000000\du}}{\pgfpoint{28.921526\du}{-6.851705\du}}{\pgfpoint{29.401907\du}{-6.406819\du}}
\pgfpathlineto{\pgfpoint{29.401907\du}{-4.034093\du}}
\pgfpathcurveto{\pgfpoint{28.921526\du}{-3.589207\du}}{\pgfpoint{28.681335\du}{-3.440911\du}}{\pgfpoint{28.200954\du}{-3.440911\du}}
\pgfpathcurveto{\pgfpoint{27.720572\du}{-3.440911\du}}{\pgfpoint{27.480381\du}{-3.589207\du}}{\pgfpoint{27.000000\du}{-4.034093\du}}
\pgfpathlineto{\pgfpoint{27.000000\du}{-6.406819\du}}
\pgfpathclose
\pgfusepath{fill,stroke}
\pgfsetbuttcap
\pgfsetmiterjoin
\pgfsetdash{}{0pt}
\definecolor{dialinecolor}{rgb}{0.000000, 0.000000, 0.000000}
\pgfsetstrokecolor{dialinecolor}
\pgfsetstrokeopacity{1.000000}
\pgfpathmoveto{\pgfpoint{27.000000\du}{-6.406819\du}}
\pgfpathcurveto{\pgfpoint{27.480381\du}{-5.961933\du}}{\pgfpoint{27.720572\du}{-5.813637\du}}{\pgfpoint{28.200954\du}{-5.813637\du}}
\pgfpathcurveto{\pgfpoint{28.681335\du}{-5.813637\du}}{\pgfpoint{28.921526\du}{-5.961933\du}}{\pgfpoint{29.401907\du}{-6.406819\du}}
\pgfusepath{stroke}
\definecolor{dialinecolor}{rgb}{0.000000, 0.000000, 0.000000}
\pgfsetstrokecolor{dialinecolor}
\pgfsetstrokeopacity{1.000000}
\definecolor{diafillcolor}{rgb}{0.000000, 0.000000, 0.000000}
\pgfsetfillcolor{diafillcolor}
\pgfsetfillopacity{1.000000}
\node[anchor=base,inner sep=0pt, outer sep=0pt,color=dialinecolor] at (28.200954\du,-4.723865\du){};
\pgfsetlinewidth{0.100000\du}
\pgfsetdash{}{0pt}
\pgfsetbuttcap
\pgfsetmiterjoin
\pgfsetlinewidth{0.100000\du}
\pgfsetbuttcap
\pgfsetmiterjoin
\pgfsetdash{}{0pt}
\definecolor{diafillcolor}{rgb}{1.000000, 1.000000, 1.000000}
\pgfsetfillcolor{diafillcolor}
\pgfsetfillopacity{1.000000}
\definecolor{dialinecolor}{rgb}{0.000000, 0.000000, 0.000000}
\pgfsetstrokecolor{dialinecolor}
\pgfsetstrokeopacity{1.000000}
\pgfpathmoveto{\pgfpoint{28.000000\du}{-5.406819\du}}
\pgfpathcurveto{\pgfpoint{28.480381\du}{-5.851705\du}}{\pgfpoint{28.720572\du}{-6.000000\du}}{\pgfpoint{29.200954\du}{-6.000000\du}}
\pgfpathcurveto{\pgfpoint{29.681335\du}{-6.000000\du}}{\pgfpoint{29.921526\du}{-5.851705\du}}{\pgfpoint{30.401907\du}{-5.406819\du}}
\pgfpathlineto{\pgfpoint{30.401907\du}{-3.034093\du}}
\pgfpathcurveto{\pgfpoint{29.921526\du}{-2.589207\du}}{\pgfpoint{29.681335\du}{-2.440911\du}}{\pgfpoint{29.200954\du}{-2.440911\du}}
\pgfpathcurveto{\pgfpoint{28.720572\du}{-2.440911\du}}{\pgfpoint{28.480381\du}{-2.589207\du}}{\pgfpoint{28.000000\du}{-3.034093\du}}
\pgfpathlineto{\pgfpoint{28.000000\du}{-5.406819\du}}
\pgfpathclose
\pgfusepath{fill,stroke}
\pgfsetbuttcap
\pgfsetmiterjoin
\pgfsetdash{}{0pt}
\definecolor{dialinecolor}{rgb}{0.000000, 0.000000, 0.000000}
\pgfsetstrokecolor{dialinecolor}
\pgfsetstrokeopacity{1.000000}
\pgfpathmoveto{\pgfpoint{28.000000\du}{-5.406819\du}}
\pgfpathcurveto{\pgfpoint{28.480381\du}{-4.961933\du}}{\pgfpoint{28.720572\du}{-4.813637\du}}{\pgfpoint{29.200954\du}{-4.813637\du}}
\pgfpathcurveto{\pgfpoint{29.681335\du}{-4.813637\du}}{\pgfpoint{29.921526\du}{-4.961933\du}}{\pgfpoint{30.401907\du}{-5.406819\du}}
\pgfusepath{stroke}
\definecolor{dialinecolor}{rgb}{0.000000, 0.000000, 0.000000}
\pgfsetstrokecolor{dialinecolor}
\pgfsetstrokeopacity{1.000000}
\definecolor{diafillcolor}{rgb}{0.000000, 0.000000, 0.000000}
\pgfsetfillcolor{diafillcolor}
\pgfsetfillopacity{1.000000}
\node[anchor=base,inner sep=0pt, outer sep=0pt,color=dialinecolor] at (29.200954\du,-3.723865\du){};
\pgfsetlinewidth{0.100000\du}
\pgfsetdash{}{0pt}
\pgfsetbuttcap
\pgfsetmiterjoin
\pgfsetlinewidth{0.100000\du}
\pgfsetbuttcap
\pgfsetmiterjoin
\pgfsetdash{}{0pt}
\definecolor{diafillcolor}{rgb}{1.000000, 1.000000, 1.000000}
\pgfsetfillcolor{diafillcolor}
\pgfsetfillopacity{1.000000}
\definecolor{dialinecolor}{rgb}{0.000000, 0.000000, 0.000000}
\pgfsetstrokecolor{dialinecolor}
\pgfsetstrokeopacity{1.000000}
\pgfpathmoveto{\pgfpoint{29.000000\du}{-4.406819\du}}
\pgfpathcurveto{\pgfpoint{29.480381\du}{-4.851705\du}}{\pgfpoint{29.720572\du}{-5.000000\du}}{\pgfpoint{30.200954\du}{-5.000000\du}}
\pgfpathcurveto{\pgfpoint{30.681335\du}{-5.000000\du}}{\pgfpoint{30.921526\du}{-4.851705\du}}{\pgfpoint{31.401907\du}{-4.406819\du}}
\pgfpathlineto{\pgfpoint{31.401907\du}{-2.034093\du}}
\pgfpathcurveto{\pgfpoint{30.921526\du}{-1.589207\du}}{\pgfpoint{30.681335\du}{-1.440911\du}}{\pgfpoint{30.200954\du}{-1.440911\du}}
\pgfpathcurveto{\pgfpoint{29.720572\du}{-1.440911\du}}{\pgfpoint{29.480381\du}{-1.589207\du}}{\pgfpoint{29.000000\du}{-2.034093\du}}
\pgfpathlineto{\pgfpoint{29.000000\du}{-4.406819\du}}
\pgfpathclose
\pgfusepath{fill,stroke}
\pgfsetbuttcap
\pgfsetmiterjoin
\pgfsetdash{}{0pt}
\definecolor{dialinecolor}{rgb}{0.000000, 0.000000, 0.000000}
\pgfsetstrokecolor{dialinecolor}
\pgfsetstrokeopacity{1.000000}
\pgfpathmoveto{\pgfpoint{29.000000\du}{-4.406819\du}}
\pgfpathcurveto{\pgfpoint{29.480381\du}{-3.961933\du}}{\pgfpoint{29.720572\du}{-3.813637\du}}{\pgfpoint{30.200954\du}{-3.813637\du}}
\pgfpathcurveto{\pgfpoint{30.681335\du}{-3.813637\du}}{\pgfpoint{30.921526\du}{-3.961933\du}}{\pgfpoint{31.401907\du}{-4.406819\du}}
\pgfusepath{stroke}
\definecolor{dialinecolor}{rgb}{0.000000, 0.000000, 0.000000}
\pgfsetstrokecolor{dialinecolor}
\pgfsetstrokeopacity{1.000000}
\definecolor{diafillcolor}{rgb}{0.000000, 0.000000, 0.000000}
\pgfsetfillcolor{diafillcolor}
\pgfsetfillopacity{1.000000}
\node[anchor=base,inner sep=0pt, outer sep=0pt,color=dialinecolor] at (30.200954\du,-2.723865\du){};
\pgfsetlinewidth{0.100000\du}
\pgfsetdash{}{0pt}
\pgfsetbuttcap
{
\definecolor{diafillcolor}{rgb}{0.000000, 0.000000, 0.000000}
\pgfsetfillcolor{diafillcolor}
\pgfsetfillopacity{1.000000}
\pgfsetarrowsend{stealth}
\definecolor{dialinecolor}{rgb}{0.000000, 0.000000, 0.000000}
\pgfsetstrokecolor{dialinecolor}
\pgfsetstrokeopacity{1.000000}
\draw (18.450171\du,-0.042044\du)--(18.125000\du,3.820644\du);
}
\definecolor{diafillcolor}{rgb}{1.000000, 1.000000, 1.000000}
\pgfsetfillcolor{diafillcolor}
\pgfsetfillopacity{1.000000}
\fill (16.000000\du,0.950000\du)--(22.252500\du,0.950000\du)--(22.252500\du,2.267500\du)--(16.000000\du,2.267500\du)--cycle;
\definecolor{dialinecolor}{rgb}{0.000000, 0.000000, 0.000000}
\pgfsetstrokecolor{dialinecolor}
\pgfsetstrokeopacity{1.000000}
\definecolor{diafillcolor}{rgb}{0.000000, 0.000000, 0.000000}
\pgfsetfillcolor{diafillcolor}
\pgfsetfillopacity{1.000000}
\node[anchor=base west,inner sep=0pt,outer sep=0pt,color=dialinecolor] at (16.000000\du,2.000000\du){withDevice};
\pgfsetlinewidth{0.100000\du}
\pgfsetdash{}{0pt}
\pgfsetbuttcap
{
\definecolor{diafillcolor}{rgb}{0.000000, 0.000000, 0.000000}
\pgfsetfillcolor{diafillcolor}
\pgfsetfillopacity{1.000000}
\pgfsetarrowsend{stealth}
\definecolor{dialinecolor}{rgb}{0.000000, 0.000000, 0.000000}
\pgfsetstrokecolor{dialinecolor}
\pgfsetstrokeopacity{1.000000}
\pgfpathmoveto{\pgfpoint{20.150176\du}{-2.799955\du}}
\pgfpatharc{285}{138}{8.180740\du and 8.180740\du}
\pgfusepath{stroke}
}
\definecolor{dialinecolor}{rgb}{0.000000, 0.000000, 0.000000}
\pgfsetstrokecolor{dialinecolor}
\pgfsetstrokeopacity{1.000000}
\definecolor{diafillcolor}{rgb}{0.000000, 0.000000, 0.000000}
\pgfsetfillcolor{diafillcolor}
\pgfsetfillopacity{1.000000}
\node[anchor=base west,inner sep=0pt,outer sep=0pt,color=dialinecolor] at (27.219440\du,1.732647\du){};
\definecolor{diafillcolor}{rgb}{1.000000, 1.000000, 1.000000}
\pgfsetfillcolor{diafillcolor}
\pgfsetfillopacity{1.000000}
\fill (9.000000\du,0.950000\du)--(13.980000\du,0.950000\du)--(13.980000\du,2.267500\du)--(9.000000\du,2.267500\du)--cycle;
\definecolor{dialinecolor}{rgb}{0.000000, 0.000000, 0.000000}
\pgfsetstrokecolor{dialinecolor}
\pgfsetstrokeopacity{1.000000}
\definecolor{diafillcolor}{rgb}{0.000000, 0.000000, 0.000000}
\pgfsetfillcolor{diafillcolor}
\pgfsetfillopacity{1.000000}
\node[anchor=base west,inner sep=0pt,outer sep=0pt,color=dialinecolor] at (9.000000\du,2.000000\du){liftmTask};
\pgfsetlinewidth{0.100000\du}
\pgfsetdash{}{0pt}
\pgfsetbuttcap
{
\definecolor{diafillcolor}{rgb}{0.000000, 0.000000, 0.000000}
\pgfsetfillcolor{diafillcolor}
\pgfsetfillopacity{1.000000}
\pgfsetarrowsend{stealth}
\definecolor{dialinecolor}{rgb}{0.000000, 0.000000, 0.000000}
\pgfsetstrokecolor{dialinecolor}
\pgfsetstrokeopacity{1.000000}
\draw (29.600477\du,-1.570670\du)--(24.000000\du,9.730208\du);
}
\definecolor{dialinecolor}{rgb}{0.000000, 0.000000, 0.000000}
\pgfsetstrokecolor{dialinecolor}
\pgfsetstrokeopacity{1.000000}
\definecolor{diafillcolor}{rgb}{0.000000, 0.000000, 0.000000}
\pgfsetfillcolor{diafillcolor}
\pgfsetfillopacity{1.000000}
\node[anchor=base west,inner sep=0pt,outer sep=0pt,color=dialinecolor] at (26.045684\du,3.144328\du){};
\definecolor{diafillcolor}{rgb}{1.000000, 1.000000, 1.000000}
\pgfsetfillcolor{diafillcolor}
\pgfsetfillopacity{1.000000}
\fill (26.000000\du,0.950000\du)--(29.337500\du,0.950000\du)--(29.337500\du,2.267500\du)--(26.000000\du,2.267500\du)--cycle;
\definecolor{dialinecolor}{rgb}{0.000000, 0.000000, 0.000000}
\pgfsetstrokecolor{dialinecolor}
\pgfsetstrokeopacity{1.000000}
\definecolor{diafillcolor}{rgb}{0.000000, 0.000000, 0.000000}
\pgfsetfillcolor{diafillcolor}
\pgfsetfillopacity{1.000000}
\node[anchor=base west,inner sep=0pt,outer sep=0pt,color=dialinecolor] at (26.000000\du,2.000000\du){liftsds};
\definecolor{dialinecolor}{rgb}{0.000000, 0.000000, 0.000000}
\pgfsetstrokecolor{dialinecolor}
\pgfsetstrokeopacity{1.000000}
\definecolor{diafillcolor}{rgb}{0.000000, 0.000000, 0.000000}
\pgfsetfillcolor{diafillcolor}
\pgfsetfillopacity{1.000000}
\node[anchor=base west,inner sep=0pt,outer sep=0pt,color=dialinecolor] at (22.357028\du,4.311980\du){};
\pgfsetlinewidth{0.100000\du}
\pgfsetdash{}{0pt}
\pgfsetmiterjoin
{\pgfsetcornersarced{\pgfpoint{0.000000\du}{0.000000\du}}\definecolor{diafillcolor}{rgb}{1.000000, 1.000000, 1.000000}
\pgfsetfillcolor{diafillcolor}
\pgfsetfillopacity{1.000000}
\fill (16.758371\du,-11.332298\du)--(16.758371\du,-8.973586\du)--(31.160128\du,-8.973586\du)--(31.160128\du,-11.332298\du)--cycle;
}{\pgfsetcornersarced{\pgfpoint{0.000000\du}{0.000000\du}}\definecolor{dialinecolor}{rgb}{0.000000, 0.000000, 0.000000}
\pgfsetstrokecolor{dialinecolor}
\pgfsetstrokeopacity{1.000000}
\draw (16.758371\du,-11.332298\du)--(16.758371\du,-8.973586\du)--(31.160128\du,-8.973586\du)--(31.160128\du,-11.332298\du)--cycle;
}
\definecolor{dialinecolor}{rgb}{0.000000, 0.000000, 0.000000}
\pgfsetstrokecolor{dialinecolor}
\pgfsetstrokeopacity{1.000000}
\definecolor{diafillcolor}{rgb}{0.000000, 0.000000, 0.000000}
\pgfsetfillcolor{diafillcolor}
\pgfsetfillopacity{1.000000}
\node[anchor=base,inner sep=0pt, outer sep=0pt,color=dialinecolor] at (23.959250\du,-9.894609\du){TOP server};
\end{tikzpicture}

%% file: 3building.tex
The following subsections are a hands-on introduction to writing more complex applications in \gls{mTask} and \gls{iTasks}.
Both \gls{mTask} and \gls{iTasks} are hosted in \gls{Clean} which has a similar syntax to \gls{Haskell}.
Peter et al.\ provide a concise overview of the syntactical differences~\cite{achten_clean_2007}.
The skeletons for the exercises are listed between brackets and can be found in the \texttt{mTask/cefp19} directory of the distribution\footnote{\url{https://ftp.cs.ru.nl/Clean/CEFP19/}}.
Section~\ref{sec:install} contains detailed setup instructions.
Solutions for all exercises are available in Section~\ref{sec:solutions}.

\subsection{Hardware \& client}
\begin{figure}
	\centering
	\includegraphics[width=.4\linewidth]{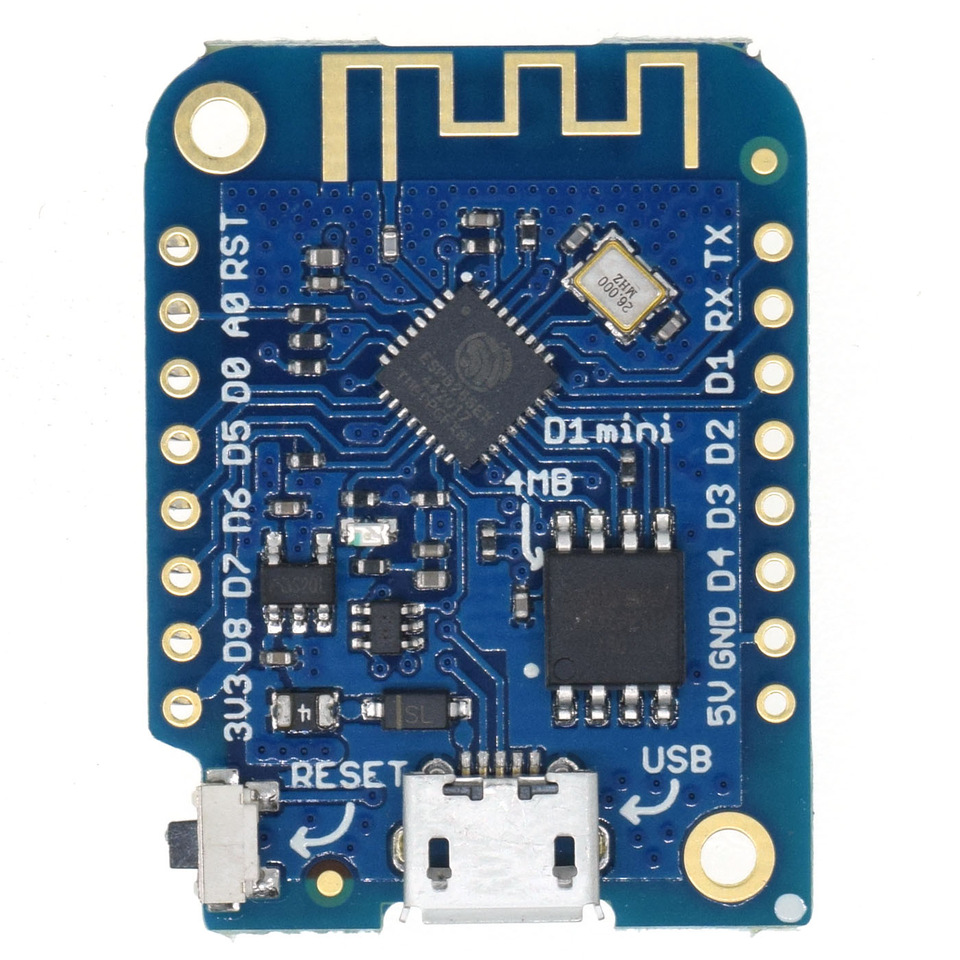}
	\caption{The mainboard of the WEMOS \gls{LOLIN} D1 mini.}\label{fig:lolin}
\end{figure}

For the examples we use the WEMOS \gls{LOLIN} D1 mini\footnote{\fmturl{https://wiki.wemos.cc/products:d1:d1_mini}} (Figure~\ref{fig:lolin}).
The D1 mini is an \gls{ESP8266} based prototyping board containing 1 analog and 11 digital \gls{GPIO} pins and a micro \gls{USB} connection for programming and debugging.
It can be programmed using MicroPython, \gls{Arduino} or \gls{LUA}.

It is assumed that they are preinstalled with the \gls{mTask} \gls{RTS} and that it has the correct shields attached.
Details on how to compile and run the \gls{mTask} \gls{RTS} on the device can be found in Section~\ref{sec:setupTheMicrocontroller}.

The devices are installed on a three-way splitter and setup with an \gls{OLED}, SHT and Matrix \gls{LED} shield.
The \gls{OLED} shield is used for displaying runtime during operation.
When booting up, it shows the WiFi status and when connected it shows the \gls{IP} address that one should enter in the device selection screen of the server application.
Furthermore, the \gls{OLED} screen contains two buttons that can be accessed from within \gls{mTask} to get some kind of feedback from the user.The SHT shield houses a \gls{DHT} sensor that can be accessed from \gls{mTask} as well.
The \gls{LED} matrix can be accessed through \gls{mTask} and can be used to display information.

\subsection{Temperature}
Reading the ambient temperature off the device is achieved using the \gls{DHT} sensor connected as a shield to the main board.
The \gls{DHT} shield contains an SHT30 sensor.
When queried via \gls{I2C}, the chip measures the temperature with a $\pm0.4\degree C$ accuracy and the relative humidity with a $\pm2\%$ accuracy.

It is accessed using the \gls{mTask} \Cl{dht} class (see Subsection~\ref{ssec:dht}).
For example, the following program will show the current temperature and humidity to the user.
The yielded values from the \Cl{temperature} and \Cl{humidity} tasks are in tenths of degrees and percents respectively instead of a floating point value.
Therefore, a lens is applied on the editor to transform them into floating point values.

\begin{lstexample}[language=Clean,caption={An \gls{mTask} program for measuring the temperature and humidity. (\texttt{tempSimple})},label={lst:temphumid},numbers=left,]
main = enterDevice >>= \spec->withDevice spec
	\dev->liftmTask temp dev >&> viewSharedInformation () [ViewAs templens]
where
	templens = maybe (0.0, 0.0) \(t, h)->(toReal t / 10.0, toReal h / 10.0)

	temp :: Main (MTask v (Int, Int)) | mtask, dht v
	temp = DHT D4 DHT22 \dht->{main=temperature dht .&&. humidity dht}
\end{lstexample}

\begin{exercise}[Show the temperature via an \acrshort{SDS}]\label{ass:tempsds}
	Modify the application so that it writes the temperature in an \gls{SDS}.
	Writing the temperature constantly in the \gls{SDS} creates a lot of network traffic.
	Therefore it is advised to create a function that will memorize the old temperature and only write the new temperature when it is different from the old one.
	Use the following template (\texttt{tempSds}):

	\begin{lstexample}[language=Clean,nolol]
main = enterDevice >>= \spec->withDevice spec
	\dev->withShared 0 \sh->
		    liftmTask (temp sh) dev
		-|| viewSharedInformation "Temperature" [ViewAs templens] sh
where
	templens t = toReal t / 10.0

	temp :: (Shared s Int) -> Main (MTask v ()) | mtask, dht, liftsds v & RWShared s
	\end{lstexample}
\end{exercise}

With the \Cl{writeD} functions from \gls{mTask} (see Subsection~\ref{ssec:gpio}) the digital \gls{GPIO} pins can be controlled.
Imagine a heater attached to a \gls{GPIO} pin that turns on when the temperature is below a given limit.

\begin{exercise}[Simple thermostat]\label{ass:thermostat}
	Modify the previous exercise so that a thermostat is mimicked.
	The user enters a temperature target and the \gls{LED} will turn on when the temperature is below the target.
	To quickly change the temperature measure, blow some air in the sensor.
	Use the following template (\texttt{thermostat}):

	\begin{lstexample}[language=Clean,nolol]
main = enterDevice >>= \spec->withDevice spec
	\dev->withShared 0 \tempShare->
	      withShared 250 \targetShare->
		    liftmTask (temp targetShare tempShare) dev
		-|| viewSharedInformation "Temperature" [ViewAs tempfro] tempShare
		-|| updateSharedInformation "Target" [UpdateAs tempfro \_->tempto] targetShare
where
	tempfro t = toReal t / 10.0
	tempto  t = toInt t * 10

	temp :: (Shared s1 Int) (Shared s2 Int)
		-> Main (MTask v ()) | mtask, dht, liftsds v & RWShared s1 & RWShared s2
	...
	\end{lstexample}
\end{exercise}

\subsection{\Acrshort{LED} matrix}
\begin{figure}
	\centering
	\includegraphics[width=.5\textwidth]{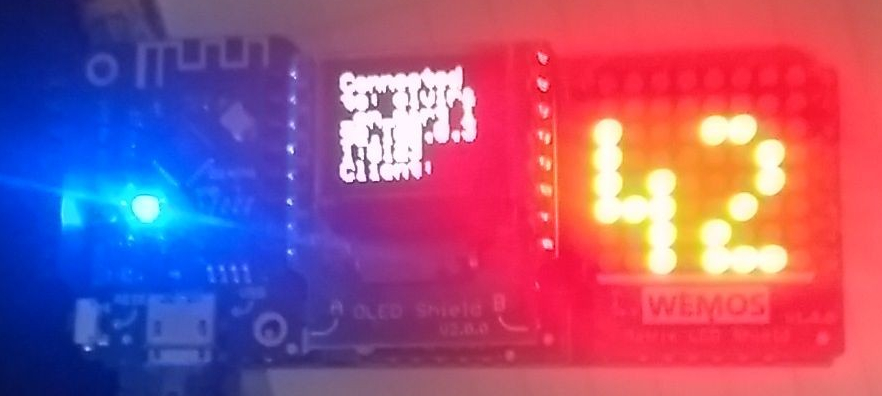}

	\caption{\emph{The Answer} printed on the \gls{LED} matrix.}%
	\label{fig:matrix42}
\end{figure}

The \gls{LED} matrix shield can be used to display information during the execution of the program.
Every \gls{LED} of the $8\times8$ matrix can be controlled individually using the functions from Subsection~\ref{ssec:dht}.
The program in Example~\ref{lst:matrixblink} shows an \gls{iTasks} program for controlling the \gls{LED} matrix.
It allows toggling the state of a given \gls{LED} and clear the display.

\begin{figure}[ht]
	\centering
	\includegraphics[width=.85\linewidth]{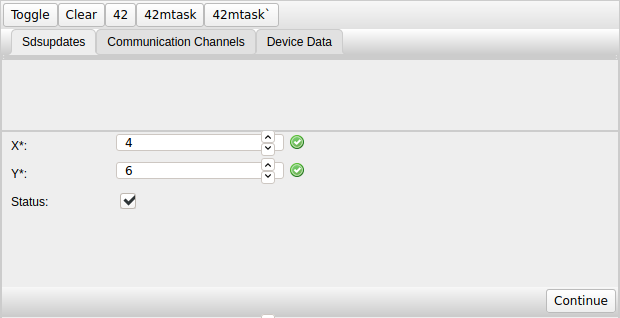}
	\caption{The user interface for the \acrshort{LED} matrix application}\label{fig:ledmatrix}
\end{figure}

To present the user with a nice interface (Figure~\ref{fig:ledmatrix}), a type is created that houses the status of an \gls{LED} in the matrix.
The main program is very similar to previous programs, only differing in the device part.
The \Cl{>^*} combinator is a special kind of \Cl{parallel} combinator that --- instead of stepping to a continuation --- forks off a continuation.
This allows the user to schedule many tasks in parallel.
Continuations can be triggered by values or by actions.
In this example, only actions are used that are always enabled.
One action is added for every operation and when the user presses the button, the according task is sent to the device.
The \Cl{toggle} and \Cl{clear} tasks are self-explanatory and only use \gls{LED} matrix \gls{mTask} functions (see Definition~\ref{lst:ledmatrix}).

\begin{lstexample}[language=Clean,caption={An interactive \gls{mTask} program for interacting with the \gls{LED} matrix. (\texttt{matrixBlink})},label={lst:matrixblink},numbers=left,]
:: Ledstatus = {x :: Int, y :: Int, status :: Bool}
derive class iTask Ledstatus

main = enterDevice >>= \spec->withDevice spec
	\dev-> viewDevice dev >^*
		[OnAction (Action "Toggle") (always (
			    enterInformation () [] >>= \s->liftmTask (toggle s) dev
			>>~ viewInformation "done" []))
		,OnAction (Action "Clear") (always (
			    liftmTask clear dev
			>>~ viewInformation "done" []))
		] @! ()
where
	dot lm s = LMDot lm (lit s.x) (lit s.y) (lit s.status)

	toggle :: Ledstatus -> Main (MTask v ()) | mtask, LEDMatrix v
	toggle s = ledmatrix D5 D7 \lm->{main=dot lm s >>|. LMDisplay lm}

	clear :: Main (MTask v ()) | mtask, LEDMatrix v
	clear = ledmatrix D5 D7 \lm->{main=LMClear lm >>|. LMDisplay lm}
\end{lstexample}

Toggling the \glspl{LED} in the matrix using the given tasks is very user intensive because for every action, a task needs to be launched.
Extend the program so that there is a new button for printing the answer to the question of life, universe and everything~\footnote{The exact question is left as an exercise to the reader but the answer is 42~\cite{adams_hitchhikers_2017}.} as seen in Figure~\ref{fig:matrix42}.
There are two general approaches possible that are presented in Assignment~\ref{ass:matrix42iTasks}~and~\ref{ass:matrix42mTask}.

\begin{exercise}[\acrshort{LED} Matrix 42 using \glsentrytext{iTasks}]\label{ass:matrix42iTasks}
	Write $42$ to the \gls{LED} matrix using only the \Cl{toggle} and the \Cl{clear} \gls{mTask} tasks and define all other logic in \gls{iTasks}
	You can add the continuations as follows (\texttt{matrixBlink}):

	\begin{lstexample}[language=Clean,postbreak=]
OnAction (Action "42") (always (iTask42 dev))
	\end{lstexample}

	The \gls{iTasks} task should then have the following type signature:

	\begin{lstexample}[language=Clean,postbreak=]
iTask42 :: MTDevice -> Task ()
	\end{lstexample}
\end{exercise}

In this situation, a whole bunch of \gls{mTask} tasks are sent to the device at once.
This strains the communication channels greatly and is a risk for running out of memory.
It is also possible to define printing $42$ in solely in \gls{mTask}.
This creates one bigger task that is sent at once.

\begin{exercise}[\acrshort{LED} Matrix 42 using \glsentrytext{mTask}]\label{ass:matrix42mTask}
	Write $42$ to the \gls{LED} matrix as a single \gls{mTask} task.
	This results in the following continuation (\texttt{matrixBlink}):

	\begin{lstexample}[language=Clean,postbreak=]
OnAction (Action "42mtask") (always (liftmTask mTask42 dev))
	\end{lstexample}

	The \gls{mTask} task should then have the following type signature:

	\begin{lstexample}[language=Clean,postbreak=]
mTask42 :: Main (MTask v ()) | mtask, LEDMatrix v
	\end{lstexample}
\end{exercise}

\subsection{Temperature plotter}
\begin{figure}
	\centering
	\begin{subfigure}[b]{.5\textwidth}
		\includegraphics[width=\textwidth]{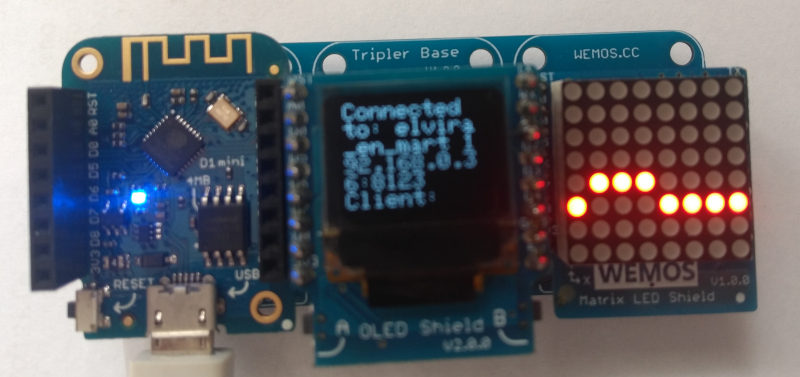}
		\caption{The temperature plotter in action.}%
		\label{fig:plotter}
	\end{subfigure}%
	\begin{subfigure}[b]{.5\linewidth}
		\centering
		\includegraphics[width=.5\textwidth]{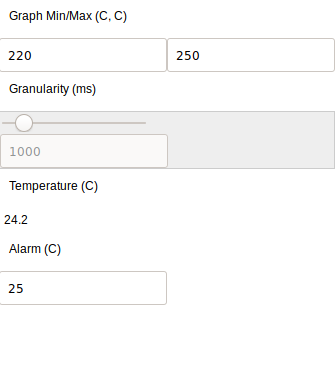}
		\caption{The temperature plotter UI.}%
		\label{fig:plotterui}
	\end{subfigure}
	\caption{The reference implementation of the plotter in action}
\end{figure}

This final exercise is about creating temperature plotter with an alarm mode.
This application uses all components of the device and communicates with the server extensively.
I.e.\ the \gls{LED} matrix to show the plot, the \gls{OLED} shield buttons to toggle the alarm, the builtin \gls{LED} to show the alarm status and the \gls{DHT} shield to measure the temperature.
Figure~\ref{fig:plotter} shows an implementation in action.
Figure~\ref{fig:plotterui} shows the user interface for it.

\begin{exercise}[Temperature plotter]\label{ass:plotter}
	There are several tasks that the plotter needs to do at the same time
	\begin{description}
		\item[Plot]

			The main task of the program is to plot the temperature over time on the \gls{LED} matrix.
			The range of the graph is specified in the \Cl{limitsShare} and may be changed by the user.
		\item[Report]

			The temperature has to be reported to the server every interval.
			This is achieved by writing the current temperature in the lifted \Cl{tempShare} \gls{SDS}.
			The server is automatically notified and the user interface will update accordingly
			Preferably it only writes to the \gls{SDS} when the temperature has changed.
		\item[Set alarm]

			When the temperature is higher than a certain limit, the builtin \gls{LED} should turn on.
			The current limit is always available in the lifted \Cl{alarmShare}.
		\item[Unset alarm]

			When the alarm went off, the user should be able to disable it by pressing the A button that resides on the \gls{OLED} shield.
	\end{description}

	The exercise is quite elaborate so please keep in mind the following tips:
	\begin{itemize}
		\item Start with the preamble and a skeleton for the tasks.

			The preamble should at least lift the \glspl{SDS} and define the peripherals (\gls{LED} matrix and \gls{DHT}).
		\item Use functions for state as much as possible.

			Especially for measuring the temperature, you do not want to write to the temperature \gls{SDS} every time you measure.
			Therefore, keep track of the old temperature using a function or alternatively a local \gls{SDS}.
		\item Write functions for routines that you do multiple times.

			For example, clearing a row on the \gls{LED} matrix is a tedious job and has to be done every cycle.
			Simplify it by either writing it as a \gls{Clean} function that generates all the code or an \gls{mTask} function that is called.
	\end{itemize}
	Create the plotter using the following template (\texttt{plotter}):

	\begin{lstexample}[language=Clean,nolol]
BUILTIN_LED :== d3
ABUTTON :== d4

main = enterDevice >>= \spec->withDevice spec
	\dev->withShared (220, 250) \limitsShare->
	      withShared 1000 \waitShare->
	      withShared 0 \tempShare->
	      withShared 250 \alarmShare->
		    liftmTask (temp limitsShare waitShare tempShare alarmShare) dev
		-|| updateSharedInformation "Graph Min/Max (C, C)" [] limitsShare
		-|| updateSharedInformation "Granularity (ms)" [updater] waitShare
		-|| viewSharedInformation "Temperature (C)" [ViewAs tempfro] tempShare
		-|| updateSharedInformation "Alarm (C)" [UpdateAs tempfro \_->tempto] alarmShare
where
	tempfro t = toReal t / 10.0
	tempto  t = toInt t * 10

	updater :: UpdateOption Int Int
	updater = UpdateUsing (\x->(x, x)) (const fst)
		(panel2
			(slider <<@ minAttr 5 <<@ maxAttr 10000)
			(integerField <<@ enabledAttr False))

temp :: (Shared s1 (Int, Int)) (Shared s2 Int) (Shared s3 Int) (Shared s4 Int)
	-> Main (MTask v ())
	| mtask, dht, liftsds, LEDMatrix v
	& RWShared s1 & RWShared s2 & RWShared s3 & RWShared s4
temp limitsShare delayShare tempShare alarmShare =
	...
	\end{lstexample}
\end{exercise}

%% file: 4related.tex
The novelties of the \gls{mTask} system can be compared to existing systems in several categories.
It is an interpreted (Subsection~\ref{ssec:related_int}) \gls{TOP} (Subsection~\ref{ssec:related_top}) language that may seem similar at first glance to \gls{FRP} (Subsection~\ref{ssec:related_frp}), it is implemented in a functional language (Subsection~\ref{ssec:related_fp}) and due to the execution semantics, multithreading is automatically supported (Subsection~\ref{ssec:related_multi}).

\subsection{Interpretation}\label{ssec:related_int}
There are a myriad of interpreted programming languages available for some of the bigger devices.
For example, for the popular \gls{ESP8266} chip there are ports of MicroPython, LUA, Basic, JavaScript and Lisp.
All of these languages except the Lisp dialect uLisp (see Subsection~\ref{ssec:related_fp}) are imperative and do not support multithreading out of the box.
They lay pretty hefty constraints on the memory and as a result do not work on smaller \glspl{MCU}.
A interpretation solution for the tiniest devices is Firmata, a protocol for remotely controlling the \gls{MCU} and using a server as the interpreter host~\cite{steiner_firmata:_2009}.
Grebe et al.\ wrapped this in a remote monad for integration with \gls{Haskell} that allowed imperative code to be interpreted on the \glspl{MCU}~\cite{grebe_haskino:_2016}.
Later this system was extended to support multithreading as well, stepping away from Firmata as the basis and using their own \gls{RTS}~\cite{grebe_threading_2019}.
It differs from our approach because continuation points need to be defined by hand there is no automatic safe data communication.

\subsection{\Acrlong{TOP}}\label{ssec:related_top}
\Gls{TOP} as a paradigm with has been proven to be effective for implementing distributed, multi-user applications in many domains.
Examples are conference management~\cite{plasmeijer_conference_2006}, coastal protection~\cite{lijnse_capturing_2011}, \gls{C2}~\cite{bolderheij_mission-driven_2018}, incident coordination~\cite{lijnse_incidone:_2012}, crisis management~\cite{jansen_towards_2010} and telemedicine~\cite{van_der_heijden_managing_2011}.
In general, \gls{TOP} results in a higher maintainability, a high separation of concerns and more effective handling of interruptions of workflow.
\Gls{IOT} applications contain a distributed and multi-user component, but the software on the device is mostly follows multiple loosely dependent workflows
A \gls{TOP} language $\mu$Tasks developed by Piers is specialized for embedded systems.
It is a non-distributed \gls{TOP} \gls{EDSL} hosted in \gls{Haskell} designed for embedded systems such as payment terminals~\cite{piers_task-oriented_2016}.
They showed that applications tend to be able to cope well with interruptions and be more maintainable.
However, the hardware requirements for running the standard \gls{Haskell} system are high.

\subsection{\Acrlong{FRP}}\label{ssec:related_frp}
The \gls{TOP} paradigm is often compared to \gls{FRP} and while they appear to be similar --- they both process events ---, in fact they are very different.
\Gls{FRP} was introduced by Elliot and Hudak~\cite{elliott_functional_1997}.
The paradigm strives to make modelling systems safer, more efficient, composable~\cite{amsden_survey_2011}.
The core concepts are behaviours and events.
A behaviour is a value that varies over time.
Events are happenings in the real world and can trigger behaviours.
Events and behaviours may be combined using combinators.
Stutterheim et al.\ showed that \gls{FRP} concepts such as events, behaviours and signal transformers can be expressed in \gls{TOP} using tasks and \glspl{SDS} as well~\cite{wang_maintaining_2018}.

The way \gls{FRP}, and for that matter \gls{TOP}, systems are programmed stays close to the design when the domain matches suits the paradigm.
The \gls{IOT} domain seems to suit this style of programming very well in just the device layer\footnote{While a bit out of scope, it deserves mention that for \gls{SN}, \gls{FRP} and stream based approaches are popular as well~\cite{sugihara_programming_2008}.} but also for entire \gls{IOT} systems.

For example, Potato is an \gls{FRP} language for building entire \gls{IOT} systems using powerful devices such as the \gls{RaspberryPi} leveraging the \gls{Erlang} \gls{VM}~\cite{troyer_building_2018}.
It requires client devices to be able to run the \gls{Erlang} \gls{VM} which makes it unsuitable for low memory environments.
The authors state that it should be possible to create lesser demanding node software using other languages such as \gls{C} or \gls{Java} but this is future work.

The emfrp language compiles a \gls{FRP} specification for a microcontroller to \gls{C} code~\cite{sawada_emfrp:_2016}.
The \gls{IO} part, the bodies of some functions, still need to be implemented.
These \gls{IO} functions can then be used as signals and combined as in any \gls{FRP} language.
Due to the compilation to \gls{C} it is possible to run emfrp programs on tiny computers.
However, the tasks are not interpreted and there is no communication with a server.

Juniper~\cite{helbling_juniper:_2016} and arduino-copilot~\cite{hess_arduino-copilot_2020} are \gls{FRP} language for creating \gls{Arduino} programs by compiling the specification to \gls{C++}.
The languages do not contain built-in interaction with the server nor do they support interpretation.

\subsection{Functional programming}\label{ssec:related_fp}
Haenisch showed that there are major benefits to using functional languages for \gls{IOT} applications.
They showed that using function languages increased the security and maintainability of the applications~\cite{haenisch_case_2016}.
Traditional implementations of general purpose functional languages have high memory requirements rendering them unusable for tiny computers.
There have been many efforts to create a general purpose functional language that does fit in small memory environments, albeit with some concessions.
For example, there has been a history of creating tiny \gls{Scheme} implementations for specific microcontrollers.
It started with {BIT}~\cite{dube_bit:_2000} that only required 64KiB of memory, followed by {PICBIT}~\cite{feeley_picbit:_2003} and {PICOBIT}~\cite{st-amour_picobit:_2009} that lowered the memory requirements even more.
More recently, Suchocki et al.\ created Microscheme, a functional language targeting \gls{Arduino} compatible microcontrollers.
The {*BIT} languages all compile to assembly while Microscheme compiles to \gls{C++}, heavily supported by \gls{C++} lambdas available even on \gls{Arduino} \gls{AVR} targets.
An interpreted Lisp implementation called uLisp also exists that runs on microcontrollers with as small as the \gls{Arduino} {UNO}~\cite{johnson-davies_lisp_2020}.

\subsection{Multitasking}\label{ssec:related_multi}
Applications for tiny computers are often parallel in nature.
Tasks like reading sensors, watching input devices, operating actuators and maintaining communication are often loosly dependent on each other and are preferably executed in parallel.
\Glspl{MCU} often do not benefit from an \gls{OS} due to memory and processing constraints.
Therefore, writing multitasking applications in an imperative language is possible but the tasks have to be interleaved by hand~\cite{feijs_multi-tasking_2013}.
This results in hard to maintain, error prone and unscalable spaghetti code.

There are many solutions to overcome this problem in imperative languages.

If the host language is a functional language (e.g.\ the aforementioned \gls{Scheme} variants) multitasking can be achieved without this burden relatively easy using continuation style multiprocessing~\cite{wand_continuation-based_1980}.
Writing in this style is complicated and converting an existing program in this \gls{CPS} results in relatively large programs.
Furthermore, there is no built-in thread-safe communication possible between the tasks.
A \gls{TOP} or \gls{FRP} based language benefits even more because the programmer is not required to explicitly define continuation points.

Regular preemptive multithreading is too memory intensive for smaller microcontrollers and therefore not suitable.
Manual interleaving of imperative code can be automated to certain extents.
Solutions often require an \gls{RTOS}, have a high memory requirement, do not support local variables, no thread-safe shared memory, no composition or no events as described in Table~\ref{tbl:multithreadingcompare} adapted from Santanna et al~\cite[p.\ 12]{santanna_safe_2013}.
The table compares the solutions in the relevant categories with \gls{mTask}.

\begin{table}[ht]
	\begin{threeparttable}
		\begin{tabular}{lc>{\columncolor[gray]{0.95}}cc>{\columncolor[gray]{0.95}}cc>{\columncolor[gray]{0.95}}cc}
			\toprule
			\multicolumn{2}{c}{Language} & \multicolumn{3}{c}{Complexity} & \multicolumn{3}{c}{Safety}\\
			\midrule
			Name         & Year & \multicolumn{1}{c}{Seq.\ ex.} & Loc.\ var. & \multicolumn{1}{c}{Par.\ comp.} & Det.\ ex.  & \multicolumn{1}{c}{Bound.\ ex.} & Safe.\  mem.\\
			\midrule
			Preemptive   & many & \cmark{} & \cmark{} &          &          & rt                 &\\
			nesC         & 2003 &          &          &          & \cmark{} & async              &\\
			OSM          & 2005 &          & \cmark{} & \cmark{} &          &                    &\\
			Protothreads & 2006 & \cmark{} &          &          & \cmark{} &                    &\\
			TinyThreads  & 2006 & \cmark{} & \cmark{} &          & \cmark{} &                    &\\
			Sol          & 2007 & \cmark{} & \cmark{} & \cmark{} & \cmark{} &                    &\\
			FlowTask     & 2011 & \cmark{} & \cmark{} &          &          &                    &\\
			Ocram        & 2013 & \cmark{} & \cmark{} &          & \cmark{} &                    &\\
			C\'eu        & 2013 & \cmark{} & \cmark{} & \cmark{} & \cmark{} & \cmark{}           & \cmark{}\\
			mTask        & 2018 & \cmark{} & \cmark{} & \cmark{} & \cmark{} & \cmark{}*\tnote{1} & \cmark{}\tnote{2}\\
			\bottomrule
		\end{tabular}
		\begin{tablenotes}
			\item [1] Only for tasks, not for expressions.
			\item [2] Using \glspl{SDS}.
		\end{tablenotes}
		\caption{%
			An overview of imperative multithreading solutions for tiny computers with their relevant characteristics.
			The characteristics are: sequential execution, local variable support, parallel composition, deterministic execution, bounded execution and safe shared memory (Adapted from Santanna et al~\cite[p.\ 12]{santanna_safe_2013}).
		}\label{tbl:multithreadingcompare}
	\end{threeparttable}
\end{table}

\subsection{\glsentrytext{mTask} history}
A first throw at a class-based shallowly \gls{EDSL} for \glspl{MCU} was made by Pieter Koopman and Rinus Plasmijer in 2016~\cite{plasmeijer_shallow_2016}.
The language was called \gls{ARDSL} and offered a type safe interface to \gls{Arduino} \gls{C++} dialect.
A \gls{C++} code generation backend was available together with an \gls{iTasks} simulation backend.
There was no support for tasks or even functions.
Some time later an unpublished extended version was created that allowed the creation of imperative tasks, \glspl{SDS} and the usage of functions.
The name then changed from \gls{ARDSL} to \gls{mTask}.

Mart Lubbers extended this in his Master's Thesis by adding integration with \gls{iTasks} and a bytecode compiler to the language~\cite{lubbers_task_2017}.
\Gls{SDS} in \gls{mTask} could be accessed on the \gls{iTasks} server.
In this way, entire \gls{IOT} systems could be programmed from a single source.
However, this version used a simplified version of \gls{mTask} without functions.
This was later improved upon by creating a simplified interface where \glspl{SDS} from \gls{iTasks} could be used in \gls{mTask} and the other way around~\cite{lubbers_task_2018}.
It was shown by Matheus Amazonas Cabral de Andrade that it was possible to build real-life \gls{IOT} systems with this integration~\cite{amazonas_cabral_de_andrade_developing_2018}.

The \gls{mTask} language as it is now was introduced in 2018~\cite{koopman_task-based_2018}.
This paper updated the language to support functions, tasks and \glspl{SDS} but still compiled to \gls{C++} \gls{Arduino} code.
Later the bytecode compiler and \gls{iTasks} integration was added to the language~\cite{lubbers_interpreting_2019}.
Moreover, it was shown that it is very intuitive to write \gls{MCU} applications in a \gls{TOP} language~\cite{lubbers_multitasking_2019}.
One reason for this is that a lot of design patterns that are difficult using standard means are for free in \gls{TOP} (e.g.\ multithreading).
Furthermore, Erin van der Veen has been working on a green computer analysis and is working on support for bounded data types.

%% file: 5conclusion.tex
These lecture notes give a complete introduction to the design and use of the \gls{mTask} system.
Furthermore it provides a hands-on tutorial for writing \gls{IOT} applications with it.

The number of \gls{IOT} devices is increasing evermore but programming them is as difficult and error-prone as it ever was.
Most programs written for \gls{IOT} devices are collections of loosely dependent parallel tasks which makes programming the devices using \gls{TOP} very natural.
The \gls{mTask} language is a multi-backend device-agnostic \gls{TOP} language specialized for \gls{IOT} tasks.
It contains a backend that will compile the program to bytecode that is then sent to the device.
The backend is fully integrated in \gls{iTasks} which means that tasks that are sent to the device act as regular \gls{iTasks} tasks, i.e.\ their task value can be observed and they can interact with \glspl{SDS} on the server.
There is no impedance problem in the \gls{mTask} ecosystem since all code is written in a single language, albeit in two \glspl{EDSL}.
The bytecode generation backend of \gls{mTask} --- and \gls{iTasks} for that matter --- make heavy use of generic programming techniques to relieve the programmer of the burden to hand-craft specifics such as the user interface, the communication protocol or serialization.
The execution semantics of the tasks makes them similar to lightweight threads --- for which there is typically no support on microcomputers due to the lack of an \gls{OS}.
This allows programmers to create multitasking applications just by using parallel combinators.
Reasonably complex \gls{IOT} applications spanning all layers of \gls{IOT} can be written in a concise and safe way using the \gls{mTask} system.

Future work may be practical topics such as extending the number of supported platforms or extending the language with more features.
For example, adding lenses and combinators to \glspl{SDS} may improve the expressiveness of the language.
Also, type errors in the \gls{DSL} are presented to the programmer as type errors in the host language.
As a result of class based shallow embedding, the type errors are quite complicated.
It would be interesting to see whether techniques for mitigating this problem can be applied to \gls{mTask} as well~\cite{serrano_type_2018}.
The execution model of the \gls{mTask} system lets the server send arbitrary code to the device to be executed.
This may pose a problem if the server, the communication technique is not to be trusted or can be snooped on.
At the time of writing a student is working on analysing this problem in his thesis.
Finally it would be interesting to allow the user instead of the programmer to write \gls{mTask} tasks from scratch.
This can be achieved by creating a type-safe editor in \gls{iTasks} that constructs tasks.

%% file: acknowledgements.tex
This paper constitutes the adapted lecture notes for the hands-on course presented at the \gls{CEFP} in Budapest between 17 and 21 June 2019.
This research is partly funded by the Royal Netherlands Navy.
Furthermore, we would like to thank the reviewers for their valuable comments.

%% file: aedsl.tex
An \gls{EDSL} is a language embedded in a host language created for a specific domain~\cite{hickey_building_2014}.
\glspl{EDSL} can have one or more backends or views.
Commonly used views are pretty printing, compiling, simulating, verifying and proving the program.
There are several techniques available for creating \glspl{EDSL}.
They all have their own advantages and disadvantages in terms of extendability, type safety and view support.
In the following subsections each of the main techniques are briefly explained.
An example expression \gls{DSL} is used as a running example.

\subsection{Deep embedding}
A deeply \gls{EDSL} is a language represented as data in the host language.
Views are functions that transform \emph{something} to the datatype or the other way around.
Definition~\ref{lst:exdeep} shows an example implementation for the expression \gls{DSL}.

\begin{lstdefinition}[language=Clean,label={lst:exdeep},caption={A deeply embedded expression \gls{DSL}.}]
:: Expr
	= LitI  Int
	| LitB  Bool
	| Var   String
	| Plus  Expr Expr
	| Eq    Expr Expr
\end{lstdefinition}

Deep embedding has the advantage that it is easy to build and views are easy to add.
On the downside, the expressions created with this language are not necessarily type-safe.
In the given language it is possible to create an expression such as \Cl{Plus (LitI 4) (LitB True)} that adds a boolean to an integer.
Extending the \gls{ADT} is easy and convenient but extending the views accordingly is tedious since it has to be done individually for all views.

The first downside of this type of \gls{EDSL} can be overcome by using \glspl{GADT}~\cite{cheney_first-class_2003}.
Example~\ref{lst:exdeepgadt} shows the same language, but type-safe with a \gls{GADT}.
\glspl{GADT} are not supported in the current version of \gls{Clean} and therefore the syntax is hypothetical.
However, it has been shown that \glspl{GADT} can be simulated using bimaps or projection pairs~\cite{cheney_first-class_2003}.
Unfortunately the lack of extendability remains a problem.
If a language construct is added, no compile time guarantee can be given that all views support it.

\begin{lstdefinition}[language=Clean,label={lst:exdeepgadt},caption={A deeply embedded expression \gls{DSL} using \glspl{GADT}.}]
:: Expr a
	=     Lit  a                        -> Expr a
	| E.e: Var   String                  -> Expr e
	|     Plus  (Expr Int)  (Expr Int)  -> Expr Int
	| E.e: Eq    (Expr e)    (Expr e)    -> Expr Bool & == e
\end{lstdefinition}

\subsection{Shallow embedding}
In a shallowly \gls{EDSL} all language constructs are expressed as functions in the host language.
An evaluator view for the example language then can be implemented as the code shown in Definition~\ref{lst:exshallow}.
Note that much of the internals of the language can be hidden using monads.

\begin{lstdefinition}[language=Clean,label={lst:exshallow}, caption={A minimal shallow \gls{EDSL}.}]
:: Env   = ...            // Some environment
:: DSL a :== (Env -> a)

Lit :: a -> DSL a
Lit x = \e->x

Var :: String -> DSL Int
Var i = \e->retrEnv e i

Plus :: (DSL Int) (DSL Int) -> DSL Int
Plus x y = \e->x e + y e

Eq :: (DSL a) (DSL a) -> DSL Bool | == a
Eq x y = \e->x e == y e
\end{lstdefinition}

The advantage of shallowly embedding a language in a host language is its extendability.
It is very easy to add functionality because the compile time checks of the host language guarantee whether or not the functionality is available when used.
Moreover, the language is type safe as it is directly typed in the host language, i.e.\ \Cl{Lit True +. Lit 4} is rejected.

The downside of this method is extending the language with views.
It is nearly impossible to add views to a shallowly embedded language.
The only way of achieving this is by reimplementing all functions so that they run all backends at the same time.
This will mean that every component will have to implement all views rendering it slow for multiple views and complex to implement.

%% file: aitasks.tex
This appendix gives a brief overview of \gls{iTasks}.
It is by far extensive but should cover all \gls{iTasks} constructions required for the exercises.
Some examples from~\cite{wang_maintaining_2018} can be found in Section~\ref{sec:itasksexamples}.

\subsection{Types}
The class collection \Cl{iTask} is used throughout the library to make sure the types used have all the required machinery for \gls{iTasks}.
This class collection contains only generic functions that can automatically be derived for any first order user defined type.
Example~\ref{lst:itasksderive} shows how to derive this class.

\begin{lstexample}[language=Clean,caption={Derive the \Cl{iTask} class for a user defined type.},label={lst:itasksderive}]
:: MyName =
	{ firstName :: String
	, lastName  :: String
	}
derive class iTask MyName
\end{lstexample}

\subsection{Editors}\label{sec:editor}
The most common basic tasks are editors for entering, viewing or update information.
For the three basic editors there are three corresponding functions to create tasks as seen in Definition~\ref{lst:itaskseditors}.

\begin{lstdefinition}[language=Clean,caption={The definitions of editors in \gls{iTasks}.},label={lst:itaskseditors}]
enterInformation  :: d [EnterOption m]      -> Task m | iTask m & toPrompt d
updateInformation :: d [UpdateOption m m] m -> Task m | iTask m & toPrompt d
viewInformation   :: d [ViewOption m]     m -> Task m | iTask m & toPrompt d
\end{lstdefinition}

The first argument of the function is something implementing \Cl{toPrompt}.
There are \Cl{toPrompt} instances for at least \Cl{String} --- for a description, \Cl{(String, String)} --- for a title and a description and \Cl{()} --- for no description.

The second argument is a list of options for modifying the editor behaviour.
This list is either empty or contains exactly one item.
The types for the options are shown in Definition~\ref{lst:itaskseditoroptions}.
Simple lenses are created using the \Cl{*As} constructor.
If an entirely different editor must be used, the \Cl{*Using} constructors can be used.

\begin{lstdefinition}[language=Clean,caption={The definitions of editors in \gls{iTasks}.},label={lst:itaskseditoroptions}]
:: ViewOption a
	= E.v: ViewAs     (a -> v)            & iTask v
	| E.v: ViewUsing  (a -> v) (Editor v) & iTask v
:: EnterOption a
	= E.v: EnterAs    (v -> a)            & iTask v
	| E.v: EnterUsing (v -> a) (Editor v) & iTask v
:: UpdateOption a b
	= E.v: UpdateAs    (a -> v) (a v -> b)            & iTask v
	| E.v: UpdateUsing (a -> v) (a v -> b) (Editor v) & iTask v
\end{lstdefinition}

Example~\ref{lst:exfahr} shows an example of such an editor using a lens.
The user enters a temperature in degrees Celsius and the editor automatically converts the result to a temperature in Fahrenheit which is in turn the observed task value.

\begin{lstexample}[language=Clean,caption={An example of an editor that converts the entered value to a different unit in \gls{iTasks}.},label={lst:exfahr}]
tempFahrenheit :: Task Real
tempFahrenheit = enterInformation "Enter the temperature in degrees Celsius"
	[EnterUsing \c->c*(9.0/5.0) + 32]
\end{lstexample}

\subsection{Task combinators}
There are two flavours of task combinators, namely parallel and sequential that are all specializations of their Swiss-army knife combinator \Cl{step} and \Cl{parallel} respectively.

\subsubsection{Parallel combinators}
The two main parallel combinators are the conjunction and disjunction combinators shown in Definition~\ref{lst:itasksparallel}.

The \Cl{-&&-} has semantics similar to the \gls{mTask} \Cl{.&&.} combinator.
The \Cl{-||-} has the same semantics as the \gls{mTask} \Cl{.||.} combinator.
The \Cl{-||} and \Cl{||-} executes both tasks in parallel but only looks at the value of the left task or the right task respectively.

\begin{lstdefinition}[language=Clean,caption={The definitions of parallel combinators in \gls{iTasks}.},label={lst:itasksparallel}]
(-&&-) infixr 4 :: (Task a) (Task b) -> Task (a,b) | iTask a & iTask b
(-|| ) infixl 3 :: (Task a) (Task b) -> Task a     | iTask a & iTask b
( ||-) infixr 3 :: (Task a) (Task b) -> Task b     | iTask a & iTask b
(-||-) infixr 3 :: (Task a) (Task a) -> Task a     | iTask a
\end{lstdefinition}

Example~\ref{lst:itasksparallelex} shows an example of a task that, using the disjunction combinator, asks the user for a temperature either in degrees Celsius or Fahrenheit using the task from Example~\ref{lst:exfahr}.
Whichever editor the user edits last, will be the observable task value.

\begin{lstexample}[language=Clean,caption={An example of parallel task combinators in \gls{iTasks}.},label={lst:itasksparallelex}]
askTemp :: Task Real
askTemp =   enterInformation "Temperature in Fahrenheit" []
       -||- tempFahrenheit
\end{lstexample}

\subsubsection{Sequential combinators}
All sequential combinators are derived from the \Cl{>>*} combinator as shown in Definition~\ref{lst:itasksseq}.
With this combinator, the task value of the left-hand side can be observed and execution continues with the right-hand side if one of the continuations yields a \Cl{Just (Task b)}.
The listing also shows many utility functions for defining task steps.

\begin{lstdefinition}[language=Clean,caption={The definitions of sequential combinators in \gls{iTasks}.},label={lst:itasksseq}]
(>>*) infixl 1 :: (Task a) [TaskCont a (Task b)] -> Task b | iTask a & ...
:: TaskCont a b
	= OnValue         ((TaskValue a) -> Maybe b)
	| OnAction Action ((TaskValue a) -> Maybe b)

:: Action = Action String //button

always       :: b                    (TaskValue a) -> Maybe b
never        :: b                    (TaskValue a) -> Maybe b
hasValue     :: (a -> b)             (TaskValue a) -> Maybe b
ifStable     :: (a -> b)             (TaskValue a) -> Maybe b
ifUnstable   :: (a -> b)             (TaskValue a) -> Maybe b
ifValue      :: (a -> Bool) (a -> b) (TaskValue a) -> Maybe b
ifCond       :: Bool b               (TaskValue a) -> Maybe b
withoutValue :: (Maybe b)            (TaskValue a) -> Maybe b
withValue    :: (a -> Maybe b)       (TaskValue a) -> Maybe b
withStable   :: (a -> Maybe b)       (TaskValue a) -> Maybe b
withUnstable :: (a -> Maybe b)       (TaskValue a) -> Maybe b
\end{lstdefinition}

Example~\ref{lst:itasksstepex} shows an example of the step combinator that forces the user to enter a number between 0 and 10.
If the user enters a different value, the continue button will remain disabled.

\begin{lstexample}[language=Clean,caption={An example of parallel task combinators in \gls{iTasks}.},label={lst:itasksstepex}]
numberBetween0and10 :: Task Int
numberBetween0and10 = enterInformation "Enter a number between 0 and 10" []
	>>* [OnAction (Action "Continue") $ ifValue (\i->i <= 10 && i >= 0) $ \i->return i]
\end{lstexample}

Derived from the \Cl{>>*} combinator are all other sequential combinators such as the ones listed in Definition~\ref{lst:itasksseqderive} with their respective documentation.

\begin{lstdefinition}[language=Clean,caption={The definitions of derived sequential combinators in \gls{iTasks}.},label={lst:itasksseqderive}]
// Combines two tasks sequentially. The first task is executed first.
// When it has a value the user may continue to the second task, which is
// executed with the result of the first task as parameter.
// If the first task becomes stable, the second task is started automatically.
(>>=) infixl 1 :: (Task a) (a -> Task b) -> Task b | iTask a & iTask b

// Combines two tasks sequentially but explicitly waits for user input to
// confirm the completion of
(>>!) infixl 1 :: (Task a) (a -> Task b) -> Task b | iTask a & iTask b

// Combines two tasks sequentially but continues only when the first task has a
// stable value.
(>>-) infixl 1 :: (Task a) (a -> Task b) -> Task b | iTask a & iTask b

// Combines two tasks sequentially but continues only when the first task has a
// stable value.
(>-|) infixl 1
(>-|) x y :== x >>- \_ -> y

// Combines two tasks sequentially but continues only when the first task has a
// value.
(>>~) infixl 1 :: (Task a) (a -> Task b) -> Task b | iTask a & iTask b

// Combines two tasks sequentially just as >>=, but the result of the second
// task is disregarded.
(>>^) infixl 1 :: (Task a) (Task b) -> Task a| iTask a & iTask b

// Execute the list of tasks one after another.
sequence :: [Task a] -> Task [a] | iTask a
\end{lstdefinition}

\subsection{\acrlongpl{SDS}}\label{sec:sds}
Data can be observer via task values but for unrelated tasks to share data, \glspl{SDS} are used.
There is an publish subscribe system powering the \gls{SDS} system that makes sure tasks are only rewritten when activity has taken place in the \gls{SDS}.
There are many types of \glspl{SDS} such as lenses, sources and combinators.
As long as they implement the \Cl{RWShared} class collection, you can use them as an \gls{SDS}.
Definition~\ref{lst:itaskssds} shows two methods for creating an \gls{SDS}, they both yield a \Cl{SimpleSDSLens} but they can be used by any task using an \gls{SDS}.

\begin{lstdefinition}[language=Clean,caption={The definitions for \glspl{SDS} in \gls{iTasks}.},label={lst:itaskssds}]
sharedStore :: String a -> SimpleSDSLens a | iTask a
withShared  :: b ((SimpleSDSLens b) -> Task a) -> Task a | iTask a & iTask b
\end{lstdefinition}

With the \Cl{sharedStore} function, a named \gls{SDS} can be created that acts as a well-typed global variable.
\Cl{withShared} is used to create an anonymous local \gls{SDS}.

There are four major operations that can be done on \glspl{SDS} that are all atomic (see Definition~\ref{lst:itaskssdstasks}).
\Cl{get} fetches the value from the \gls{SDS} and yields it as a stable value.
\Cl{set} writes the given value to the \gls{SDS} and yields it as a stable value.
\Cl{upd} applies an update function to the \gls{SDS} and returns the written value as a stable value.
\Cl{watch} continuously emits the value of the \gls{SDS} as an unstable task value.
The implementation uses a publish subscribe system to evaluate the watch task only when the value of the \gls{SDS} changes.

\begin{lstdefinition}[language=Clean,caption={The definitions for \gls{SDS} access tasks in \gls{iTasks}.},label={lst:itaskssdstasks}]
get   ::          (sds () r w) -> Task r | iTask r & iTask w & RWShared sds
set   :: w        (sds () r w) -> Task w | iTask r & iTask w & RWShared sds
upd   :: (r -> w) (sds () r w) -> Task w | iTask r & iTask w & RWShared sds
watch ::        (sds () r w) -> Task r | iTask r & iTask w & RWShared sds
\end{lstdefinition}

For all editors, there are shared variants available as shown in Definition~\ref{lst:itaskssharededitors}.
This allows a user to interact with the \gls{SDS}.

\begin{lstdefinition}[language=Clean,caption={The definitions for \gls{SDS} editor tasks in \gls{iTasks}.}]
updateSharedInformation :: d [UpdateOption r w] (sds () r w) -> Task r | ...
viewSharedInformation   :: d [ViewOption r]     (sds () r w) -> Task r | ...
\end{lstdefinition}

\begin{lstexample}[language=Clean,caption={An example of multiple tasks interacting with the same \gls{SDS} in \gls{iTasks}.},label={lst:itaskssharededitors}]
sharedUpdate :: Task Int
sharedUpdate = withShared 42 \sharedInt->
	     updateSharedInformation "Left"  [] sharedInt
	-||- updateSharedInformation "Right" [] sharedInt
\end{lstexample}

\subsection{Extra task combinators}
Not all workflow patterns can be described using only the derived combinators.
Therefore, some other task combinators have been invented that are not truly sequential nor truly parallel.
Definition~\ref{lst:itaskshybrid} shows some combinators that might be useful in the exercises.

\begin{lstdefinition}[language=Clean,caption={The definitions for hybrid combinators in \gls{iTasks}.},label={lst:itaskshybrid}]
//Feed the result of one task as read-only shared to another
(>&>) infixl 1 :: (Task a) ((SDSLens () (Maybe a) ()) -> Task b) -> Task b | ...

// Sidestep combinator. This combinator has a similar signature as the >>*
// combinator, but instead of moving forward to a next step, the selected step is
// executed in parallel with the first task. When the chosen task step becomes
// stable, it is removed and the actions are enabled again.
(>^*) infixl 1 :: (Task a) [TaskCont a (Task b)] -> Task a | iTask a & iTask b

// Apply a function on the task value while retaining stability
(@) infixl 1 :: (Task a) (a -> b) -> Task b
// Map the task value to a constant value while retaining stability
(@) infixl 1 :: (Task a) b -> Task b

// Repeats a task indefinitely
forever :: (Task a) -> Task a | iTasks a
\end{lstdefinition}

\subsection{Examples}\label{sec:itasksexamples}
Some workflow task patterns can easily be created using the builtin combinator as shown in Examples~\ref{lst:taskpatterns}.

\begin{lstexample}[language=Clean,caption={Some workflow task patterns.},label={lst:taskpatterns}]
maybeCancel :: String (Task a) -> Task (Maybe a) | iTask a
maybeCancel panic t = t >>*
	[ OnValue (ifStable (return o Just))
	, OnAction (Action panic) (always (return Nothing))
	]

:: Date //type from iTasks.Extensions.DateTime
currentDate :: SDSLens () Date () // Builtin SDS

waitForDate :: Date -> Task Date
waitForDate d = viewSharedInformation ("Wait until" +++ toString d) [] currentDate
	>>* [OnValue (ifValue (\now -> date < now) return)]

deadlineWith :: Date a (Task a) -> Task a | iTask a
deadlineWith d a t = t -||- (waitForDate d >>| return a)

reminder :: Date String -> Task ()
reminder d m = waitForDate d >>| viewInformation ("Reminder: please " +++ m) [] ()
\end{lstexample}

%% file: ainstall.tex
This section will give detailed instructions on how to install \gls{mTask} on your system.
The distribution used also includes the example skeletons.

\subsection{Fetch the \acrshort{CEFP} distribution}
Download the \gls{CEFP} version of \gls{mTask} distribution for your operating system as given in Table~\ref{tbl:downloadLinks} and decompress the archive.
The archives is all you need since it contains a complete clean distribution.
The windows version contains an \gls{IDE} and \gls{CPM}.
Mac and Linux only have a project manager called \gls{CPM}.

\begin{table}[ht]
	\centering
	\begin{tabular}{lll}
		\toprule
		\gls{OS} & Arch & URL\\
		\midrule
		Linux   & x64 & \fmturl{https://ftp.cs.ru.nl/Clean/CEFP19/mtask-linux-x64.tar.gz}\\
				&     & {\footnotesize Requires GCC}\\
		Windows & x64 & \fmturl{https://ftp.cs.ru.nl/Clean/CEFP19/mtask-windows-x64.zip}\\
		MacOS   & x64 & \fmturl{https://ftp.cs.ru.nl/Clean/CEFP19/mtask-macos-x64.tar.gz}\\
				&     & {\footnotesize Requires XCode}\\
		\bottomrule
	\end{tabular}
	\caption{Download links for the \gls{CEFP} builds of \gls{mTask}.}\label{tbl:downloadLinks}
\end{table}

\subsection{Setup}
\subsubsection{Linux}
Assuming you uncompressed the archive in \path{~/mTask}, run the following commands in a terminal.

\begin{lstexample}[language=bash]
# Add the bin directory of the clean distribution to $PATH
echo 'export PATH=~/mTask/clean/bin:$PATH' >> ~/.bashrc
# Correctly set CLEAN_HOME
echo 'export CLEAN_HOME=~/mTask/clean' >> ~/.bashrc
# Source it for your current session
source ~/.bashrc
\end{lstexample}

\subsubsection{Windows}
You do not need to setup anything on windows.
However, if you want to use \gls{CPM} as well, you need to add the \path{;C:\Users\frobnicator\mTask\clean} to your \texttt{\%PATH\%}\footnote{Instructions from \fmturl{https://hmgaudecker.github.io/econ-python-environment/paths.html}}.

\subsubsection{MacOS}
Assuming you uncompressed the archive in \path{~/mTask}, run the following commands in a terminal.

\begin{lstexample}[language=bash]
# Add the bin directory of the clean distribution to $PATH
echo 'export PATH=~/mTask/clean/bin:$PATH' >> ~/.bash_profile
# Correctly set CLEAN_HOME
echo 'export CLEAN_HOME=~/mTask/clean' >> ~/.bash_profile
# Source it for your current session
source ~/.bashrc
\end{lstexample}

\subsection{Compile the test program}
Note that the first time compiling everything can take a while and will consume quite some memory.

\subsubsection{Windows}
Assuming you uncompressed the archive in \path{C:\Users\frobnicator\mTask}.
Connect a device or start the local \gls{TCP} client by executing \path{C:\Users\frobnicator\mTask\client.exe}

\paragraph{\acrshort{IDE}}
\begin{itemize}
	\item Open the \gls{IDE} by starting \path{C:\Users\frobnicator\mTask\clean\CleanIDE.exe}.
	\item Click on \menu{File > Open} or press \keys{\ctrl+O} ond open \path{C:\Users\frobnicator\mTask\mTask\cefp19\blink.prj}.
	\item Click on \menu{Project > Update and Run} or press \keys{\ctrl+R}.
\end{itemize}

\paragraph{\acrshort{CPM}}
Enter the following commands in a command prompt or PowerShell session:

\begin{lstexample}[language=command.com]
cd C:\Users\frobnicator\mTask\mTask\cefp19
cpm blink.prj
blink.exe
\end{lstexample}

\subsubsection{Linux \& MacOS}
Assuming you uncompressed the archive in \path{~/mTask}.
Connect a device or start the local \gls{TCP} client by executing \path{~/mTask/client}.
In a terminal enter the following commands:

\begin{lstexample}[language=bash]
cd ~/mTask/cefp19
cpm blink.prj
./blink
\end{lstexample}

\subsection{Setup the \acrlong{MCU}}\label{sec:setupTheMicrocontroller}
For setting up the \gls{RTS} for the \gls{MCU}, the reader is kindly referred to here\footnote{\url{https://gitlab.science.ru.nl/mlubbers/mTask/blob/cefp19/DEVICES.md}}.

%% file: asolutions.tex
\setcounter{lstlisting}{2}
\begin{lstsolution}[language=Clean,caption={Blink the builtin \acrshort{LED} with two patterns}]
main :: Task Bool
main = enterDevice
	>>= \spec->enterInformation "Enter the intervals (ms)"
	>>= \(i1, i2)->withDevice spec
		\dev->liftmTask (blink i1 i2) dev -|| viewDevice dev
where
	blink :: Int Int -> Main (MTask v Bool) | mtask v
	blink x y
		= fun \blink = (\(p, x, y)->
			     delay y
			>>|. writeD p x
			>>=. \x->blink (p, Not x, y))
		In {main = blink (d4, true, lit x)
		      .||. blink (d4, true, lit y)}
\end{lstsolution}

\begin{lstsolution}[language=Clean,caption={Blink the builtin \acrshort{LED} on demand}]
main :: Task Bool
main = enterDevice >>= \spec->withDevice spec
	\dev->withShared True \blinkOk->
		    liftmTask (blink blinkOk) dev
		-|| updateSharedInformation "Blink Enabled" [] blinkOk
where
	blink :: (Shared s Bool) -> Main (MTask v Bool) | mtask, liftsds v & RWShared s
	blink blinkShare = liftsds \blinkOk=blinkShare
		In fun \blink = (\x->
			     writeD d2 x
			>>|. delay (lit 500)
			>>|. getSds blinkOk
			>>*. [IfValue (\x->x) (\_->blink (Not x))])
		In {main = blink (lit True)}
\end{lstsolution}

\clearpage
\begin{lstsolution}[language=Clean,caption={Show the temperature via an \acrshort{SDS}}]
temp :: (Shared s Int) -> Main (MTask v ()) | mtask, dht, liftsds v & RWShared s
temp tempShare =
	DHT D4 DHT22 \dht->
	liftsds \sTemp = tempShare
	In fun \monitor = (\x->temperature dht
		>>*. [IfValue ((!=.)x) (setSds sTemp)]
		>>=. monitor)
	In {main = monitor (lit 0)}
\end{lstsolution}

\begin{lstsolution}[caption={Simple thermostat},language=Clean]
temp :: (Shared s1 Int) (Shared s2 Int) -> Main (MTask v ())
	| mtask, dht, liftsds v & RWShared s1 & RWShared s2
temp targetShare tempShare =
	DHT D4 DHT22 \dht->
	   liftsds \sTemp = tempShare
	In liftsds \sTarget = targetShare
	In fun \monitor = (\x->temperature dht
		>>*. [IfValue ((!=.)x) (setSds sTemp)]
		>>=. monitor)
	In fun \heater = (\st->getSds sTemp .&&. getSds sTarget
		>>*. [IfValue (tupopen \(temp, target)->temp <. target &. Not st)
				\_->writeD d4 (lit True)
		     ,IfValue (tupopen \(temp, target)->temp >. target &. st)
				\_->writeD d4 (lit False)]
		>>=. heater)
		In {main = monitor (lit 0) .||. heater (lit True)}
\end{lstsolution}

\begin{lstsolution}[caption={\Acrshort{LED} Matrix 42 using \glsentrytext{iTasks}},language=Clean]
iTask42 :: MTDevice -> Task ()
iTask42 dev = liftmTask clear dev
	>-| sequence [liftmTask (toggle {x=x,y=y,status=True}) dev\\(x,y)<-fourtytwo] @! ()
            //Four
fourtytwo = [(0, 5), (0, 4), (0, 3), (0, 2) ,(1, 2), (2, 2), (2, 3) ,(2, 1), (2, 0)
            //Two
            ,(4, 5), (5, 5), (6, 4), (6, 3), (5, 2), (4, 1), (4, 0), (5, 0), (6, 0)]
\end{lstsolution}

\begin{lstsolution}[caption={\Acrshort{LED} Matrix 42 using \glsentrytext{mTask}},language=Clean]
mTask42 :: Main (MTask v ()) | mtask, LEDMatrix v
mTask42 = ledmatrix D5 D7 \lm->{main = LMClear lm >>|.
	foldr (>>|.) (LMDisplay lm) [dot lm {x=x, y=y, status=True} \\ (x,y) <- fourtytwo]}
\end{lstsolution}

\clearpage
\begin{lstsolution}[caption={Temperature plotter},language=Clean]
temp :: (Shared s1 (Int, Int)) (Shared s2 Int) (Shared s3 Int) (Shared s4 Int)
	-> Main (MTask v ()) | ...
temp limitsShare delayShare tempShare alarmShare =
	DHT D4 DHT22 \dht->
	ledmatrix D5 D7 \lm->
	   liftsds \sLimits = limitsShare
	In liftsds \sDelay  = delayShare
	In liftsds \sTemp   = tempShare
	In liftsds \sAlarm  = alarmShare
	In fun \print = (\(targety, currentx, currenty)->
		If (currenty ==. lit 8)
			(LMDisplay lm)
			(LMDot lm currentx currenty (targety ==. currenty)
				>>|. print (targety, currentx, currenty +. lit 1)))
	In fun \min = (\(x, y)->If (x <. y) x y)
	In fun \calcy = (\(up, down, val)->
		min (down, (val -. down) /. ((up -. down) /. lit 7)))
	In fun \plot = (\x->
		     getSds sLimits
		>>~. tupopen \(gmin, gmax)->temperature dht
		>>~. \y->print (min (lit 7, calcy (gmin, gmax, y)), x, lit 0)
		>>|. setSds sTemp y
		>>|. getSds sDelay
		>>~. delay
		>>|. plot (If (x ==. lit 7) (lit 0) (x +. lit 1))
	)
	In {main = plot (lit 0)
		  .||. rpeat (readD BUILTIN_LED >>*. [IfValue Not (writeD ABUTTON o Not)])
		  .||. rpeat (getSds sAlarm .&&. getSds sTemp
			>>*. [IfValue (tupopen \(a, t)->t >. a) \_->writeD ABUTTON (lit False)]
	)}
\end{lstsolution}

%% file: main.bbl
\begin{thebibliography}{10}
\providecommand{\url}[1]{\texttt{#1}}
\providecommand{\urlprefix}{URL }
\providecommand{\doi}[1]{https://doi.org/#1}

\bibitem{achten_clean_2007}
Achten, P.: Clean for {Haskell98} {Programmers} (Jul 2007)

\bibitem{adams_hitchhikers_2017}
Adams, D.: The {Hitchhiker}'s {Guide} to the {Galaxy} {Omnibus}: {A} {Trilogy}
  in {Four} {Parts}, vol.~6. Pan Macmillan (2017)

\bibitem{alimarine_generic_2005}
Alimarine, A.: Generic {Functional} {Programming}. {PhD}, Radboud University,
  Nijmegen (2005)

\bibitem{amazonas_cabral_de_andrade_developing_2018}
Amazonas Cabral De~Andrade, M.: Developing {Real} {Life}, {Task} {Oriented}
  {Applications} for the {Internet} of {Things}. Master's thesis, Radboud
  University, Nijmegen (2018)

\bibitem{amsden_survey_2011}
Amsden, E.: A {Survey} of {Functional} {Reactive} {Programming}. Tech. rep.
  (2011)

\bibitem{baccelli_reprogramming_2018}
Baccelli, E., Doerr, J., Jallouli, O., Kikuchi, S., Morgenstern, A., Padilla,
  F.A., Schleiser, K., Thomas, I.: Reprogramming {Low}-end {IoT} {Devices} from
  the {Cloud}. In: 2018 3rd {Cloudification} of the {Internet} of {Things}
  ({CIoT}). pp.~1--6. IEEE (2018)

\bibitem{baccelli_scripting_2018}
Baccelli, E., Doerr, J., Kikuchi, S., Padilla, F., Schleiser, K., Thomas, I.:
  Scripting {Over}-{The}-{Air}: {Towards} {Containers} on {Low}-end {Devices}
  in the {Internet} of {Things}. In: {IEEE} {PerCom} 2018 (2018)

\bibitem{bolderheij_mission-driven_2018}
Bolderheij, F., Jansen, J., Kool, A., Stutterheim, J.: A {Mission}-{Driven}
  {C2} {Framework} for {Enabling} {Heterogeneous} {Collaboration}. In: {NL}
  {ARMS} {Netherlands} {Annual} {Review} of {Military} {Studies} 2018, pp.
  107--130. Springer (2018)

\bibitem{brus_clean_1987}
Brus, T., van Eekelen, M., Van~Leer, M., Plasmeijer, M.: Clean – a language
  for functional graph rewriting. In: Conference on {Functional} {Programming}
  {Languages} and {Computer} {Architecture}. pp. 364--384. Springer (1987)

\bibitem{carette_finally_2009}
Carette, J., Kiselyov, O., Shan, C.C.: Finally tagless, partially evaluated:
  {Tagless} staged interpreters for simpler typed languages. Journal of
  Functional Programming  \textbf{19}(05), ~509 (Sep 2009).
  \doi{10.1017/S0956796809007205}

\bibitem{cheney_first-class_2003}
Cheney, J., Hinze, R.: First-class phantom types. Tech. rep., Cornell
  University (2003)

\bibitem{da_xu_internet_2014}
Da~Xu, L., He, W., Li, S.: Internet of things in industries: a survey.
  Industrial Informatics, IEEE Transactions on  \textbf{10}(4),  2233--2243
  (2014)

\bibitem{domoszlai_parametric_2014}
Domoszlai, L., Lijnse, B., Plasmeijer, R.: Parametric lenses: change
  notification for bidirectional lenses. In: Proceedings of the 26nd 2014
  {International} {Symposium} on {Implementation} and {Application} of
  {Functional} {Languages}. p.~9. ACM (2014)

\bibitem{dube_bit:_2000}
Dubé, D.: {BIT}: {A} very compact {Scheme} system for embedded applications.
  Proceedings of the Fourth Workshop on Scheme and Functional Programming
  (2000)

\bibitem{elliott_functional_1997}
Elliott, C., Hudak, P.: Functional reactive animation. In: {ACM} {SIGPLAN}
  {Notices}. vol.~32, pp. 263--273. ACM (1997)

\bibitem{feeley_picbit:_2003}
Feeley, M., Dubé, D.: {PICBIT}: {A} {Scheme} system for the {PIC}
  microcontroller. In: Proceedings of the {Fourth} {Workshop} on {Scheme} and
  {Functional} {Programming}. pp. 7--15. Citeseer (2003)

\bibitem{feijs_multi-tasking_2013}
Feijs, L.: Multi-tasking and {Arduino} : why and how? In: Chen, L.L.,
  Djajadiningrat, T., Feijs, L.M.G., Fraser, S., Hu, J., Kyffin, S., Steffen,
  D. (eds.) Design and semantics of form and movement. 8th {International}
  {Conference} on {Design} and {Semantics} of {Form} and {Movement} ({DeSForM}
  2013). pp. 119--127. Wuxi, China (2013)

\bibitem{grebe_haskino:_2016}
Grebe, M., Gill, A.: Haskino: {A} remote monad for programming the arduino. In:
  International {Symposium} on {Practical} {Aspects} of {Declarative}
  {Languages}. pp. 153--168. Springer (2016)

\bibitem{grebe_threading_2019}
Grebe, M., Gill, A.: Threading the {Arduino} with {Haskell}. In: Van~Horn, D.,
  Hughes, J. (eds.) Trends in {Functional} {Programming}. pp. 135--154.
  Springer International Publishing, Cham (2019)

\bibitem{haenisch_case_2016}
Haenisch, T.: A case study on using functional programming for internet of
  things applications. Athens Journal of Technology \& Engineering
  \textbf{3}(1) (2016)

\bibitem{helbling_juniper:_2016}
Helbling, C., Guyer, S.Z.: Juniper: a functional reactive programming language
  for the {Arduino}. In: Proceedings of the 4th {International} {Workshop} on
  {Functional} {Art}, {Music}, {Modelling}, and {Design}. pp. 8--16. ACM (2016)

\bibitem{hess_arduino-copilot_2020}
Hess, J.: arduino-copilot: {Arduino} programming in haskell using the {Copilot}
  stream {DSL} (2020), \url{http://hackage.haskell.org/package/arduino-copilot}

\bibitem{hickey_building_2014}
Hickey, P.C., Pike, L., Elliott, T., Bielman, J., Launchbury, J.: Building
  embedded systems with embedded {DSLs}. In: {ACM} {SIGPLAN} {Notices}.
  vol.~49, pp.~3--9. ACM Press (2014). \doi{10.1145/2628136.2628146}

\bibitem{jansen_towards_2010}
Jansen, J.M., Lijnse, B., Plasmeijer, R.: Towards dynamic workflows for crisis
  management  (2010)

\bibitem{johnson-davies_lisp_2020}
Johnson-Davies, D.: Lisp for microcontrollers (2020), \url{https://ulisp.com}

\bibitem{koopman_task-based_2018}
Koopman, P., Lubbers, M., Plasmeijer, R.: A {Task}-{Based} {DSL} for
  {Microcomputers}. In: Proceedings of the {Real} {World} {Domain} {Specific}
  {Languages} {Workshop} 2018 on - {RWDSL2018}. pp. 1--11. ACM Press, Vienna,
  Austria (2018). \doi{10.1145/3183895.3183902}

\bibitem{lijnse_capturing_2011}
Lijnse, B., Jansen, J.M., Nanne, R., Plasmeijer, R.: Capturing the netherlands
  coast guard's sar workflow with itasks  (2011)

\bibitem{lijnse_incidone:_2012}
Lijnse, B., Jansen, J.M., Plasmeijer, R., {others}: Incidone: {A} task-oriented
  incident coordination tool. In: Proceedings of the 9th {International}
  {Conference} on {Information} {Systems} for {Crisis} {Response} and
  {Management}, {ISCRAM}. vol.~12 (2012)

\bibitem{lijnse_itasks_2009}
Lijnse, B., Plasmeijer, R.: {iTasks} 2: {iTasks} for {End}-users. In:
  International {Symposium} on {Implementation} and {Application} of
  {Functional} {Languages}. pp. 36--54. Springer (2009)

\bibitem{lubbers_multitasking_2019}
Lubbers, M., Koopman, P., Plasmeijer, R.: Multitasking on {Microcontrollers}
  using {Task} {Oriented} {Programming}. In: 2019 42nd {International}
  {Convention} on {Information} and {Communication} {Technology}, {Electronics}
  and {Microelectronics} ({MIPRO}). pp. 1587--1592. Opatija, Croatia (May
  2019). \doi{10.23919/MIPRO.2019.8756711}

\bibitem{lubbers_task_2017}
Lubbers, M.: Task {Oriented} {Programming} and the {Internet} of {Things}.
  Master's thesis, Radboud University, Nijmegen (2017)

\bibitem{lubbers_task_2018}
Lubbers, M., Koopman, P., Plasmeijer, R.: Task {Oriented} {Programming} and the
  {Internet} of {Things}. In: Proceedings of the 30th {Symposium} on the
  {Implementation} and {Application} of {Functional} {Programming} {Languages}.
  p.~12. ACM, Lowell, MA (2018). \doi{10.1145/3310232.3310239}

\bibitem{lubbers_interpreting_2019}
Lubbers, M., Koopman, P., Plasmeijer, R.: Interpreting {Task} {Oriented}
  {Programs} on {Tiny} {Computers}. In: Proceedings of the 31st {Symposium} on
  {Implementation} and {Application} of {Functional} {Languages}. {IFL} '19,
  Association for Computing Machinery, New York, NY, USA (2019).
  \doi{10.1145/3412932.3412936}, \url{https://doi.org/10.1145/3412932.3412936},
  event-place: Singapore, Singapore

\bibitem{michels_uniform_2012}
Michels, S., Plasmeijer, R.: Uniform data sources in a functional language. In:
  Unpublished {Manuscript}. p.~16 (2012)

\bibitem{piers_task-oriented_2016}
Piers, J.: Task-{Oriented} {Programming} for developing non-distributed
  interruptible embedded systems. Master's thesis, Radboud University, Nijmegen
  (2016)

\bibitem{plasmeijer_conference_2006}
Plasmeijer, R., Achten, P.: A conference management system based on the {iData}
  toolkit. In: Symposium on {Implementation} and {Application} of {Functional}
  {Languages}. pp. 108--125. Springer (2006)

\bibitem{plasmeijer_itasks:_2007}
Plasmeijer, R., Achten, P., Koopman, P.: {iTasks}: executable specifications of
  interactive work flow systems for the web. ACM SIGPLAN Notices
  \textbf{42}(9),  141--152 (2007)

\bibitem{plasmeijer_shallow_2016}
Plasmeijer, R., Koopman, P.: A {Shallow} {Embedded} {Type} {Safe} {Extendable}
  {DSL} for the {Arduino}. In: Trends in {Functional} {Programming}, Lecture
  {Notes} in {Computer} {Science}, vol.~9547. Springer International
  Publishing, Cham (2016). \doi{10.1007/978-3-319-39110-6}

\bibitem{plasmeijer_task-oriented_2012}
Plasmeijer, R., Lijnse, B., Michels, S., Achten, P., Koopman, P.: Task-oriented
  programming in a pure functional language. In: Proceedings of the 14th
  symposium on {Principles} and practice of declarative programming. pp.
  195--206. ACM (2012)

\bibitem{santanna_safe_2013}
Sant'Anna, F., Rodriguez, N., Ierusalimschy, R., Landsiedel, O., Tsigas, P.:
  Safe system-level concurrency on resource-constrained nodes. In: Proceedings
  of the 11th {ACM} {Conference} on {Embedded} {Networked} {Sensor} {Systems}.
  p.~11. ACM (2013)

\bibitem{sawada_emfrp:_2016}
Sawada, K., Watanabe, T.: Emfrp: a functional reactive programming language for
  small-scale embedded systems. In: Companion {Proceedings} of the 15th
  {International} {Conference} on {Modularity}. pp. 36--44. ACM (2016)

\bibitem{serrano_type_2018}
Serrano, A.: Type {Error} {Customization} for {Embedded} {Domain}-{Specific}
  {Languages}. {PhD} {Thesis}, Utrecht University (2018)

\bibitem{st-amour_picobit:_2009}
St-Amour, V., Feeley, M.: {PICOBIT}: a compact scheme system for
  microcontrollers. In: International {Symposium} on {Implementation} and
  {Application} of {Functional} {Languages}. pp. 1--17. Springer (2009)

\bibitem{steiner_firmata:_2009}
Steiner, H.C.: Firmata: {Towards} {Making} {Microcontrollers} {Act} {Like}
  {Extensions} of the {Computer}. In: {NIME}. pp. 125--130 (2009)

\bibitem{wang_maintaining_2018}
Stutterheim, J., Achten, P., Plasmeijer, R.: Maintaining {Separation} of
  {Concerns} {Through} {Task} {Oriented} {Software} {Development}. In: Wang,
  M., Owens, S. (eds.) Trends in {Functional} {Programming}, vol. 10788, pp.
  19--38. Springer International Publishing, Cham (2018).
  \doi{10.1007/978-3-319-89719-6}

\bibitem{sugihara_programming_2008}
Sugihara, R., Gupta, R.K.: Programming models for sensor networks: {A} survey.
  ACM Transactions on Sensor Networks  \textbf{4}(2),  1--29 (Mar 2008).
  \doi{10.1145/1340771.1340774}

\bibitem{troyer_building_2018}
Troyer, de, C., Nicolay, J., Meuter, de, W.: Building {IoT} {Systems} {Using}
  {Distributed} {First}-{Class} {Reactive} {Programming}. In: 2018 {IEEE}
  {International} {Conference} on {Cloud} {Computing} {Technology} and
  {Science} ({CloudCom}). pp. 185--192 (Dec 2018).
  \doi{10.1109/CloudCom2018.2018.00045}

\bibitem{van_der_heijden_managing_2011}
Van Der~Heijden, M., Lijnse, B., Lucas, P.J., Heijdra, Y.F., Schermer, T.R.:
  Managing {COPD} exacerbations with telemedicine. In: Conference on
  {Artificial} {Intelligence} in {Medicine} in {Europe}. pp. 169--178. Springer
  (2011)

\bibitem{wand_continuation-based_1980}
Wand, M.: Continuation-based multiprocessing. In: Proceedings of the 1980 {ACM}
  conference on {LISP} and functional programming - {LFP} '80. pp. 19--28. ACM
  Press, Stanford University, California, United States (1980).
  \doi{10.1145/800087.802786}

\end{thebibliography}
